\documentclass[twocolumn]{aastex63}

\usepackage{commath} 
\usepackage{gensymb} 
\usepackage{eucal} 
\usepackage{graphicx}
\usepackage{subfigure}
\usepackage{natbib}
\usepackage{hyperref}
\usepackage{tabu}

\usepackage{movie15}

\usepackage{mathrsfs,amssymb,amstext}
\usepackage{ulem}
\usepackage{url}
\usepackage{bm}

\usepackage{color}
\usepackage{algorithm}
\usepackage{algorithmic}
\usepackage{listings}
\usepackage{color}
\definecolor{dkgreen}{rgb}{0,0.6,0}
\definecolor{gray}{rgb}{0.5,0.5,0.5}
\definecolor{mauve}{rgb}{0.58,0,0.82}
\def\mean#1{\left< #1 \right>}

\newcommand{\h}{\ensuremath{\, {\rm h}}}

\newcommand{\cm}{\ensuremath{\,{\rm cm}}}
\newcommand{\m}{\ensuremath{\,{\rm m}}}

\newcommand{\kpc}{\ensuremath{\,{\rm kpc}}}

\newcommand{\K}{\ensuremath{\, {\rm K}}}

\newcommand{\MHz}{\ensuremath{\, {\rm MHz}}}

\def\CII{\hbox{[C$\scriptstyle\rm II$]\,}}

\begin{document}

\title{An Ultra-long Wavelength Sky Model with Absorption Effect}

\correspondingauthor{Bin Yue; Yidong Xu; Xuelei Chen}
\email{yuebin@nao.cas.cn; xuyd@nao.cas.cn; xuelei@cosmology.bao.ac.cn}
\author{Yanping Cong}
\affiliation{National Astronomical Observatories, Chinese Academy of Sciences, 20A, Datun Road, Chaoyang District, Beijing 100101, China}
\affiliation{School of Astronomy and Space Science, University of Chinese Academy of Sciences, Beijing 100049, China}

\author{Bin Yue$^*$}
\affiliation{National Astronomical Observatories, Chinese Academy of Sciences, 20A, Datun Road, Chaoyang District, Beijing 100101, China}

\author{Yidong Xu$^*$}
\affiliation{National Astronomical Observatories, Chinese Academy of Sciences, 20A, Datun Road, Chaoyang District, Beijing 100101, China}

\author{Qizhi Huang}
\affiliation{National Astronomical Observatories, Chinese Academy of Sciences, 20A, Datun Road, Chaoyang District, Beijing 100101, China}

\author{Shifan Zuo}
\affiliation{Department of Astronomy, Tsinghua University, Beijing 100084, China}
\affiliation{National Astronomical Observatories, Chinese Academy of Sciences, 20A, Datun Road, Chaoyang District, Beijing 100101, China}
 
\author{Xuelei Chen$^*$}
\affiliation{National Astronomical Observatories, Chinese Academy of Sciences, 20A, Datun Road, Chaoyang District, Beijing 100101, China}
\affiliation{School of Astronomy and Space Science, University of Chinese Academy of Sciences, Beijing 100049, China}
\affiliation{Center of High Energy Physics, Peking University, Beijing 100871, China}

\begin{abstract}

The radio sky at frequencies below $\sim$10 MHz is still largely unknown, this remains the last unexplored part
of the electromagnetic spectrum in astronomy. The upcoming space experiments aiming at such low frequencies (ultra-long wavelength or ultra-low frequency)  would benefit from reasonable expectations of the sky brightness distribution at relevant frequencies. In this work, we develop a radio sky model that is valid down to $\sim 1 \MHz$. In addition to the discrete HII objects, we take into account the free-free absorption by thermal electrons in the Milky Way's warm ionized medium (WIM). This absorption effect becomes obvious at $\lesssim10$ MHz, and could make the global radio spectrum turn over at $\sim3$ MHz. Our sky map shows unique features at the ultra-long wavelengths, including a darker Galactic plane in contrast to the sky at higher frequencies, and the huge shadows of the spiral arms on the sky map. It would be a useful guidance for designing the future ultra-long wavelength observations. Our Ultralong-wavelength
Sky Model with Absorption (ULSA) model could be downloaded at \href{https://doi.org/10.5281/zenodo.4454153}{DOI:10.5281/zenodo.4454153}.

\end{abstract}


\section{introduction}
 
The radio sky has been surveyed from $\sim$10 MHz up to $\sim$THz.
Based on data of surveys at the relevant frequencies, full-sky maps were produced at 408 MHz \citep{haslam1982408,remazeilles2015improved}, 1.42 GHz \citep{Reich1982,Reich1986,Reich2001}, and a number of higher frequencies
 from 22.8 GHz to 857 GHz by WMAP and Planck satellites,  see \citet{WMAP5yr,Planck2015_I}. In addition, there are also surveys which cover a large portion but not the 
 full sky. In order to produce sky maps at frequencies in which the full sky have not be surveyed,  
 a number of sky models have been developed, based essentially on 
 interpolation or extrapolation of the available data, and statistical modeling of the sky intensity distribution.  Over much of the radio 
 wave band, the sky intensity has a nearly power law spectrum, making the interpolation/extrapolation relatively simple and accurate. 
These includes  the Global Sky Model (GSM) \citep{deOliveira-Costa2008GSM}, its improvements \citep{PyGSM2016,Zheng2017,GMOSS2017,eGSM2018}, and the Self-consistent Sky Model (SSM) \citep{Huang2019SSM}. 
They are very useful in the design of new instruments, study of observation strategies,  and testing foreground removal/mitigation methods (e.g., \citealt{Shaw2014,Zuo2019}).

However,  the extrapolation becomes trickier at frequencies $\lesssim10$ MHz, where the absorption of the interstellar medium (ISM) becomes significant, and at the same time there is a dearth of observation data. 
Observations below 10 MHz is severely hampered by the absorption and 
distortions of the ionosphere \citep{Jester2009}. Until now there are only a few ground-based observations performed during 
$\sim$1950s $-$ 1960s \citep{George2015,Cane1977}, and a few space-based low-resolution observations 
performed during $\sim$1960s $-$ 1970s \citep{Alexander1969, Brown1973,Alexander1974,Alexander1975,Cane1979}. 
Below we shall call this wave band the ultra-long wavelength or ultra-low frequency band, though these are not the name used by the 
radio engineers. The ultra-long wavelength sky is still a largely unexplored regime in the astronomical observations.
To open this new window in the electromagnetic spectrum, a number of space missions, e.g. the
lunar-orbit array for Discovering the Sky at the Longest wavelength (DSL) 
\citep{ChenXL2019DSL}, and the lunar-surface based FARSIDE \citep{FARSIDE}  
have been proposed, with the aim of obtaining high resolution\footnote{At such low frequencies, the resolution is limited by the scattering in the interstellar medium and interplanetary medium. This angular scattering limit is roughly arc-minute level at 1 MHz and roughly scales with $\propto \nu^{-2}$, see \citet{Jester2009}.  } full-sky maps at ultra-long wavelengths.

 However, to design such low frequency experiments and observe the sky at this largely unexplored part of electromagnetic spectrum, 
one needs a reasonable estimate of the sky brightness at these frequencies, and a fair expectation of the structures, 
so that the required system gain and dynamic range can be derived. More importantly, at present or in the near future, a lunar-orbit or 
lunar-surface based array necessarily has a limited number of antenna elements (though the orbital precession could improve the {\it uvw} coverage at the price of longer observation time), and the antennas are all-sky sensitive, these make the imaging process quite challenging, especially when the sky brightness itself varies dramatically with 
frequency and direction, as we will show below. It is therefore crucial to have a reasonable full-sky model for these 
unexplored frequencies, which could serve as a starting point for the deconvolution process, 
for both end-to-end simulations in designing such experiments and the upcoming real observations.

At $\lesssim 10$ MHz, the existing sky models based on extrapolation are grossly inadequate, 
because the free-free absorption and synchrotron self-absorption \citep{Orlando2013,Ghisellini2013} become
very significant. In the Galactic diffuse ISM the former is much more important \citep{scheuer1953,shain1959iau,hurley2017galactic}. Free-free absorption is the inverse process of bremsstrahlung (free-free) radiation, so it is proportional to the square of free electron density.  The absorption level depends not only on the smoothed distribution, but also on the small-scale turbulence.

Since as early as 1960s, it has been discovered that the global radio background (the sum of Galactic and extragalactic radiation) spectrum has a downturn at $\sim3-5$ MHz, by ground-based telescope (e.g. \citealt{Ellis1966}), by radio-astronomy rocket (e.g.  \citealt{Alexander1965}), and by  space satellites (e.g. Ariel II \citealt{Smith1965}; RAE-1 \citealt{Alexander1969}; IMP-6 \citealt{Brown1973};  and RAE-2 \citealt{Novaco1978}), and free-free absorption is probably the reason of this downturn in spectrum.
However, it is also possible that the intrinsic  energy spectrum of the cosmic ray particles becomes softer at low energy \citep{strong1998propagation}. More references about early low frequency observations could be found in \citet{Cane1979}.

For dense electron clumps, such as those in discrete HII regions, the free-free absorption could make these objects opaque even for radio signal at higher frequencies (e.g. \citealt{Odegard1986,Kassim1988}). This effect can be used to separate the synchrotron emissivity behind and in front of such regions. If there are sufficient number of such HII regions with known distance, the 3D distribution of the synchrotron emissivity could be reconstructed. This has been successfully put into practice, see e.g. \citet{Nord2006,Hindson2016,Su2017,Su2018,Polderman2019} and references therein. Since the synchrotron radiation is produced by the cosmic ray electrons in magnetic field, the cosmic ray electron distribution can be inferred if the magnetic field is known by other means \citep{Polderman2020}. The degeneracy between the cosmic ray and the magnetic field could also be broken by combining with gamma-ray observations  \citep{Nord2006}, as the Galactic gamma-ray is mainly from the collisions between the cosmic ray particles and ISM particles \citep{Nava2017}, and the ISM density profile could be constructed from 21cm, CO, and dust thermal emission observations \citep{Ackermann2012}.

A sky model for the ultra-long wavelengths has to take into account the free-free absorption so as to give a reasonable prediction for the upcoming observations \citep{Reynolds1990}.
In this work, we  develop an observation-based sky model that is still valid  below $\sim$ 10 MHz, 
by including the free-free absorption effect.
We start with a description of our methods in Sec. \ref{method}, which incorporate a fitting result for the 
Galactic synchrotron spectral index, a fitted Galactic synchrotron emissivity, and an adopted electron density distribution that
is responsible for the free-free absorption. The main results are given in Sec. \ref{result},
including the model with constant spectral index, the model with frequency-dependent spectral index and 
the one with direction-dependent spectral index. At last, we summarize our main conclusions in Sec. \ref{conclusion}. 
A  brief description of our software is presented in Appendix \ref{sec:software}.

\section{methods}\label{method}
 
The free-free absorption becomes significant at low frequencies. 
The optical depth can be written as the 
integration of the absorption coefficient, $\kappa_{\nu}$, along the line of sight, 
\begin{equation}
\tau(\nu,s) = \int_{0}^{s}\kappa_{\nu}\left(n_{\rm e},T_{\rm e}\right)ds^{\prime},
\end{equation}
where $n_{\rm e}$ and $T_{\rm e}$ are the electron density and temperature at $s'$ respectively.
For constant electron temperature, the following approximation holds \citep{condon2016essential},
\begin{equation}
\tau_{\nu} \approx 3.28 \times 10^{-7}\left(\frac{T_e}{10^4\ {\rm K}}\right)^{-1.35}\left(\frac{\nu}{\rm GHz}\right)^{-2.1}{\rm \left(\frac{EM}{pc\ cm^{-6}}\right)},
\label{optical_depth_eq}
\end{equation}
where the emission measure (EM) is the integration of electron density squared at {\it all scales},
\begin{equation}
{\rm \frac{EM}{pc\ cm^{-6}}} = \int \left (\frac{n_e}{\rm cm^{-3}}\right)^{2}\left(\frac{d s}{\rm pc}\right).
\end{equation}
The ISM inside the Milky Way, the circum-galactic medium (CGM), and the intergalactic medium (IGM) can all absorb low frequency 
radio waves. We can make some simple estimates to assess the absorption on the 
various astrophysical scales. 

Regarding the CGM, \citet{Voort2018} has investigated the distribution  of the hydrogen surrounding 
Milky Way-size galaxies from high resolution zoom-in cosmological hydrodynamic simulations with star formation. They presented 
the column number density of the hydrogen as a function of $r$ in their Figure 2. 
We adopt a similar radial density profile, and assume that the free electron density follows the same distribution, 
with a clumping factor of $\sim 3$ and CGM electron temperature of $\sim10^5$ K. We find $\tau_{\rm CGM}=0.9, 0.09$ and 0.007 for 1, 3 and 10 MHz respectively. If, however, there are some cooler dense clumps in the CGM, the absorption would be stronger. Thus, the CGM is nearly transparent above 1 MHz, but becomes opaque below 1 MHz.

The EM of the IGM up to a redshift $z$ is  
\begin{equation}
{\rm EM}= \int_0^z \frac{\kappa_{\nu'}}{\kappa_\nu}C_{\rm IGM}(z') \bar{n}^2_{\rm e}(z')dr_p',
\end{equation}
where $C_{\rm IGM}(z')$ is the clumping factor at $z'$, and $dr_p'=cdz'/[(1+z')H(z')]$ is the proper distance element at $z'$. Also 
$\bar{n}_{\rm e}(z)=f_{\rm e}(z)\bar{n}_{\rm H}(1+z)^3$ is the mean free electrons density at $z$, with  
$\bar{n}_{\rm H}\approx1.89\times10^{-7}$ cm$^{-3}$ the mean hydrogen number density at present.
 Note that to use the EM as in Eq. (\ref{optical_depth_eq}),
we add the factor $\kappa_{\nu'}/\kappa_\nu$ to correct for the redshift-dependence of the absorption for a given observing 
frequency at $\nu$. Since $\kappa_\nu\propto \nu^{-2.1}$ \citep{condon2016essential} and $\nu'=\nu(1+z')$, we have $\kappa_{\nu'}/\kappa_\nu=(1+z')^{-2.1}$, hence
\begin{align}
{\rm EM}&= \int_0^z   ( 1+z')^{-2.1} C_{\rm IGM}(z') \bar{n}^2_e(z')  \frac{cdz'}{(1+z')H(z')} .   \nonumber \\
\end{align}

\begin{figure}
	\centering
	\includegraphics[width=0.4\textwidth]{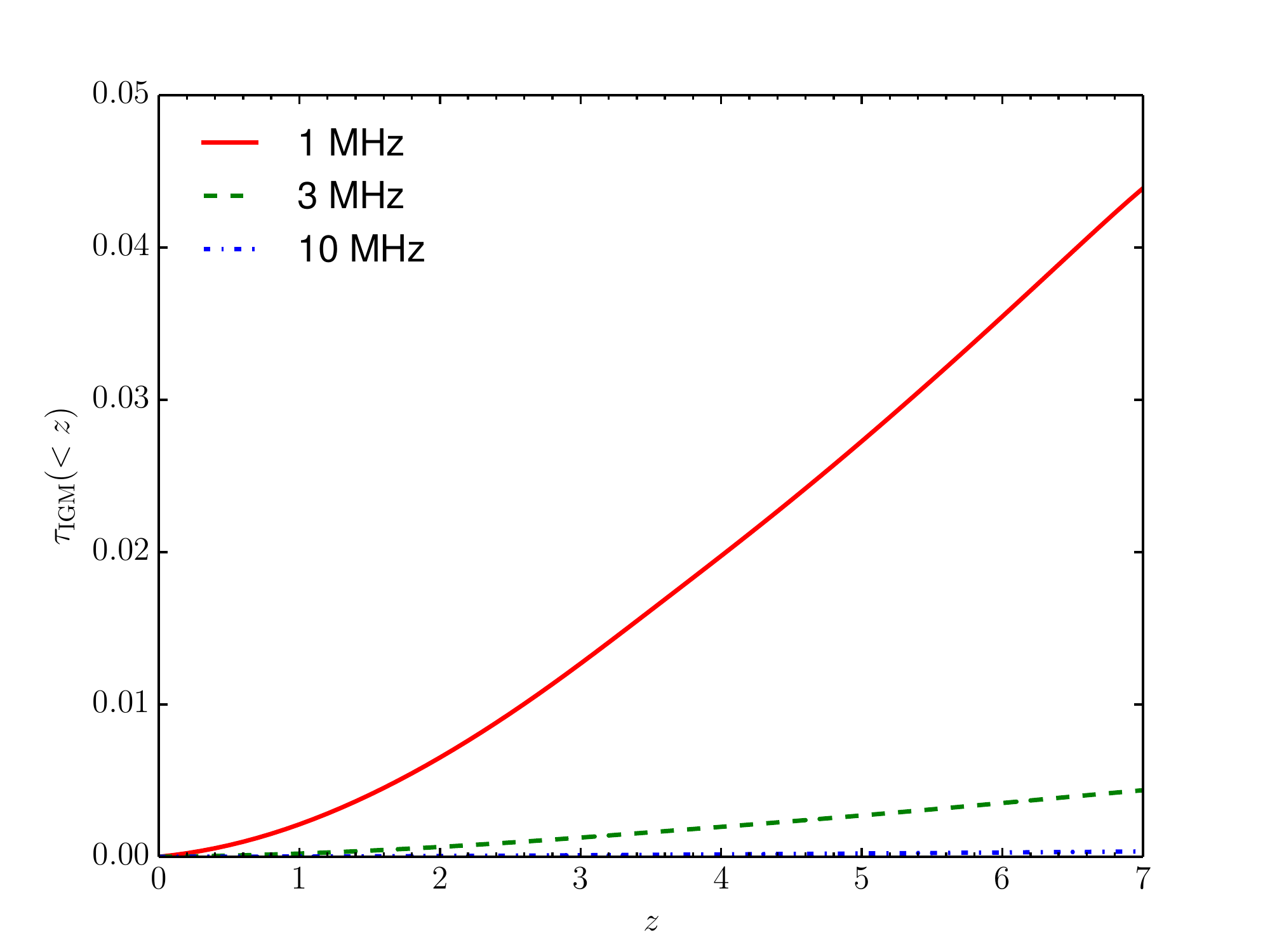}
	\caption{The IGM mean optical depth for $\nu = 1, 3, 10 \MHz$ to redshift $z$. }
	\label{fig:tauigm}
\end{figure}

 We use the tanh model to describe the reionization history for both HI and HeI, and HeII, i.e.
\begin{align}
f_{\rm e}^{\rm HI,HeI}=\frac{1+n_{\rm He}/n_{\rm H}}{2} \left[1+\tanh\left( \frac{ y_{\rm re}-y}{\Delta y} \right)\right],
\end{align}
\begin{align}
f^{\rm HeII}_{\rm e}=0.5\left[1+\tanh\left( \frac{ y_{\rm re}-y}{\Delta y} \right)\right],
\end{align}
and
\begin{align}
f_{\rm e}&=f_{\rm e}^{\rm HI,HeI}+(n_{\rm He}/n_{\rm H})f_{\rm e}^{\rm HeII}.
\end{align}
Here $y(z)=(1+z)^{3/2}$, $y_{\rm re}=y(z_{\rm re})$, $\Delta y=3/2(1+z_{\rm re})^{1/2}\Delta z$, 
and $n_{\rm He}/n_{\rm H}\approx0.083$.
For HI and HeI, we assume $z_{\rm re}=7.68$ and $\Delta z =0.5$, and for HeII, we adopt $z_{\rm re}=3.5$ and $\Delta z=0.5$ \citep{Planck2018_VI}.
We use the clumping factor model that fits reionization simulations presented in \citet{Iliev2007}, but renormalize it to be 3.0 at $z=5$, i.e.
\begin{equation}
C_{\rm IGM}(z)=6.8345\exp(-0.1822z+0.003505z^2).
\end{equation}
The IGM free-free absorption optical depths at frequencies 1, 3 and 10 MHz are shown in Fig. \ref{fig:tauigm} as a function of $z$, 
it is not large even at 1 MHz. 

Next we consider the ISM, which is much denser than the IGM and CGM. The observed sky brightness temperature in the direction given by Galactic coordinates $(l,b)$ is the 
sum of the Galactic radiation $T_{\rm G}(\nu,l,b)$ and the extragalactic background $T_{\rm E}(\nu)$,
\begin{align}
T(\nu,l,b)&=T_{\rm G}(\nu,l,b) +T_{\rm E}(\nu)    \nonumber \\
&=\int_0^{s_{\rm G}} \epsilon(\nu, R, Z, \phi) e^{-\tau(\nu,s)}ds +   T^{\rm iso}_{\rm E}(\nu)e^{-\tau(\nu,s_{\rm G})}   ,
\label{eq:specific_intensity}
\end{align}
 where $s_{\rm G}$ is the maximum distance to the edge of the Milky Way along any line of sight, $\epsilon$ is the  
three-dimensional
radiation emissivity in Galactic-centric cylindrical coordinates $(R,Z,\phi)$,
and $\tau(\nu,s)$ is the free-free absorption optical depth integrated up to a distance of $s$.
Here we set $s_{\rm G} = 50 \kpc$. 
$T^{\rm iso}_{\rm E}$ is the extragalactic radio background for which we assume to be isotropic in the first approximation.  
With models of an extragalactic radio background $T^{\rm iso}_{\rm E}$ at ultra-long wavelengths, a three-dimensional 
emissivity $\epsilon(\nu, R, Z, \phi)$, and a three-dimensional distribution of free electrons in the Milky Way $n_{\rm e}$, we 
can derive the sky map at the required frequencies from Eq. (\ref{eq:specific_intensity}).
In the following subsections, we describe the modeling of $T^{\rm iso}_{\rm E}$, $\epsilon$, and $n_e$ respectively.

\subsection{The isotropic extragalactic background}

The extragalactic background is assumed to be largely isotropic, and a summary of some past results 
is given in \citealt{guzman2011all}. Two methods are commonly used to extract it from the observed sky maps \citep{Kogut2011,seiffert2011interpretation}. In the first approach the Galactic radiation is described by a plane-parallel model, so the total radiation is 
\begin{equation}
T(\nu)=T^{\rm iso}_{\rm E}(\nu) + T_{\rm G}(\nu) \times {\rm csc}|b|,
\end{equation}
 where both $T^{\rm iso}_{\rm E}$ and $T_{\rm G}$ depend only on frequency. In the second approach, the Galactic radiation is derived from a tracer, usually the Galactic \CII emission, say 
 \begin{equation}
 T(\nu)=\sum_{i=1}^2 T^{\rm iso}_{{\rm E},i}(\nu) + a_i(\nu) \times \sqrt{I_{\rm CII}},
 \end{equation}
 where $I_{\rm CII}$ is from the CORE/FIRAS measurements. In practice, all coefficients are fitted from observed radio maps and these two methods give consistent results \citep{Kogut2011,Dowell2018}.  
However,  whether the extracted isotropic background really originates from extragalactic sources is still  being debated. \citet{seiffert2011interpretation} extracted a background which is larger than the integrated radio emission of external galaxies \citep{Gervasi2008}. The excess could be from unknown extragalactic source populations, or could be unaccounted Galactic foreground  (see also \citealt{Subrahmanyan2013}).

Here we assume that the isotropic background is from extragalactic sources. We adopt the following simple model of 
isotropic extragalactic background,
\begin{equation}
T^{\rm iso}_{\rm E}(\nu)=1.2\left(\frac{\nu}{\rm 1~GHz}\right)^{-2.58}~[{\rm K}].
\label{eq:iso1}
\end{equation}
This is the fit obtained by the ARCADE-2 balloon experiment, which observed the absolute brightness temperature of the sky between 3 and 90 GHz \citep{seiffert2011interpretation},  minus the cosmic microwave background (CMB). This is in good agreement with the fit obtained from the Long Wavelength Array (LWA) \citep{Dowell2018} below 100 MHz. Strictly speaking, ``background" is the cumulative radiation from unresolved sources, so it depends on the sensitivity and angular resolution of the telescope. For simplicity we directly use the background fit given by the ARCADE-2 experiment, since in our model we do not specify any particular instrument.

In addition to the absorption by our Milky Way, the extragalactic radio background would also be absorbed by the ISM in the host galaxies. Moreover, the spectrum of relativistic electrons could become softer at lower energy,  resulting in softer extragalactic radio background \citep{Protheroe1996}. 
In a recent estimate for the extragalactic radio background, \citet{Nitu2021} found the extragalactic radio background deviates from power-law below $\sim 1$ MHz.  In Sec. \ref{freq-depend} we investigate the case that the extragalactic radio background is not a power-law form.

\subsection{The Galactic free-free emission}\label{sec:ff}

\begin{figure}
	\centering
	\includegraphics[width=0.5\textwidth]{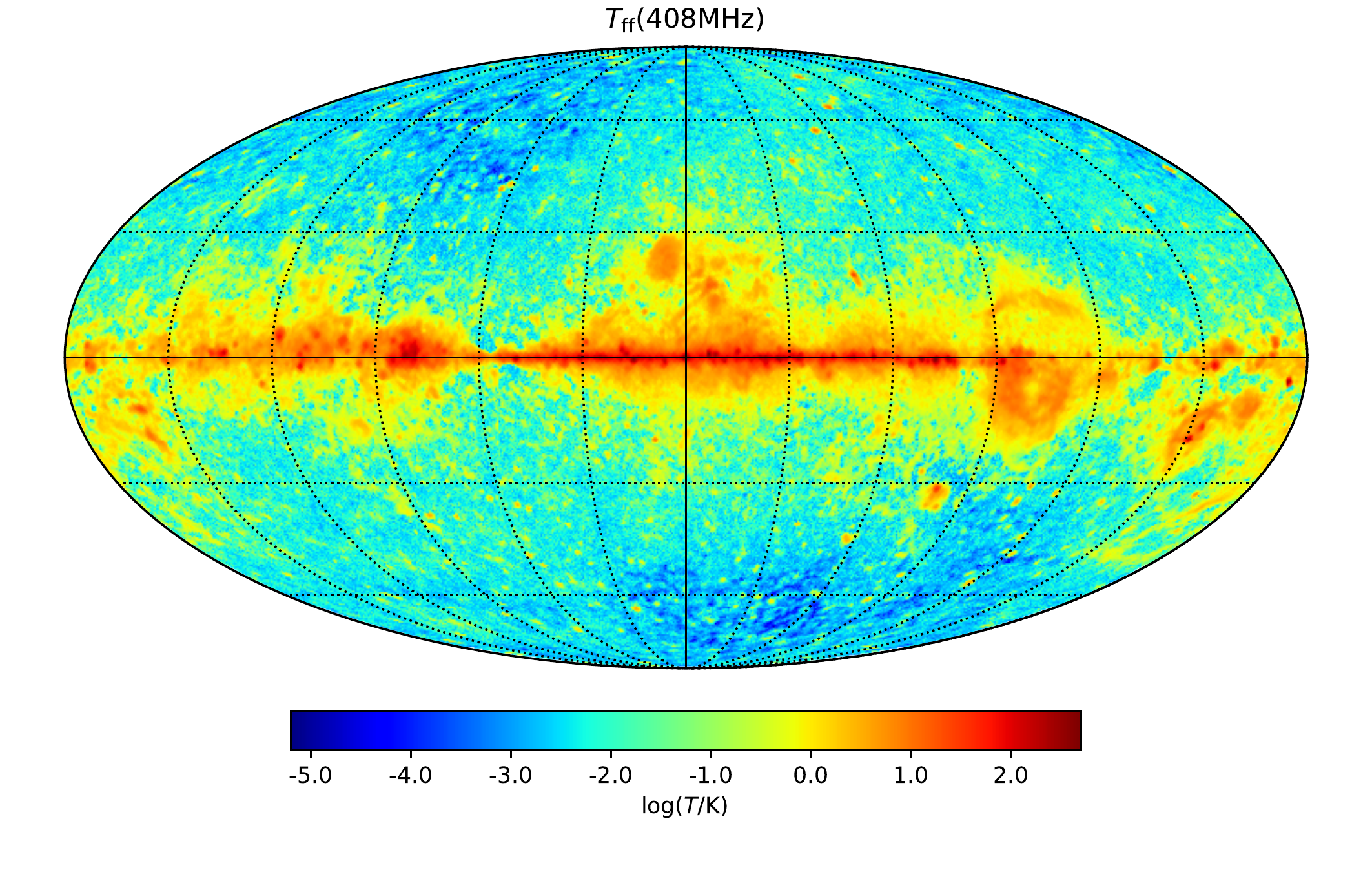}
	\includegraphics[width=0.4\textwidth]{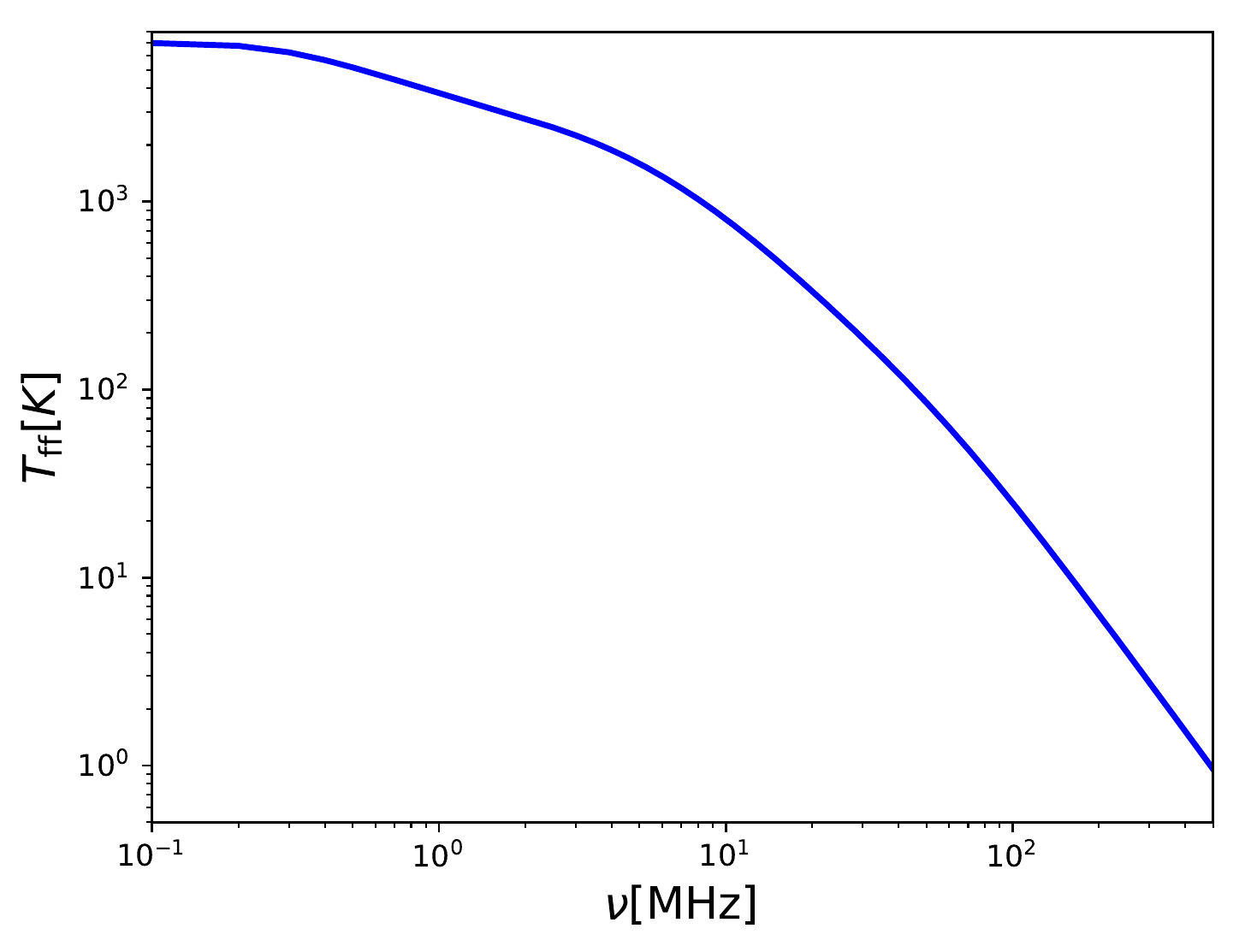}	
	\caption{ {\it Top:} the Galactic free-free emission map at 408 MHz. {\it Bottom:} the spectrum of the mean Galactic free-free emission. }
	\label{fig:ff}
\end{figure}

The free-free emission also contributes to the low frequency sky brightness and should be subtracted from the observed maps for the purpose of assessing Galactic synchrotron. A free-free template could be constructed by using the H$\alpha$ emission line as a tracer. After correcting the absorption and scattering, the intrinsic H$\alpha$ flux could be converted into free-free flux \citep{Dickinson2003,Lian2020}. Moreover, it can also be constructed from the multi-frequency microwave observations. In this work, we use the EM and $T_e$ maps constructed from {\tt Planck} multi-frequency observations \citep{Planck2015_X}\footnote{\url{http://pla.esac.esa.int/pla/aio/product-action?MAP.MAP_ID=COM_CompMap_freefree-commander_0256_R2.00.fits}} to derive the free-free map, say
\begin{align}
g_{\rm ff}(\nu)&=\ln\left\{\exp\left[5.960-\frac{\sqrt{3}}{\pi}\ln\left(  \nu_9\, T_4^{-1.5} \right)\right]+\exp(1)\right\} \nonumber \\
\tau_{\rm ff}(\nu)&=0.05468\, T_e^{-1.5}\,\nu_9^{-2}\,{\rm EM}\, g_{\rm ff} \nonumber \\
T_{\rm ff}(\nu)&=10^6\, T_e[1-\exp(-\tau_{\rm ff})],
\label{eq:free-free}
\end{align}
where $\nu_9$ is the frequency in units $10^9$ Hz and $T_4$ is the electron temperature in units $10^4$ K \citep{Draine2011}. 
In the top panel of Fig. \ref{fig:ff}, we show the 408 MHz Galactic free-free emission map constructed from the {\tt Planck} observations. 
The bottom panel shows the Galactic free-free spectrum. Clearly, most of the free-free emission is from the Galactic plane.

Comparing the above free-free map with observations at 408 MHz (see next subsection), we find that the free-free emission contributes 
$4\%$ of the mean observed sky brightness. At $|b|>10\degree$, the free-free fraction decreases to $1\%$. If we smooth both maps with a resolution of 10$\degree$, then the r.m.s. contributed by the free-free emission is $15\%$ of the total, while at $|b|>10\degree$  the r.m.s. fraction decreases to $6\%$. This is consistent with \citet{Dickinson2003}.

\subsection{The absorption-free sky map}

\label{sec:absorption-free}

We first derive the synchrotron spectral index from the observed sky maps at frequencies where absorption is assumed to be negligible.
At low frequencies, there are full-sky maps at 408 MHz \citep{haslam1974further,haslam1981408,haslam1982408,remazeilles2015improved}
and 1.42 GHz \citep{Reich1982,Reich1986,Reich2001}. 
Here we use the free-free subtracted Haslam 408 MHz map as the base, assuming that in the absence of absorption, the brightness temperature of the Galactic synchrotron radiation follows a power-law form \citep{platania1998determination,Kogut2012,mckinley2018measuring}, 
\begin{equation}
T_{\rm G}(\nu)=T_{\rm G}(\nu_*)\left( \frac{\nu}{\nu_*} \right)^{\beta_{\rm G}},
\label{eq:T_G_extrapolated}
\end{equation}
where $\nu_*=408$ MHz.
We determine $\beta_{\rm G}$ by minimizing
\begin{align}
\chi^2&=\sum_{\nu_j}\sum_i \frac{1}{T^2_i(\nu_j)} \nonumber \\
&\times \left[T_{\rm E}(\nu_j) + T_{{\rm G},i}(\nu_*)\left( \frac{\nu_j}{\nu_*} \right)^{\beta_{\rm G}}+T_{{\rm ff},i} (\nu)- T_i(\nu_j)\right]^2,
\label{eq:Y2}
\end{align}
where  $T_i(\nu_j)$ is the observed brightness temperature of the $i$-th pixel in the sky-map at frequency $\nu_j$, 
$T_{{\rm G},i}(\nu_*)$ is the Galactic radiation for the $i$-th pixel at $\nu_*$. 
The summation is performed for all available pixels at all observed frequencies of our selected data.

Below several hundred MHz, there are observed sky maps, each of which covers a part of the sky. 
In determining the spectral index, in addition to the Haslam 408 MHz map, we use a selection of observations that have large sky coverages:  
the 45 MHz sky map in combination of two observations \citep{guzman2011all}, and the 35 MHz\footnote{Using Eq. (\ref{optical_depth_eq}) and electron model in Sec. \ref{sec:electron-model}, we check that at this frequency, for most of the sky regions the free-free absorption is negligible, except at a very thin plane with $b\sim0\degree$ and near the Galactic center where the optical depth could be up to $\sim1$ , so it is safe to include this low frequency observations when constructing the absorption-free map.}, 38 MHz, 40 MHz, 50 MHz, 60 MHz, 70 MHz, 74 MHz, and 80 MHz maps observed by the LWA \citep{dowell2017lwa1}. 
The Guzman 45 MHz map covers 96\% of the sky, with resolutions of 
$4.6\degree \times2.4\degree$ and  $3.6\degree\times3.6\degree$ for the southern and northern sky respectively. 
It combines two surveys performed by devices of similar characteristics. 
All of the LWA maps cover a total of about 80\% of the sky, with resolutions ranging from $4.8\degree \times 4.5\degree$ to $2.1\degree \times 2.0 \degree$.
As the different maps have different angular resolutions, we  smooth all the maps with a beam with FWHM $= 5\degree$  when deriving the power-law index.  For all the maps we subtract the free-free emission given in Sec. \ref{sec:ff}.
Our fiducial model assumes that the spectrum index is a constant, but we will also investigate a model with a frequency-dependent spectral index in Sec. \ref{freq-depend} and a model with direction dependence in the spectral index in Sec. \ref{sec:spatial-variation}.

\begin{figure}
	\includegraphics[width=0.49\textwidth]{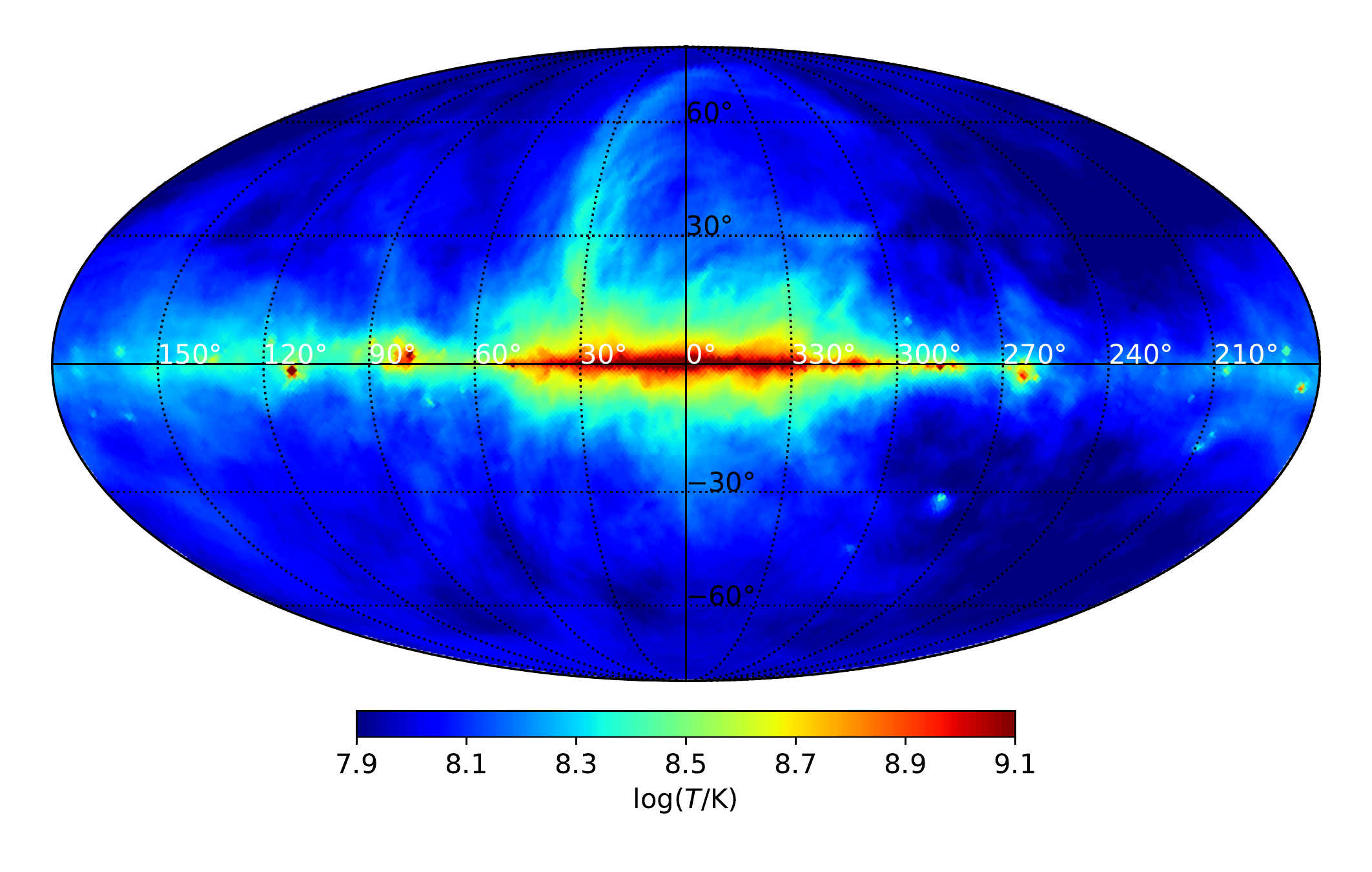}
	\caption{Simple extrapolation of the free-free subtracted Haslam 408 MHz map to 1 MHz without absorption. 
	}
	\label{fig_absorptionfreeskymap1mhz}
\end{figure}

Using the above method, we derive a spectral index $\beta_{\rm G}=-2.51$ for the Galactic synchrotron radiation. 
This is actually quite similar to the spectral index of the extragalactic background, $\beta_{\rm E} = -2.58$.  
In Fig. \ref{fig_absorptionfreeskymap1mhz} we plot the 1 MHz sky map that is extrapolated from free-free subtracted Haslam 408 MHz map using $\beta_{\rm G}=-2.51$ for the Galactic component and $\beta_{\rm E} = -2.58$ for the extragalactic component. As this is a simple power-law extrapolation, its structure is the same as the Haslam 408 MHz map, except for different temperature scale. This map could be compared with the absorption-included maps presented later in this paper.

\subsection{The Galactic emissivity model}

We tentatively adopt an axisymmetric form for the Galactic emissivity $\epsilon$.
In the cylindrical coordinates, the emissivity can be written as
\begin{equation}
	\epsilon(\nu,R,Z) = A \left(\frac{R+r_1}{R_0}\right)^\alpha
	e^{-R/R_0} e^{-\abs{Z/Z_0}^{\gamma}}\left( \frac{\nu}{\nu_*} \right)^{\beta_{\rm G}}.
	\label{eq:emissivity}
\end{equation}
Here $r_1$ is a small cut off radius to avoid singularity at the Galactic center, and we take $r_1=0.1 \kpc$.
The five frequency-independent free parameters: $A$, $R_0$, $\alpha$, $Z_0$ and $\gamma$ will be obtained by fitting the free-free subtracted Haslam 408 MHz map. The parameter $\beta_{\rm G}$ is generally a function of frequency and spatial position. 
Below we take the case of a constant $\beta_{\rm G}$
as our fiducial model, but will also discuss the cases of a frequency-dependent $\beta_{\rm G}$
and a direction-dependent $\beta_{\rm G}$.

Before deriving the emissivity parameters, we mask the Loop I and North Polar Spur (NPS) regions. They  
are generally believed to be nearby objects (see e.g., \citealt{wolleben2007new}), which do not suffer from the 
absorption on the Galaxy scale, hence one should not use these to derive the parameters of the  
Galactic emission and absorption model. We fit the emissivity parameters from free-free subtracted Haslam 408 MHz map and the fitted parameters are listed in Table \ref{table_params}.

 \begin{table}[ht] 
	\caption{Fitted Galactic emissivity model parameters}  
	\centering    
	\begin{tabular}{l c l l l l l}   
		\hline\hline 
		$A$  & $43.10$ $\rm K \,kpc^{-1}$\\
		$R_0$  & 3.41  kpc \\
		$\alpha$ & 0.46   \\
		$Z_0$ 	&1.12 kpc	\\
		$\gamma$	&1.23 		\\
 		\hline
	\end{tabular}
	\label{table_params}
\end{table}

\subsection{The Galactic free electron distribution}\label{sec:electron-model}

Most free electrons in the ISM are either in the warm ionized medium (WIM), with typical density of
$\sim0.01-0.1$ cm$^{-3}$ and typical temperature of $\sim 10^4 \K$ \citep{gaensler2008vertical,deAvillez2012}, 
or in the hot ionized medium (HIM), with typical density of $\sim 10^{-3} \cm^{-3}$ and typical temperature of
$\sim 10^5-10^6 \K$ \citep{Ferriere2001}. The free-free absorption is dominated
by the WIM. In addition, there are some dense HII regions with typical density of $\sim10^2 \cm^{-3}$ and typical temperature of several thousand K \citep{Hindson2016}. They are almost opaque to the low-frequency radio radiation, and hence good for separating the contributions to the synchrotron radiation from different distances along lines of sight \citep{Su2017,Su2018,Polderman2019}. Moreover, around the dense HII regions, there could be some extended HII region envelopes (EHEs), as inferred from the 
observations \citep{Anantharamaiah1986,Kassim1989b}. They have smaller density ($\sim0.5-10$ cm$^{-3}$) but larger size ($\sim0.05-0.2$ kpc) than the classical HII regions. These could also potentially contribute more to the absorption.

\begin{figure}[tbp]
	\centering
	\includegraphics[width=0.49\textwidth]{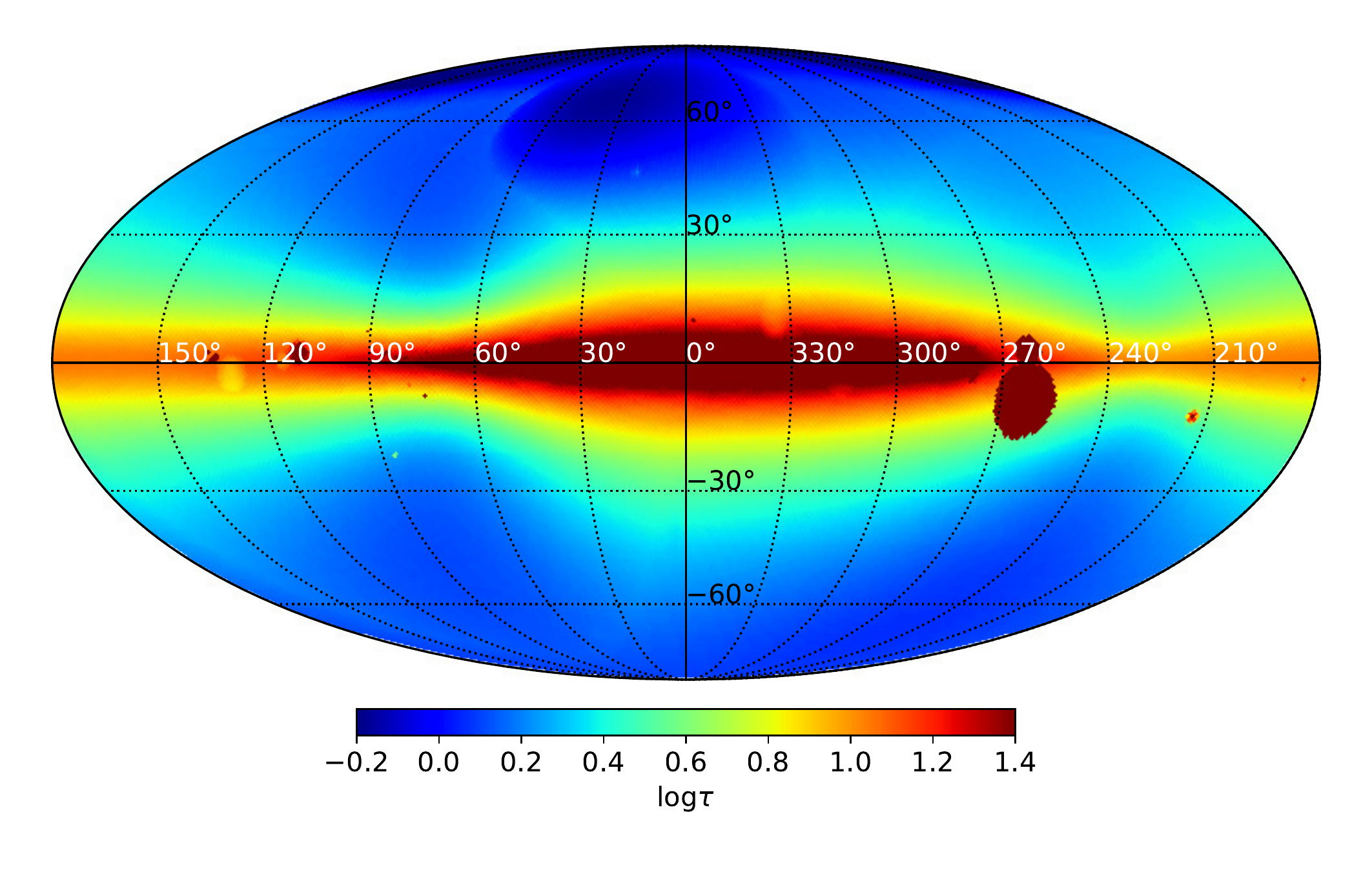}
	\caption{ The free-free absorption optical depth toward extragalactic sources at 1 MHz based on the {\tt NE2001} model. }
	\label{fig:fixed_distance_optical_depth}
\end{figure}

There are a variety of models for the Galactic electron distribution (e.g.
\citealt{gaensler2008vertical,gomez2001reexamination,taylor1993pulsar,cordes2002ne2001,cordes2003ne2001};\citealt{YMW16} (YMW16); \citealt{schnitzeler2012modelling}). 
In the present work, we mainly focus on the general effect of the absorption,
so we adopt the {\tt NE2001} model \citep{cordes2002ne2001,cordes2003ne2001} for the electron density,
which is one of the most well-accepted models. 
In {\tt NE2001}, the Galactic free electrons have five components, i.e., smooth components including a thick disk, 
a thin disk and five spiral arms; the Galactic center component; the local ISM; a list of known dense clumps;
and a list of voids.  The dense clumps are mainly HII regions around massive OB stars or  supernova remnants (SNRs).
The voids are low density regions found  between the Sun and some pulsars. 
The smooth components contain lots of free electron clouds and their volume filling factor is $\eta$. The could-by-cloud density fluctuations is described by a parameter 
\begin{equation}
\zeta=\frac{ \mean{n^2_{\rm c}}}{n_{\rm smooth}^2},
\end{equation}
where $n_{\rm c}$ is the free electron  density of the cloud, $\mean{}$ denotes cloud-by-cloud average, $n_{\rm smooth}$ is the density of the smooth component.  Inside each cloud, the free electrons also have small scale structures below a largest scale $L_0$.
It is assumed that the power spectrum of such electron density fluctuations follows a power-law distribution 
with the  index $-11/3$  as in the classical Kolmogorov turbulence model \citep{cordes2002ne2001}. 
Inside a cloud the small scale free electron fluctuations is described by the fractional variance
\begin{equation}
\omega^2=\frac{ \overline{ (\delta n_e)^2} }{ n^2_c},
\end{equation}
where the bar represents the average inside the cloud for all scales below $L_0$.
The emission measure is then
\begin{eqnarray}
{\rm EM}&=&\int \eta^{-1}\zeta \mean{\omega^2} n^2_{\rm smooth}{\rm d}s.\\
&=&\int  L_0^{2/3} F n^2_{\rm smooth}{\rm d}s.
\end{eqnarray}
where the fluctuation parameter $F$ is defined as 
\begin{equation}
F = L_0^{-2/3}\omega^2 \zeta \eta ^{-1},
\end{equation}
Hence, given the smooth density distribution by {\tt NE2001}, the fluctuation parameter determines the strength of absorption/emission related to the density squared.
 The fluctuation parameter can be derived from the observed scattering measure (SM) of Galactic pulsars 
 \citep{cordes1985small,cordes1991interstellar,cordes1991galactic,taylor1993pulsar} and extragalactic AGNs and GRBs \citep{cordes2003ne2001}. In {\tt NE2001}, the fluctuation parameter is given for each of the three smooth components. 
 The default values for the thick disk, thin disk and the spiral arms are $F_1=0.18$, $F_2=120$ and $F_a=5$ respectively. 
 $F_1$ is the most relevant one for our purpose,  while $F_2$ and $F_a$ are for components on the Galactic plane.

The free-free absorption also depends on the electron temperature (see Eq. \ref{optical_depth_eq}). 
The temperature of the WIM is generally $\sim 8000$ K \citep{gaensler2008vertical}. 
Throughout this paper we adopt a constant of $T_e =$ 8000 K for all Galactic electron components.  The result is actually not 
sensitive to this.
In Fig. \ref{fig:fixed_distance_optical_depth} we plot the optical depth of the free-free absorption 
toward extragalactic sources, i.e., $\tau(\nu,s_{\rm G})$ in Eq. (\ref{eq:specific_intensity}) at 1 MHz, 
based on the {\tt NE2001} model. As expected, the optical depth is large near the Galactic plane, especially in 
the direction of the Galactic center, while outside the Galactic plane the optical depth is small. 
At Galactic plane around $l\sim90\degree$ and around $l\sim240\degree$, the optical depth is small because there are gaps between two spiral arms.

\section{Results}\label{result}
\subsection{The cylindrical model}\label{sec:smooth_sky}

\begin{figure}
	\centering
\includegraphics[width=0.49\textwidth]{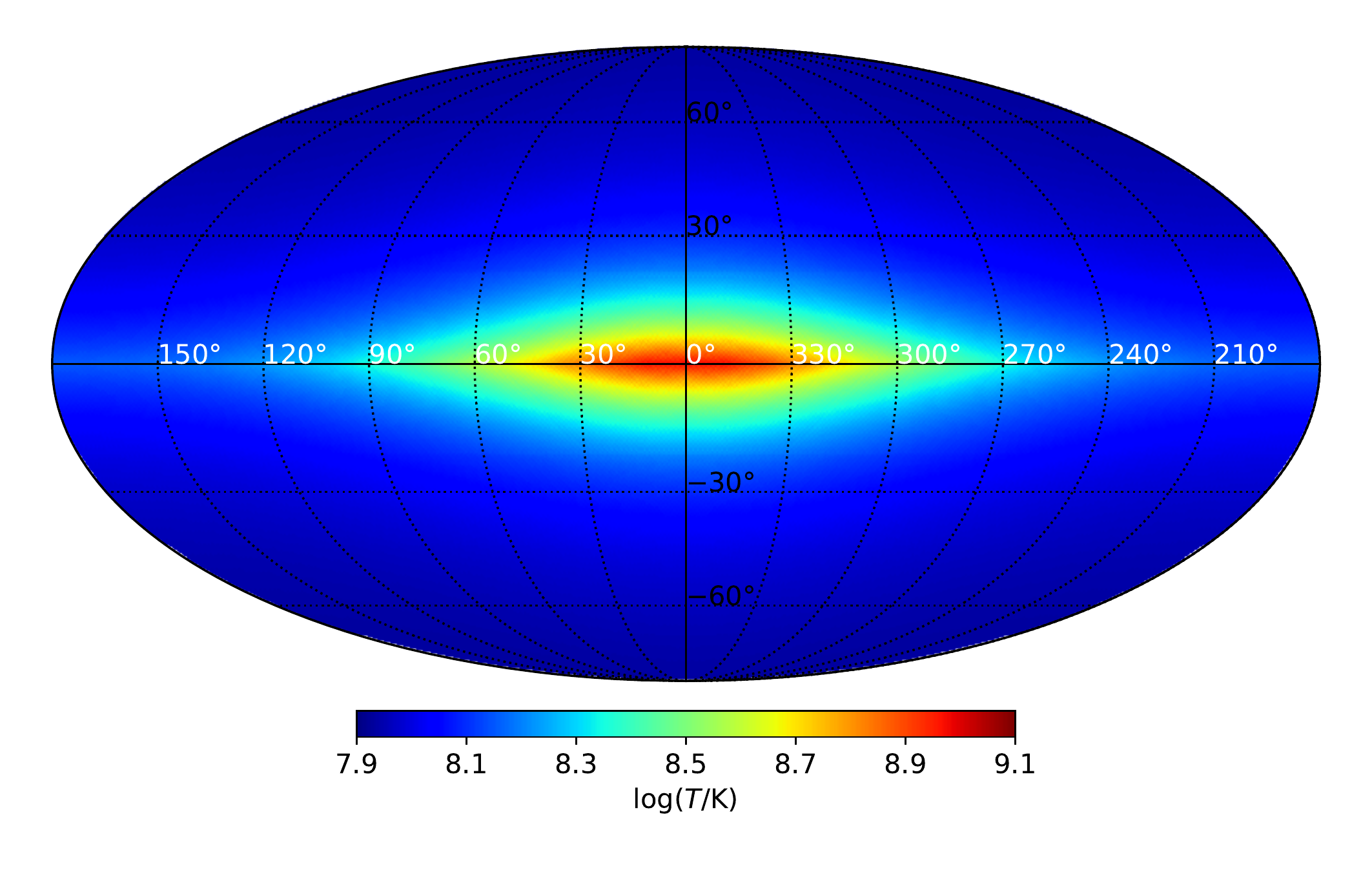}
	\caption{
The sky map at 1 MHz without absorption. 	}
	\label{fig:map_fit}
\end{figure}

In  Fig. \ref{fig:map_fit}, we plot the absorption-free sky map at 1 MHz, obtained by integration of our fitted emissivity  
Eq. (\ref{eq:emissivity}) without considering absorption. This is a symmetric distribution, as it is derived from a model with cylindrical 
symmetry. 
If we mask out the Loop I and NPS regions where the bright nearby feature dominates, the relative difference with the actual map is 
at the level of  $\sim$16\%.

\begin{figure}[tbp]
	\centering
	\includegraphics[width=0.45\textwidth]{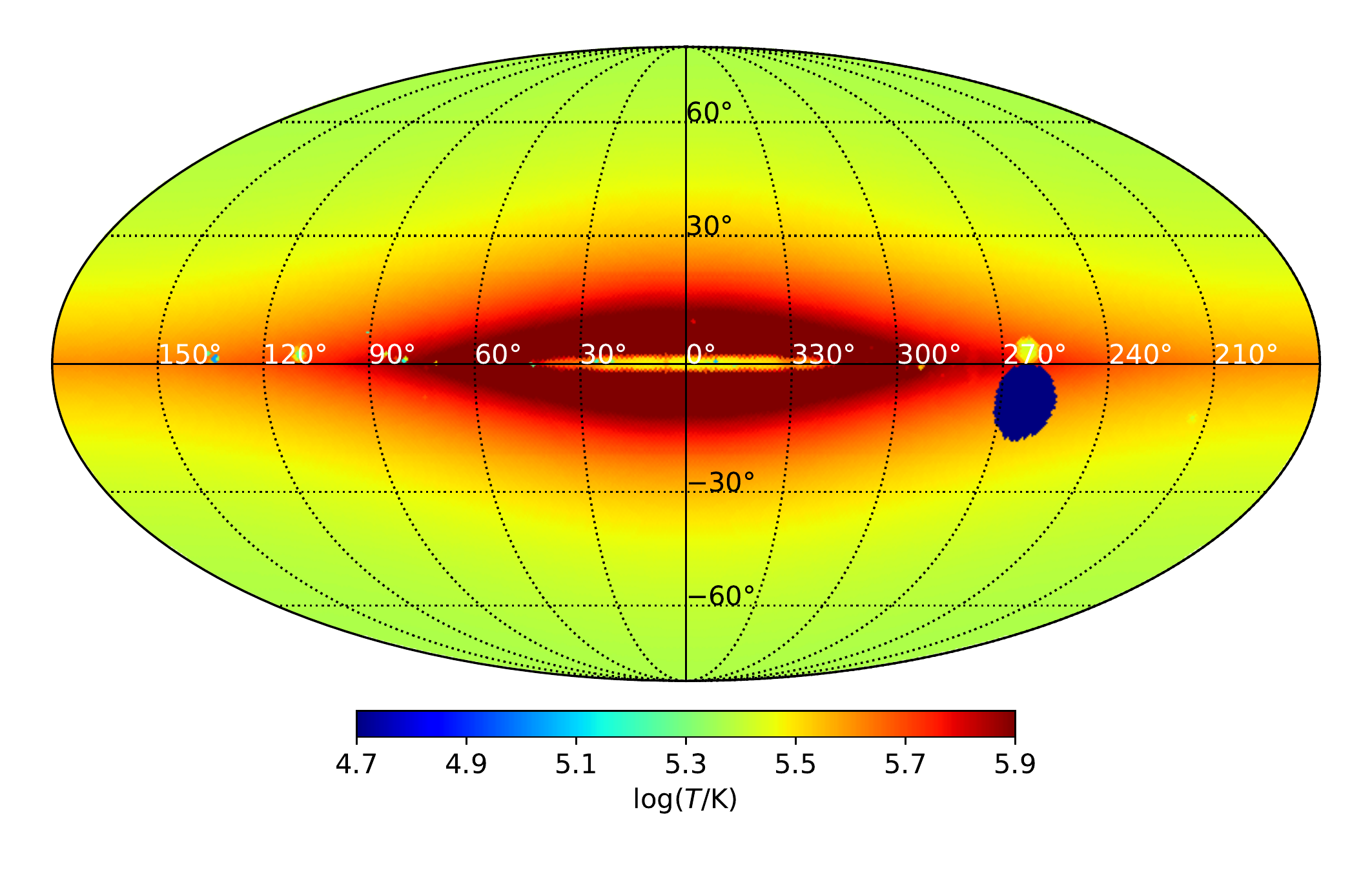}
	\includegraphics[width=0.45\textwidth]{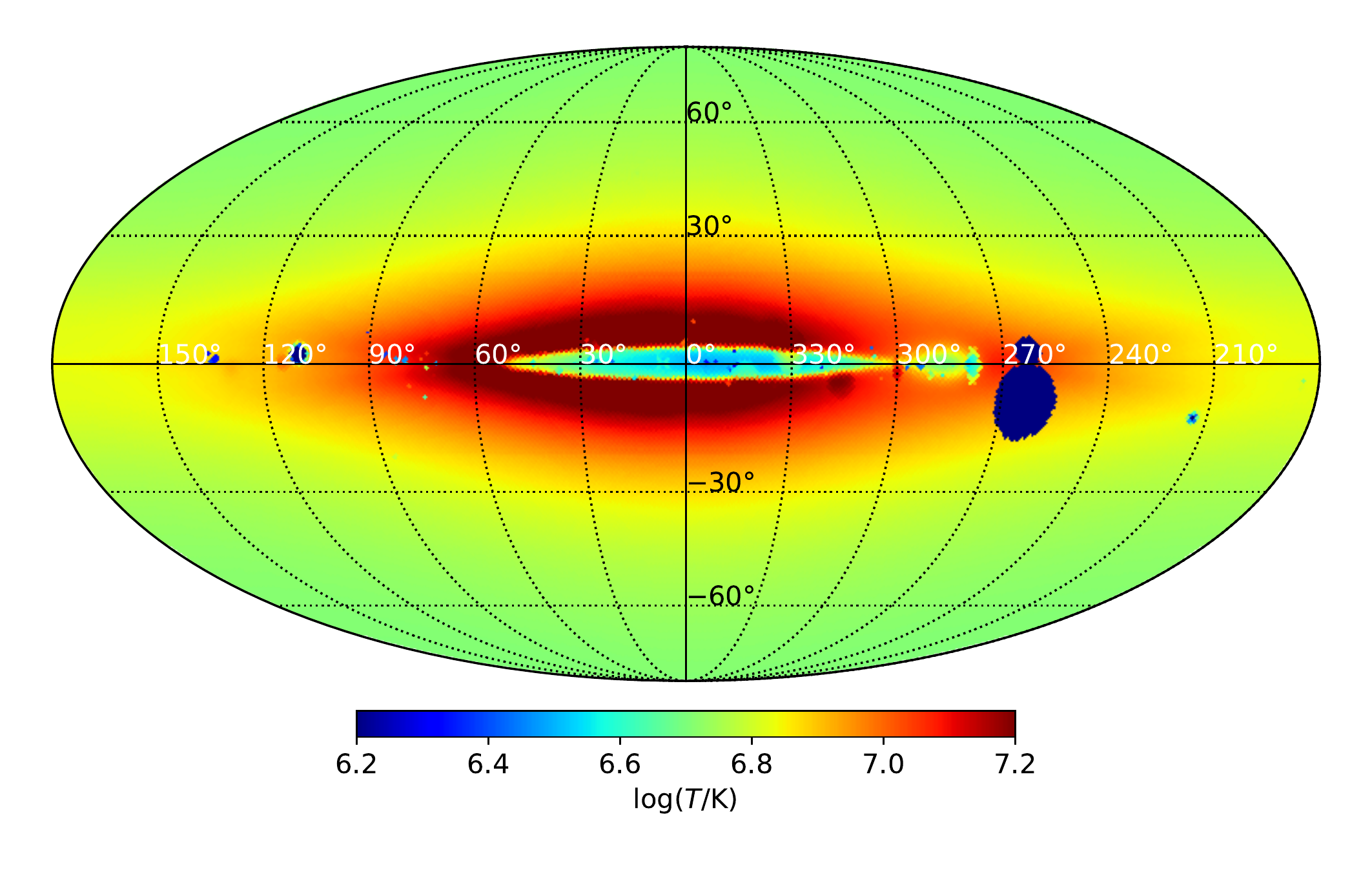}
		\includegraphics[width=0.45\textwidth]{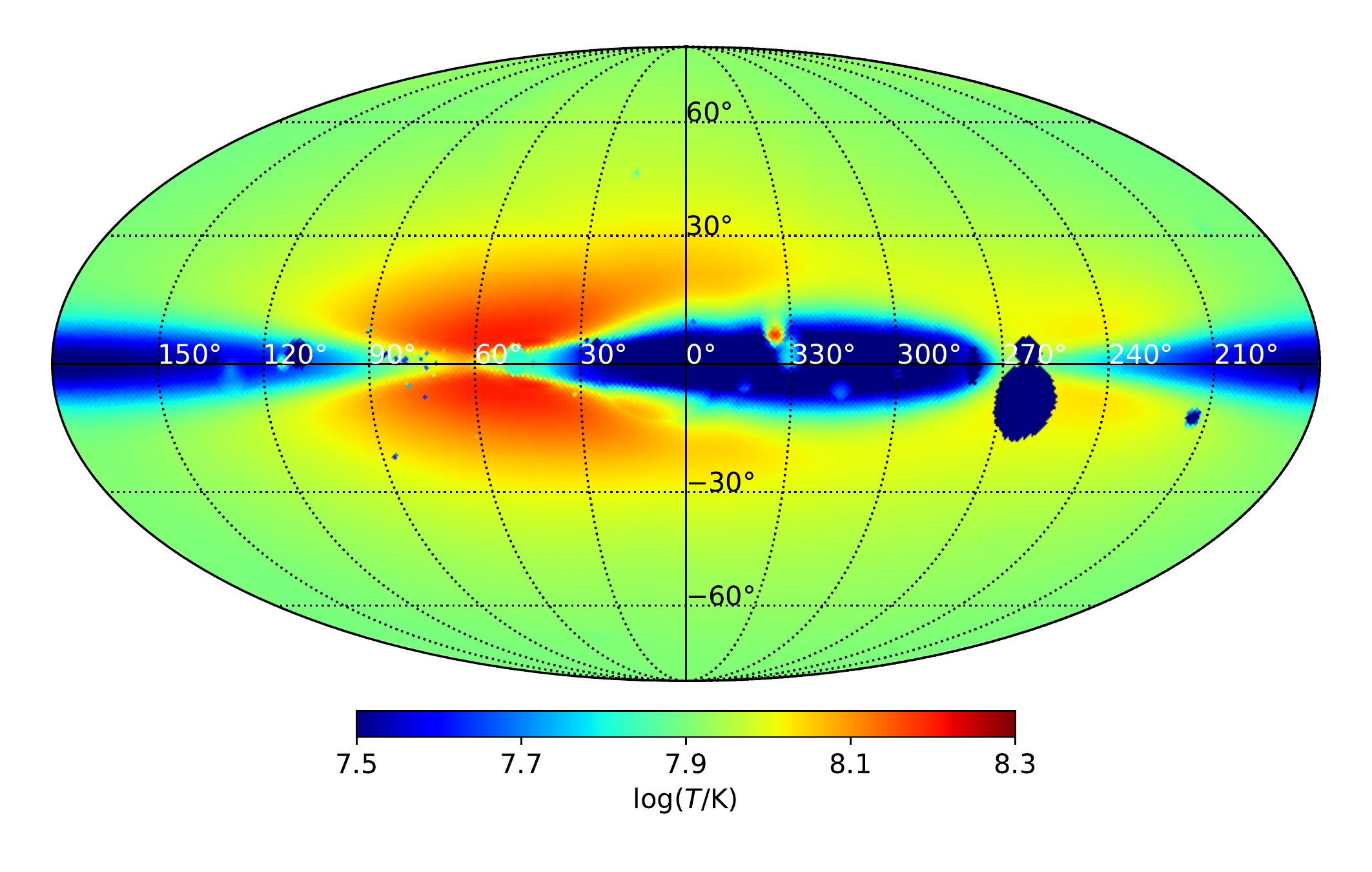}
	\caption{The ultra-long wavelength sky maps at 10, 3 and 1 MHz respectively (from top to bottom), for the fiducial {\tt NE2001} parameters ($F_1=0.18$). Note in this cylindrical model the features from nearby sources such as the Loop I and Northern Spur are not included.
	\label{fig:standard_smooth_absorption_cf_1}}
\end{figure}

With the cylindrical emissivity model and the {\tt NE2001} free electron distribution, one can construct maps with free-free
absorption. In Fig. \ref{fig:standard_smooth_absorption_cf_1}, we plot the sky map with absorption using the {\tt NE2001} fiducial parameters.  As one can see in these figures, at the lower frequencies the absorption becomes very significant. In particular, 
 the absorption is strongest along the Galactic plane, and as a result the {\it Galactic disk is darker than high Galactic latitude regions}
 at the lower frequencies. This is in stark contrast with observations at  higher frequencies or the model 
 map without absorption (see Fig.~\ref{fig_absorptionfreeskymap1mhz}), where the Galactic disk is the brightest part
 of the sky.

\begin{figure}[ht]
	\includegraphics[width=0.49\textwidth]{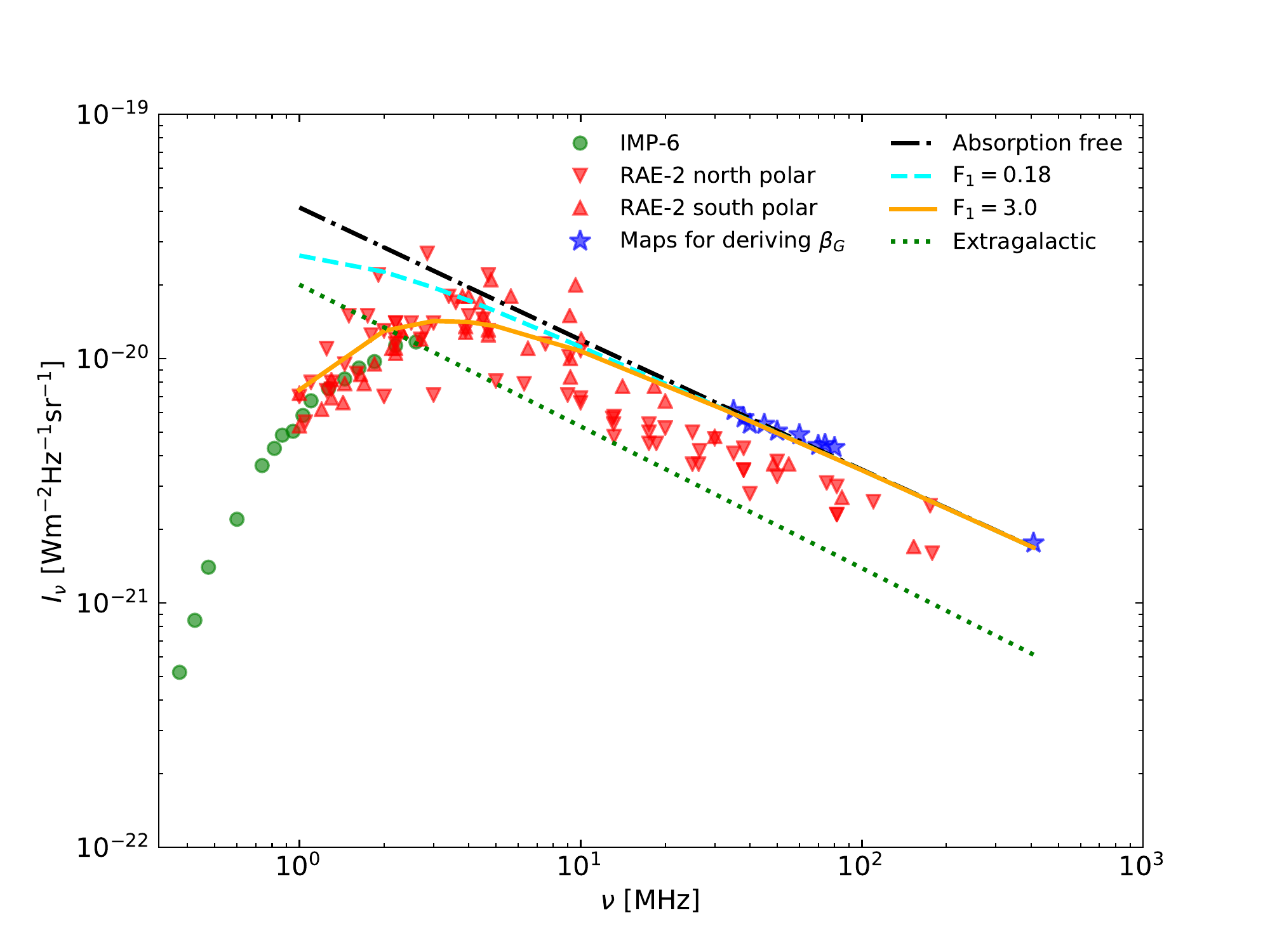}
	\caption{The mean sky brightness in our models (curves) and in observations (points). dashed-dotted line is the absorption-free sky brightness, the dashed line is the absorption-included sky brightness with $F_1=0.18$ in the {\tt NE2001} model, while solid line with $F_1=3.0$. The data points are from the IMP-6 (filled circles) , the RAE-2 (triangles),  and the maps used to derive spectral index (stars). As a comparison we plot the extragalactic radiation Eq. (\ref{eq:iso1}) by dotted line. 
 	}
	\label{fig_globalspectrum}
\end{figure}

How do these maps compare with the observation? The data at this low frequency range are scarce, but there are some global spectrum
measurements, i.e. the average spectrum for the whole or a large part of the sky. 
In Fig.~\ref{fig_globalspectrum} we plot the global spectrum derived from averaging the brightness of the maps of our model, 
 together with some observational data points. The data were taken  by the IMP-6 satellite (filled circles, \citealt{Brown1973},) and the RAE-2 satellite (upward and downward triangles, \citealt{Cane1979}). Besides instrumental differences, the differences between the measurements may be due partly to the different sky area being averaged over for the two satellites \citep{2004ApJ...617..281K}. 
 The models are plotted as curves, including the absorption-free case (dot-dashed line), the extragalactic source (dots), the fiducial {\tt NE2001} model ($F_1=0.18$, cyan dash line), and a model with enhanced fluctuations ($F_1=3.0$, orange solid line, see discussions below). 
Compared with the data, we find that the sky brightness below $\sim3$ MHz is over-predicted
when using the {\tt NE2001} fiducial fluctuation parameter for thick disk ($F_1=0.18$), as shown by the cyan dashed curve. This has already been noticed 
by several earlier works (e.g., \citealt{peterson2002interstellar,webber2008limits}). 

\begin{figure}[tbp]
	\centering
	\includegraphics[width=0.45\textwidth]{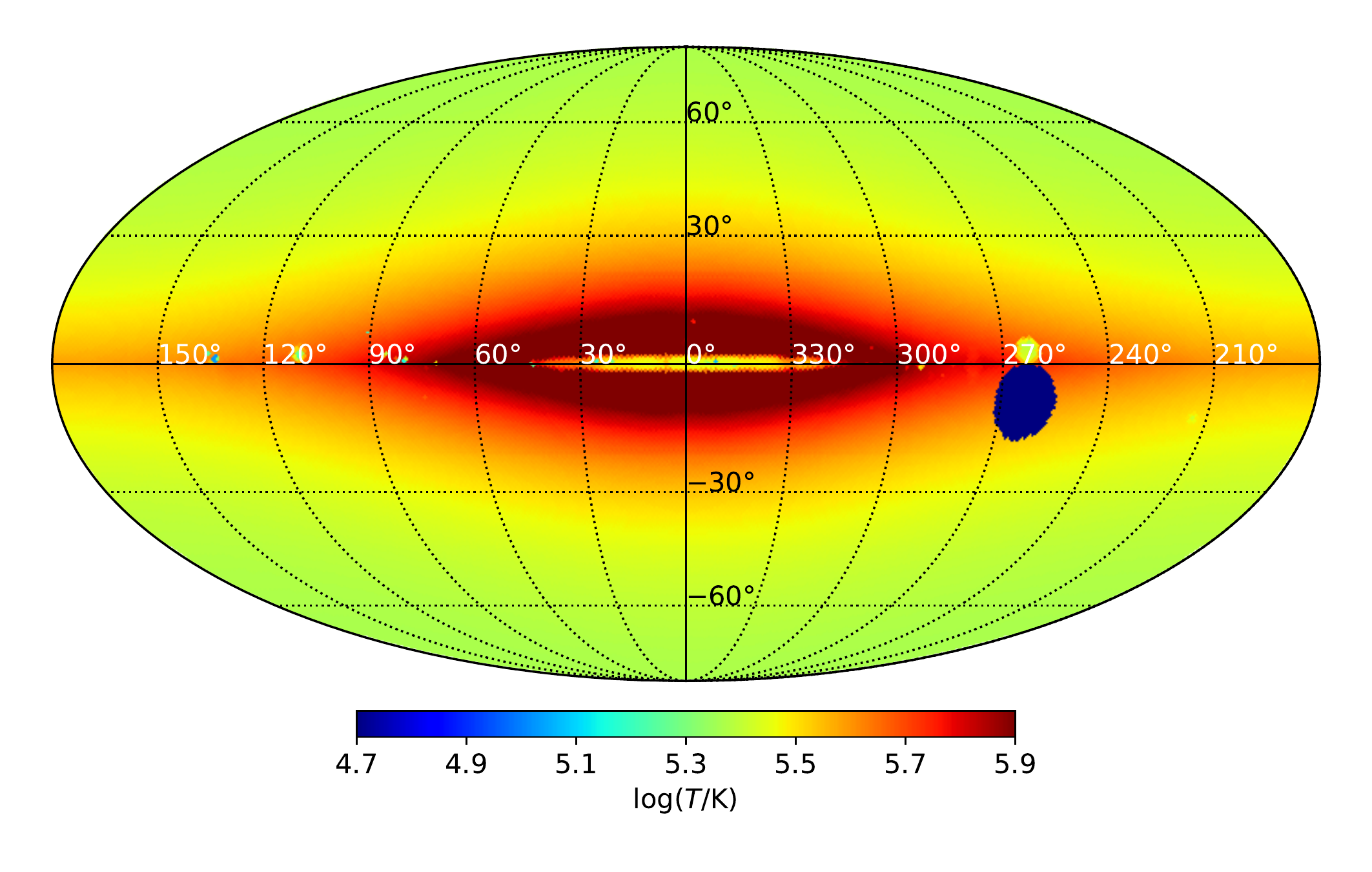}
	\includegraphics[width=0.45\textwidth]{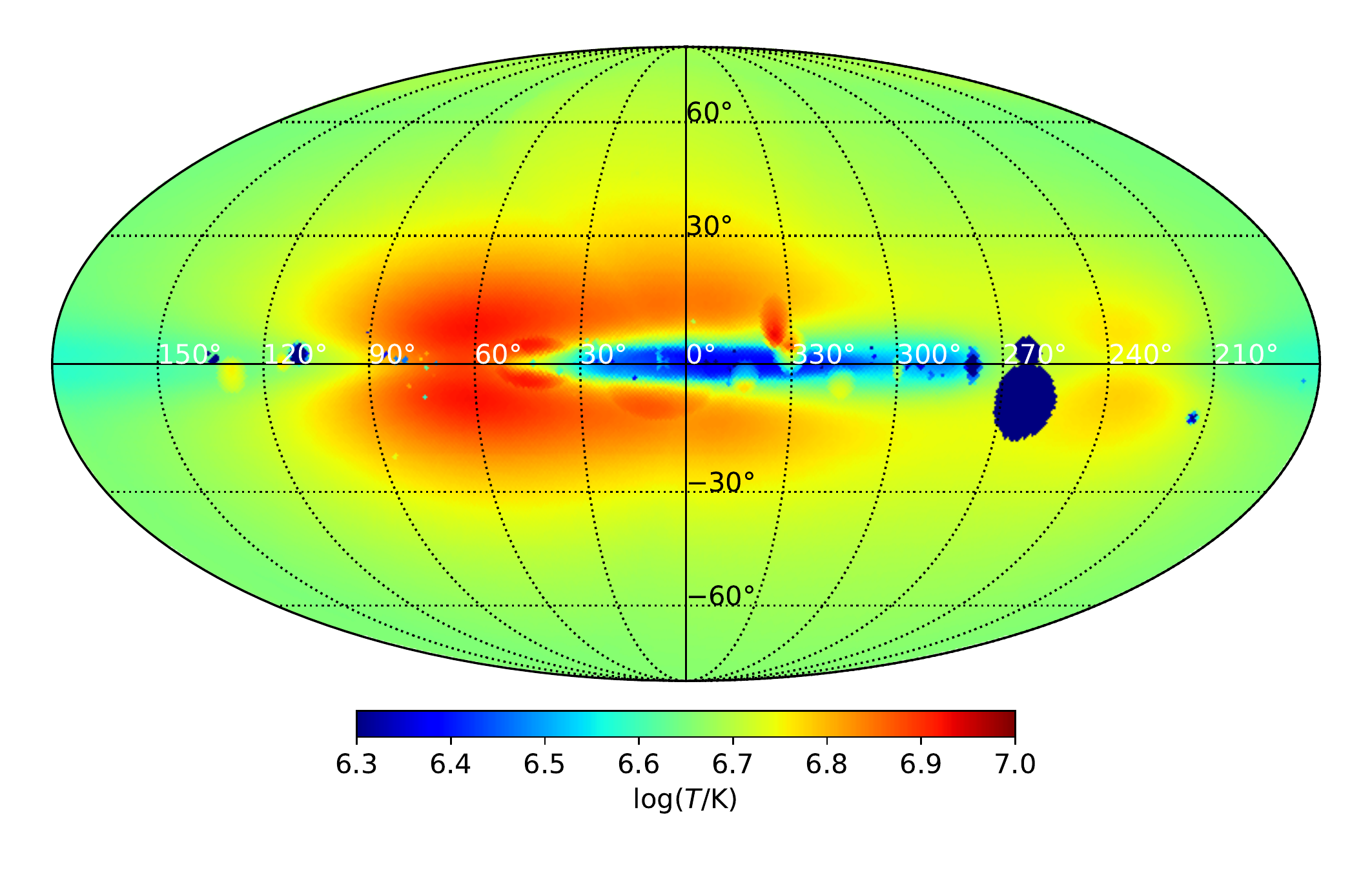}
	\includegraphics[width=0.45\textwidth]{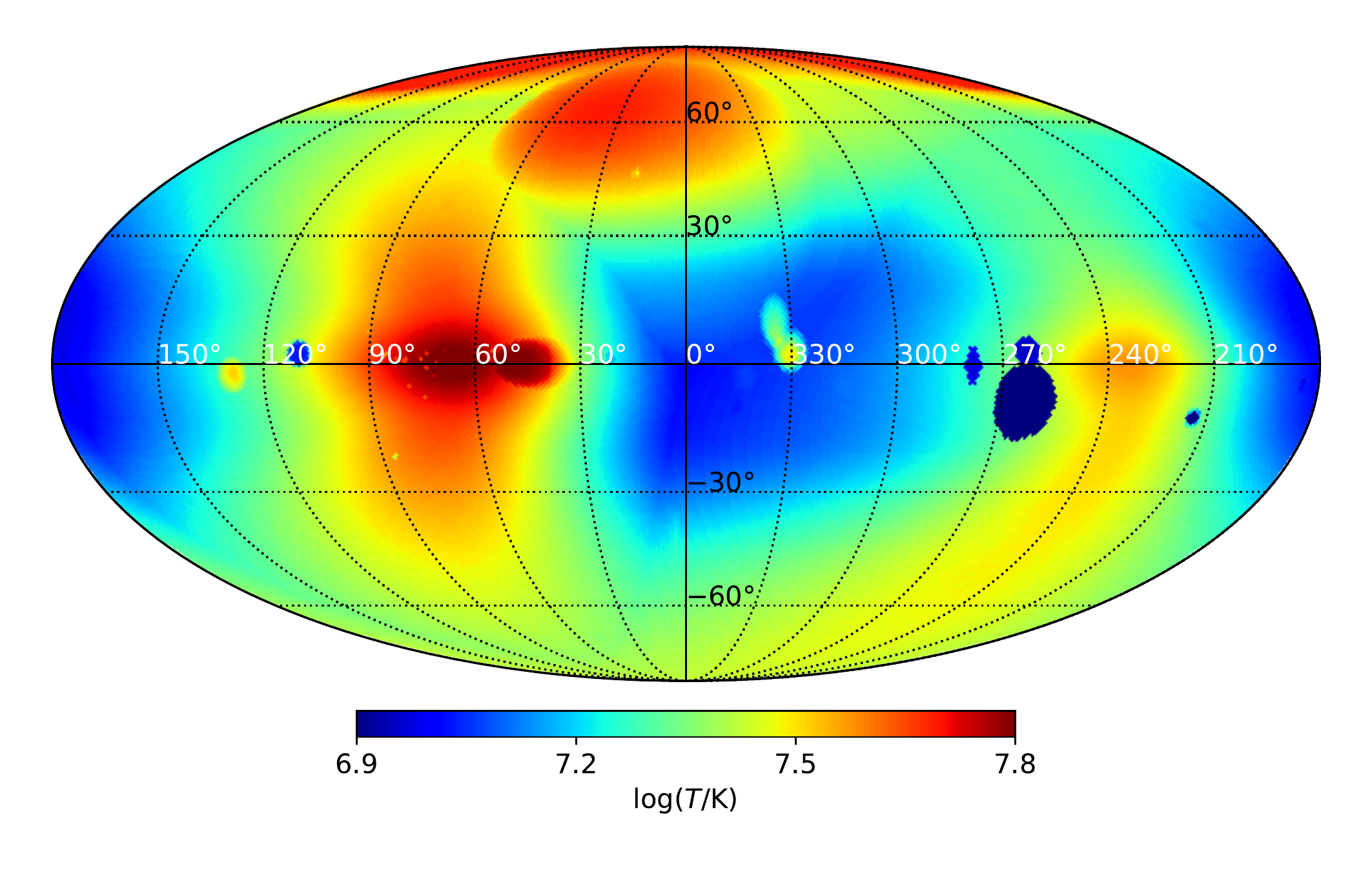}
	\caption{Same as Fig. \ref{fig:standard_smooth_absorption_cf_1}, but for higher fluctuation parameter
	 $F_1=3.0$.		 
	 \label{fig:smooth_absorption_cf_1}}
\end{figure}

Some works (\citealt{gaensler2008vertical,sun2008radio}) show that the scale height of the thick disk could be twice 
 the default value used in {\tt NE2001}. This would in principle enhance the free-free absorption. 
 However, adopting a larger scale height of 1.8 kpc, as compared to the default 0.97 kpc, results in a 
 decrement of the mean sky brightness by only 11\% at 1 MHz,
because a large fraction of the sky radiation is from high Galactic latitude regions that are not influenced much
by the increase of the disk thickness.

 \begin{figure*}[htbp]
	\centering
	
	\subfigure{
		\begin{minipage}[t]{.31\linewidth}
			\centering
			\includegraphics[width=2.3in]{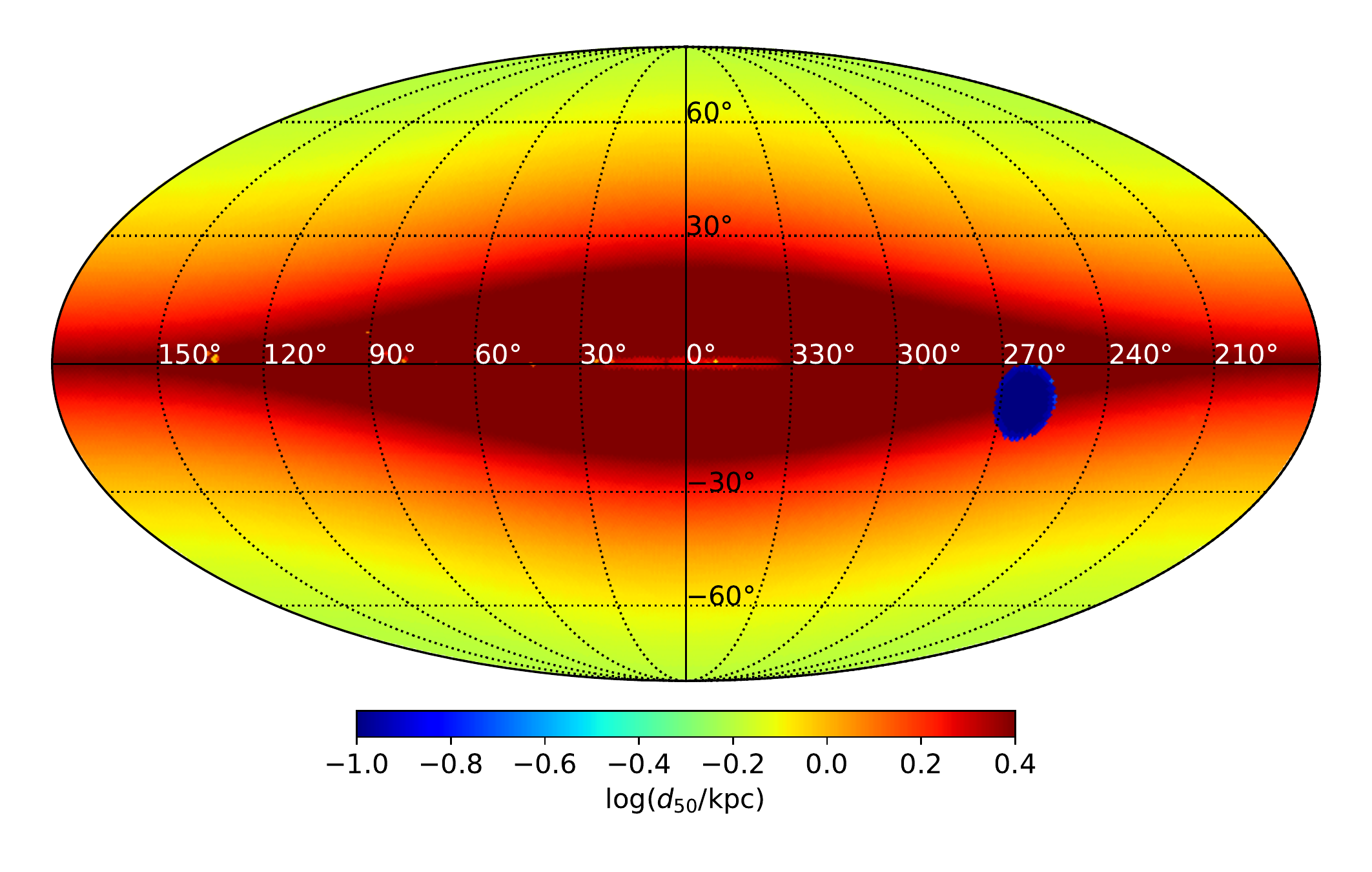}
	\end{minipage} 
	}
	\subfigure{
		\begin{minipage}[t]{.31\linewidth}
			\centering
			\includegraphics[width=2.3in]{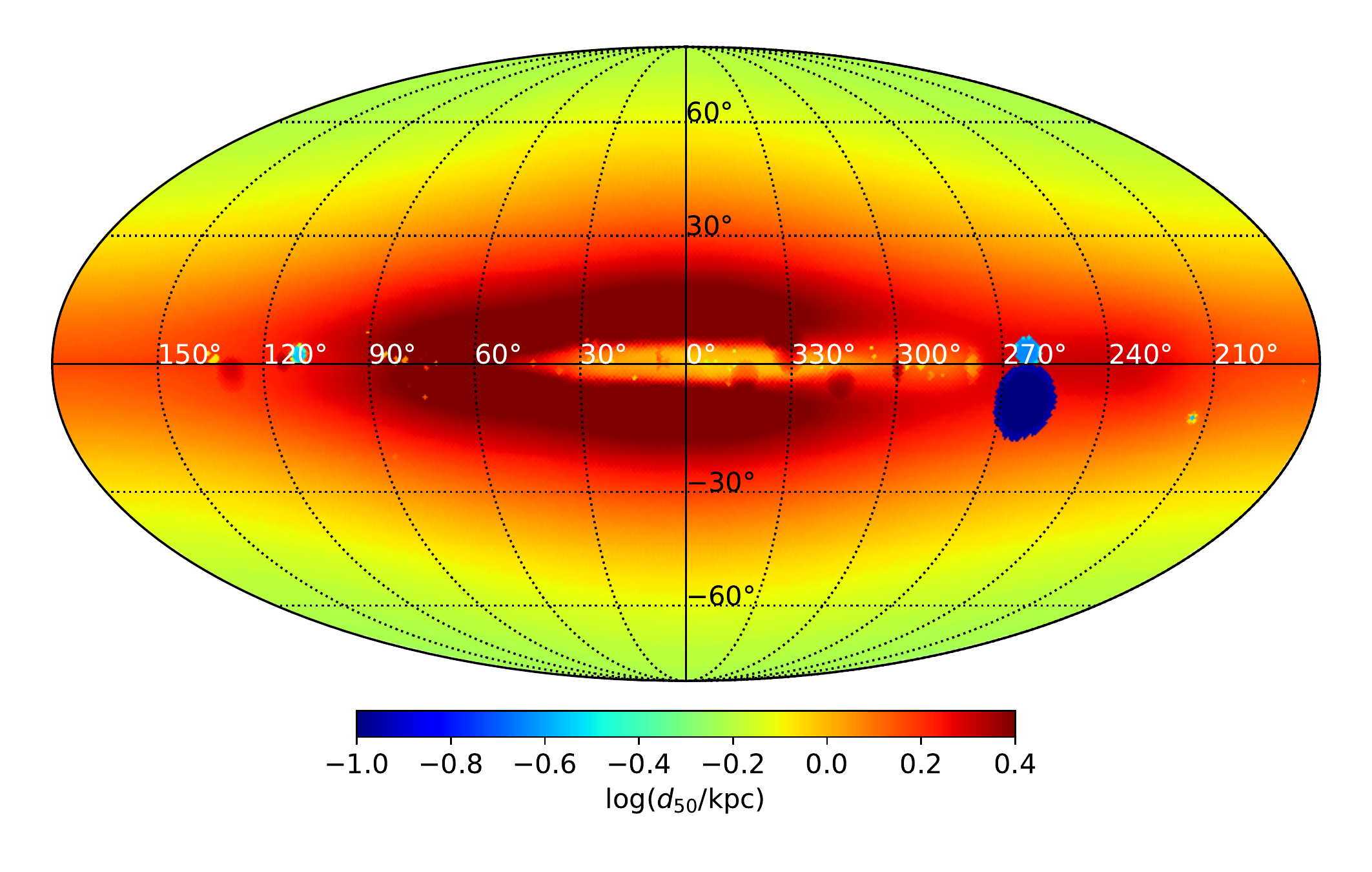}
		\end{minipage}%
	}
	\subfigure{
		\begin{minipage}[t]{.31\linewidth}
			\centering
			\includegraphics[width=2.3in]{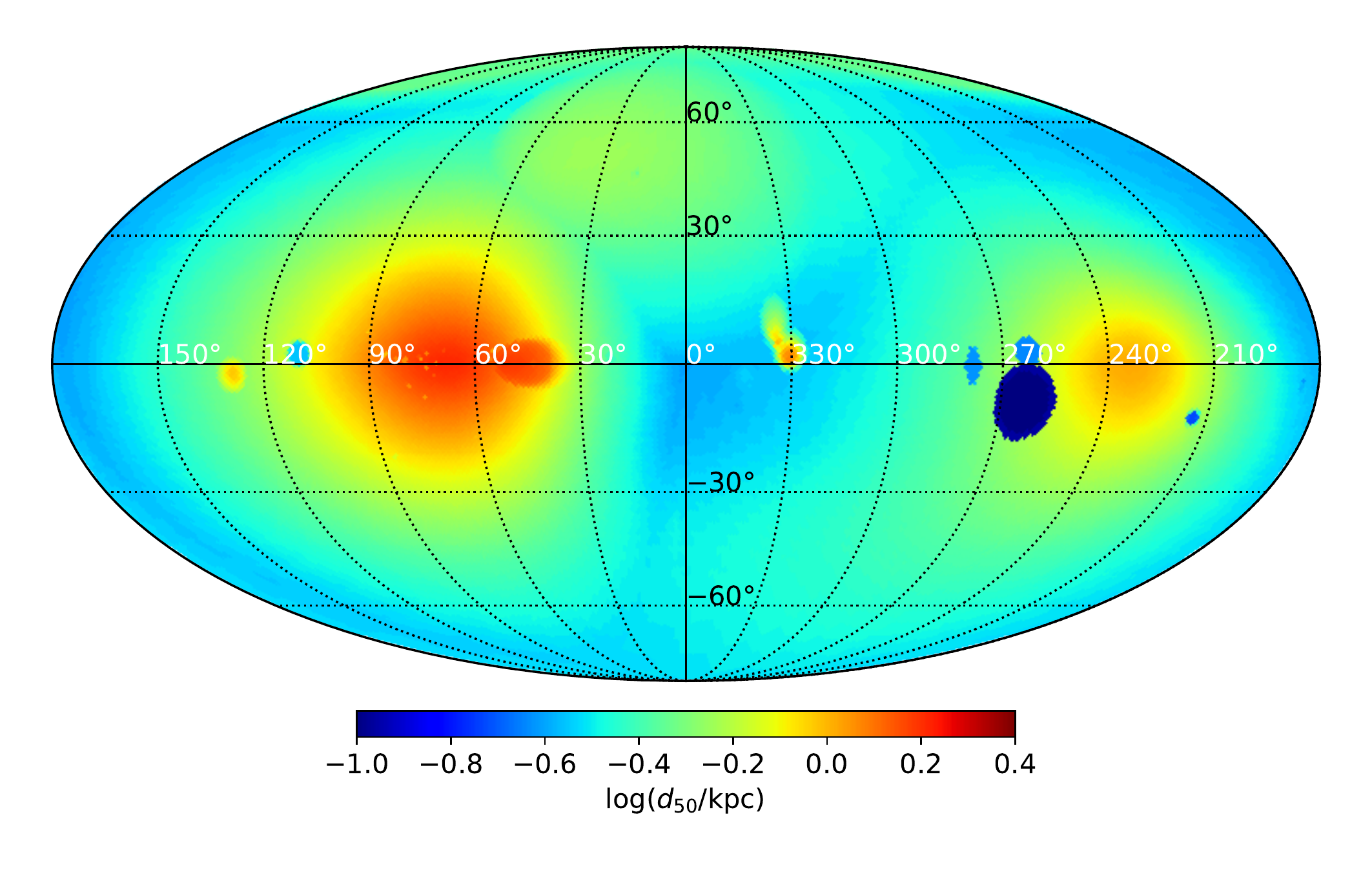}
		
		\end{minipage}%
	}%

	\caption{The half brightness distance for 10, 3 and 1 MHz respectively (from left to right).
			\label{fig:critical_distance}}
\end{figure*}

\begin{figure*}[htbp]
	\centering
	\includegraphics[width=0.45\textwidth]{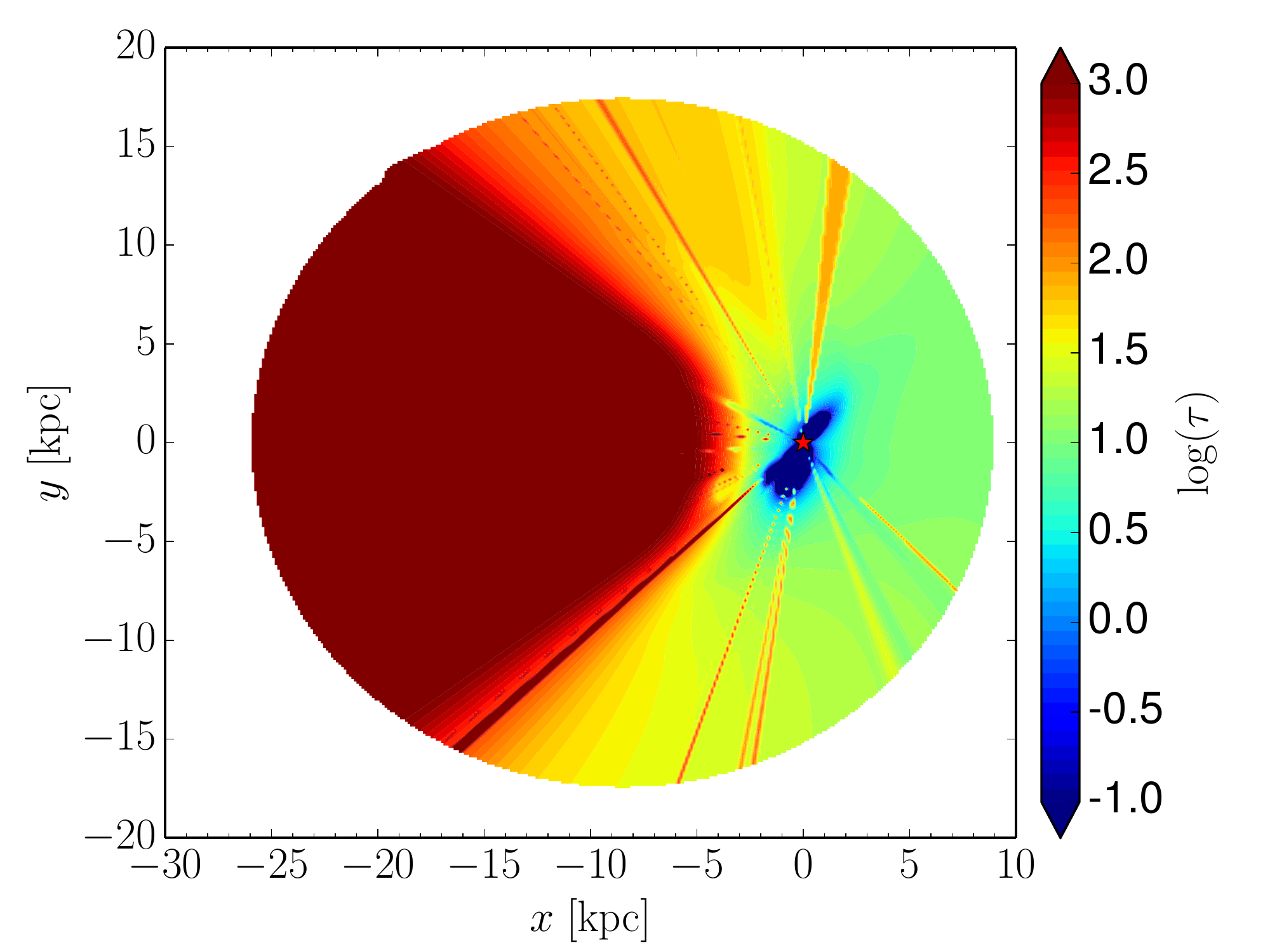}
	\includegraphics[width=0.45\textwidth]{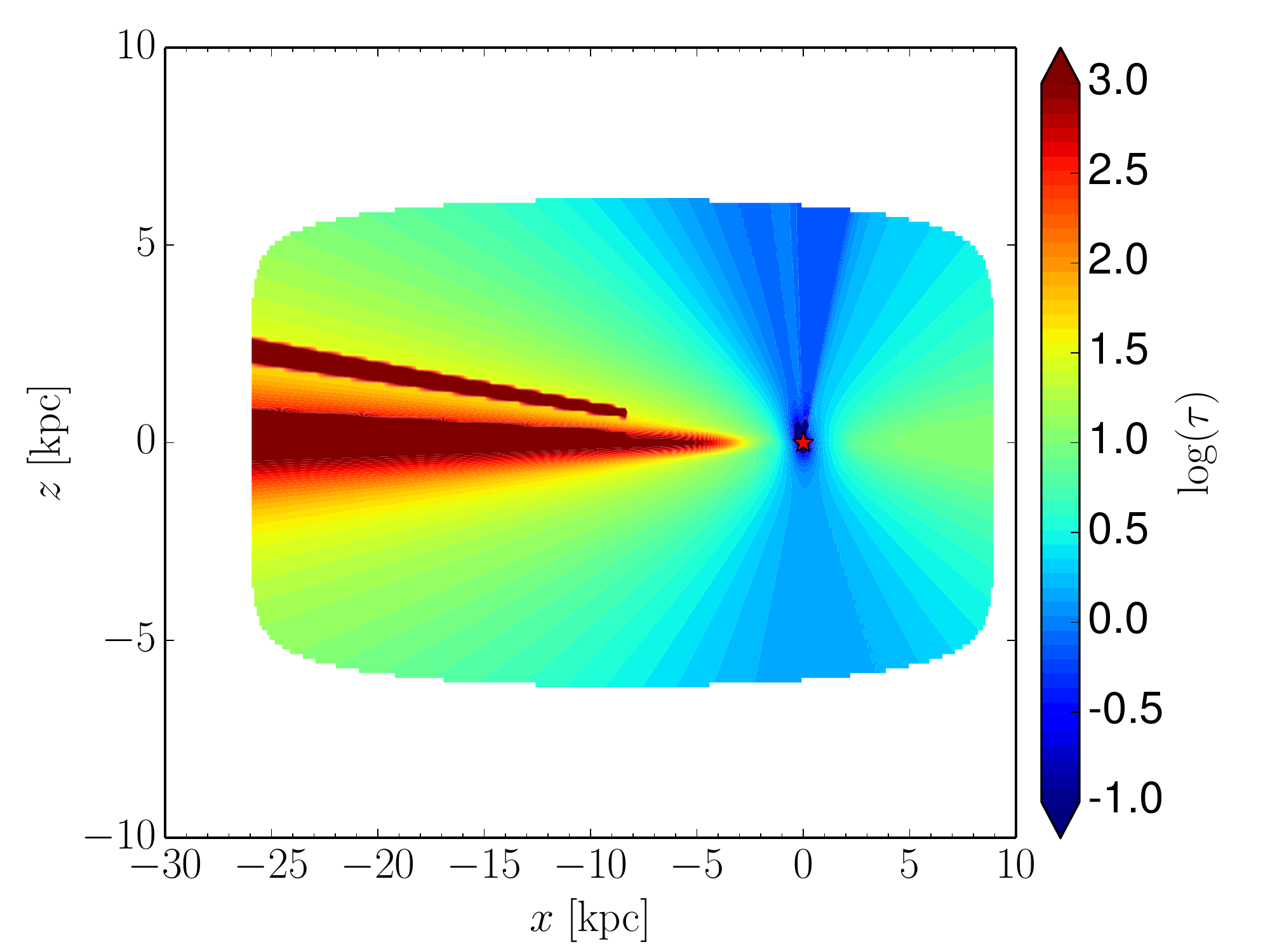}
	\caption{The optical depth distribution at 1 MHz as viewed from the Earth, project along the Z(left) and Y(right) axis of the 
	Galactic coordinates respectively. The location of our Sun is marked by a star symbol.}
\label{fig:optical_depth_YZ}
\end{figure*}

Now there are some large uncertainties in the data itself,  but the down-turn of the spectrum below $\sim3$ MHz 
seems robust. If we take the data points below 3 MHz at face value,  a 
possible solution of this discrepancy is that there may be small-scale structures not modeled in 
the current electron distribution model which enhances the absorption.  
In order to reach consistent results with the low-frequency observations 
at $\lesssim3$ MHz, one can enhance the absorption by adopting larger fluctuation parameters 
for the Galactic diffuse components, or alternatively, adopt a softer spectrum index at lower frequencies. 
Since the thick disk has the largest influence on the 
global mean sky brightness, we test the model in which the thick disk 
has a larger fluctuation parameter of $F_1=3.0$, while keeping other parameters to be the same as the 
{\tt NE2001} fiducial values. Note that the fiducial value of this fluctuation parameter is $F_1=0.18$ in {\tt NE2001}, 
so the optical depth is enhanced by more than fifteen times. The results is the solid curve in 
Fig. \ref{fig_globalspectrum}, which now agrees with the observational data below 3 MHz. 
The corresponding maps are shown in Fig. \ref{fig:smooth_absorption_cf_1}, for frequencies 10, 3 and 1 MHz respectively.

The maps in Fig. \ref{fig:standard_smooth_absorption_cf_1} and Fig.  \ref{fig:smooth_absorption_cf_1} begin to show difference at 3 MHz, and the two 1 MHz maps are quite different. For the 1 MHz map, in Fig. \ref{fig:smooth_absorption_cf_1} one can see huge shadows of the spiral arms and Galactic center component in the sky map (the blue-cyan regions at the leftmost  and rightmost, and near the center). 
Note that the spiral arms do not only have higher WIM density than the gap between arms, but also are more crowed with discrete HII regions \citep{Paladini2004} and their extended envelopes. They produce especially prominent large-scale shadows even at higher frequencies, as has been recognized much earlier
\citep{Bridle1969,roger1999radio,guzman2011all}. 
On the other hand, for regions near the Galactic pole, and at some voids, the absorption is much weaker. Such directions would be recommended for detecting extragalactic sources. There are some dark spots, most of which are near the Galactic plane. They are HII regions around massive stars or SNRs, with enhanced absorption.  For example near the Galactic plane at around $l\sim270\degree$, there is the Gum Nebula.  It is an ancient SNR with a distance between 200 pc and 500 pc \citep{woermann2001kinematics}.  
 So far, we have not taken into account the radiation from these sources themselves, so the brightness at their locations may be underestimated in such models.

The IMP-6 satellite has detected maximum radiation from the direction near Galactic poles at frequencies between 0.13 and 2.6 MHz \citep{Brown1973}, and the modulation index (defined as the ratio between the maximum flux minus minimum flux and maximum flux plus minimum flux) is $\sim$15\% at 1 MHz. It has a poor resolution of $\sim 100\degree$ however.
We checked that if we smooth our 1 MHz map by a window with FWHM$\sim100\degree$, the modulation index is $\sim20\%$.
Qualitatively, the IMP-6 is in line with our prediction that at ultra-long wavelengths,
the Galactic plane becomes darker because of the absorption, leaving brighter regions at high Galactic latitudes. 
However, our smoothed map shows a darker Galactic plane and the brighter Galactic poles also significantly modulated by the projected spiral arms, so the direction of the maximum flux pole is shifted from the Galactic poles. It is not clear if the anisotropy seen by the IMP-6 also have other reasons.

Because of the free-free absorption, the sky radiation observed at ultra-long wavelength would be dominated by nearby sources.
As the frequency increases, more and more contribution would be from further sources. 
Therefore, by combining observations at various frequencies, the ultra-long wavelength observations 
provide a 3D tomography tool for measuring the electrons distribution and emissivity distribution in our Milky Way. 
Moreover, with the synchrotron emissivity distribution, one can further derive the cosmic ray electron spectrum and the magnetic field distribution
if the gamma-ray observations are analyzed jointly, see \citet{Nord2006,Polderman2020}.
In Fig. \ref{fig:critical_distance}, we show the maps of the critical distance $d_{50}$, 
within which about 50\% of the radiation we observe is emitted.
Here we adopt a thick disk fluctuation parameter of $F_1=3.0$, and
from left to right, we shown results for 10, 3 and 1 MHz, respectively.
For each line-of-sight, by comparing the observed specific intensity at different frequencies, one can derive the contribution from different distances.

We also show the optical depth from the Z-axis and Y-axis perspectives. On the Galactic plane, the shape of the local optical depth distribution is irregular, with obvious anisotropy, as well as ray-like features  due to nearby ISM distribution. On larger scales, 
the electron density increases rapidly toward the inner Galaxy region, forming a fan-like structure. From the side view,  we can see
there are high transparency cones above and below the Galactic plane. The overall structure is what we would expect: there is more 
opacity towards the inner region of the Galaxy, but local ISM could have significant impact.

\subsection{Non-symmetric variations}\label{sec:tem_fluct}

The model discussed above is based on an emissivity model with cylindrical symmetry with respect to the Galactic center.  
For practical usage, it is desirable to have mock sky maps which have non-symmetrical brightness temperature  variations. 
We therefore consider a model which includes such variations, given by
\begin{align}
\Delta T_{\rm G}(\nu,l,b) &= \int_0^{s_{\rm G}} \Delta \epsilon(\nu,R,Z,\phi,s)  e^{-\tau(l,b)}   ds.  
\label{eq:delta_T_G}
\end{align} 
The emissivity fluctuations $\Delta \epsilon(\nu,R,Z,\phi,s)$ is of course very poorly known 
at present, so we adopt an approximated approach as below. 
The cylindrical absorption-free sky brightness temperature is
\begin{equation}
T_{\rm G}^{\rm af}(\nu,l,b)=\int_0^{s_{\rm G}} \epsilon(\nu,R,Z) ds,
\end{equation}
and the absorption-included sky brightness temperature is
\begin{align}
T_{\rm G}(\nu,l,b) &=\int_0^{s_{\rm G}} \epsilon(\nu,R,Z) e^{-\tau(l,b)} ds.
\end{align}
A ``transmittance" for each line of sight is defined as
\begin{equation}
R_1(\nu,l,b)=\frac{ T_{\rm G}(\nu,l,b)}{ T^{\rm af}_{\rm G}(\nu,l,b)}.
\end{equation}
We  then assume that the non-symmetric temperature variations have the same ``transmittance" as $R_1$ defined above, therefore,
instead of looking for a $\Delta \epsilon$ in Eq. (\ref{eq:delta_T_G}), we directly derive the non-symmetric variations by
\begin{align}
\Delta T_{\rm G}(\nu,l,b) =R_1  \Delta T^{\rm af}_G(\nu,l,b),
\end{align}  
where $\Delta T^{\rm af}_{\rm G}(\nu,l,b)$ is the difference between the extrapolated sky map  (using Eq. \ref{eq:T_G_extrapolated}) and the cylindrical $T_{\rm G}^{\rm af}$. Note that for some large scale features on the sky, this may be not a good approximation. For example the Loop I is more likely a nearby object \citep{Dickinson2018}, therefore the $R_1$ should be higher than the average. We leave this for future work.

\begin{figure}[htbp]
	\centering
	\includegraphics[width=0.45\textwidth]{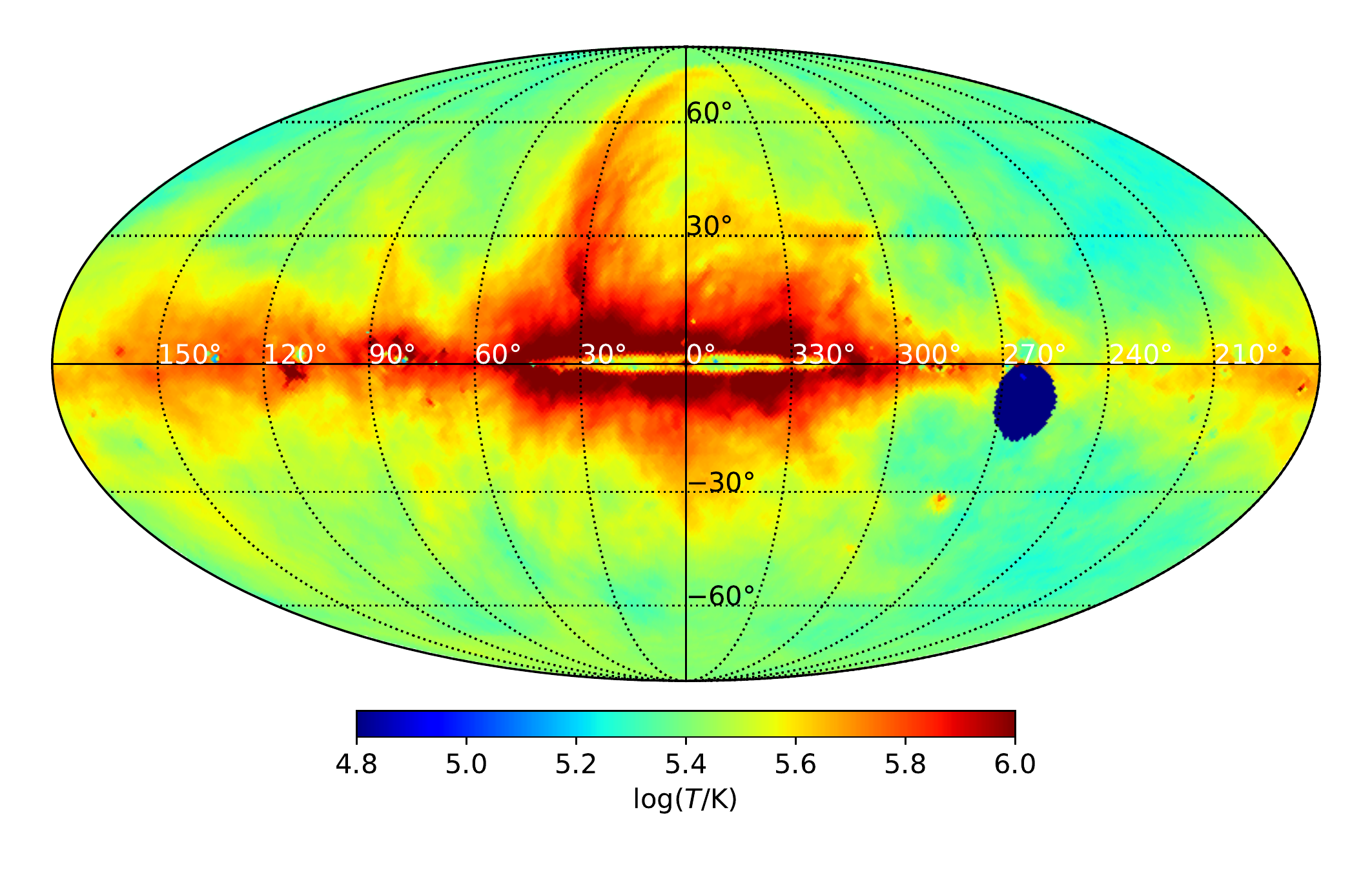}
	\includegraphics[width=0.45\textwidth]{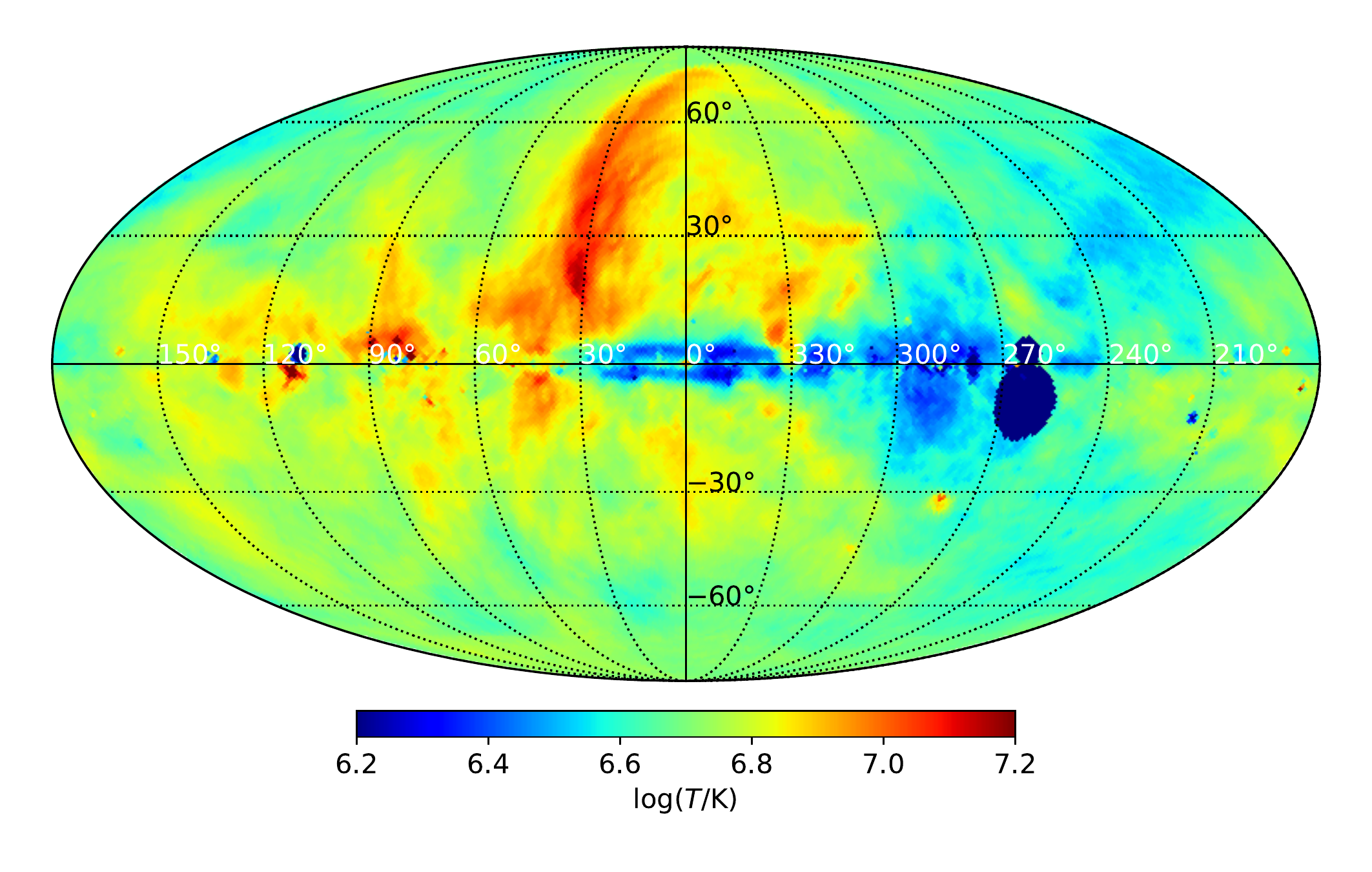}
	\includegraphics[width=0.45\textwidth]{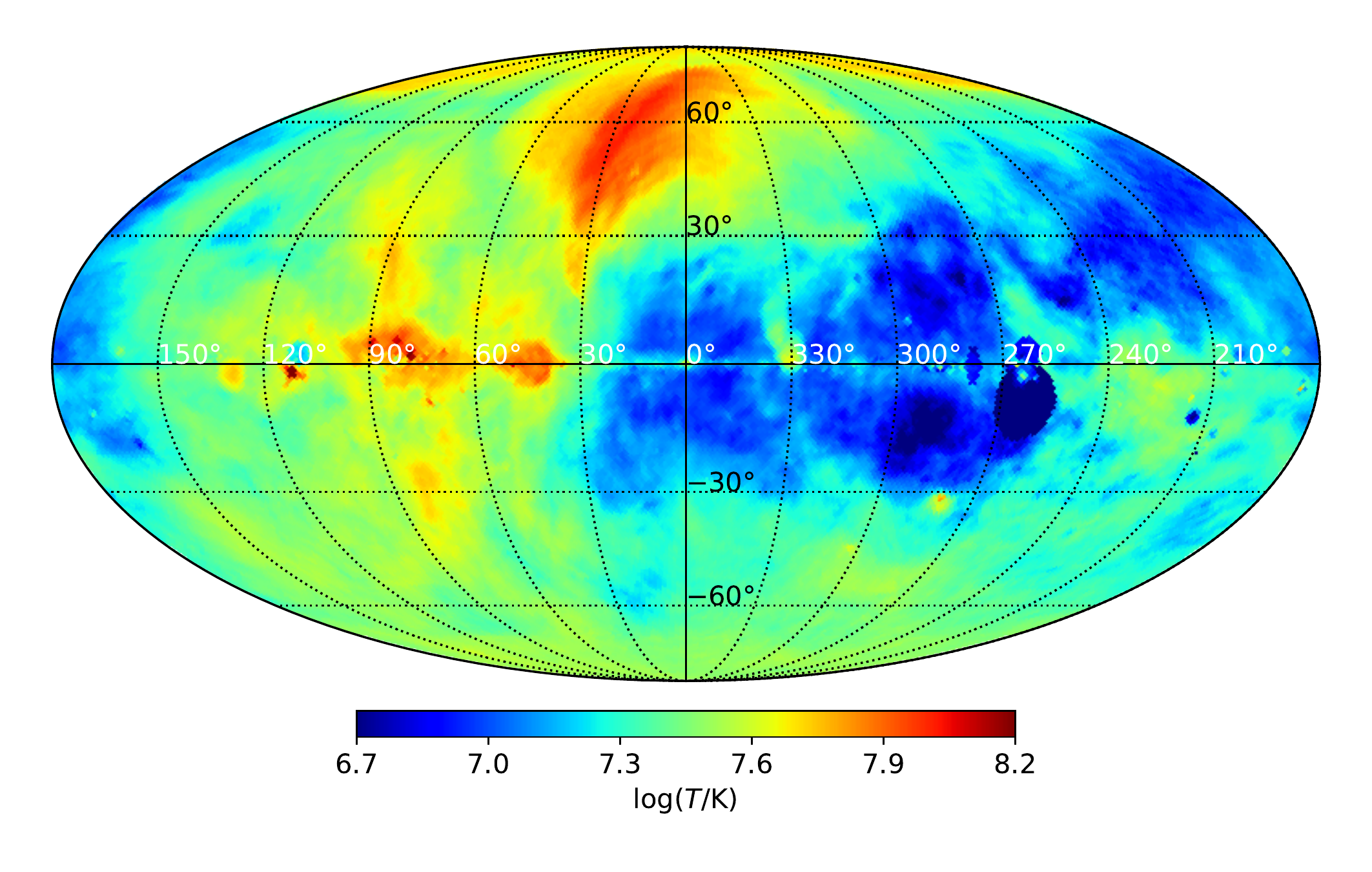}
	\caption{The sky maps at 10, 3 and 1 MHz respectively from top to bottom. Here we add fluctuations to 
	the cylindrical symmetry model, and $F_1=3.0$ is adopted.
		\label{fig:fluctuation_absorption}}
\end{figure}

We add these fluctuations to the cylindrical symmetry model sky maps and the results are shown 
in Fig. \ref{fig:fluctuation_absorption}. 
In  Fig. \ref{fig:map_evolution}, we also show evolution of the Galactic absorption in frequencies from 10 to 0.1 MHz.
We compare our generated map at 10 MHz with the observed map by \citet{Caswell1976_10MHz} at the same frequency. In the common sky region, the relative mean deviation is 18.6\%. 
We also compare our generated map at 22 MHz with the observed one by \citet{roger1999radio}, the mean relative difference is  19.2\%.
 
\begin{figure*}[ht!]
	\begin{center} 
		\includemovie[poster={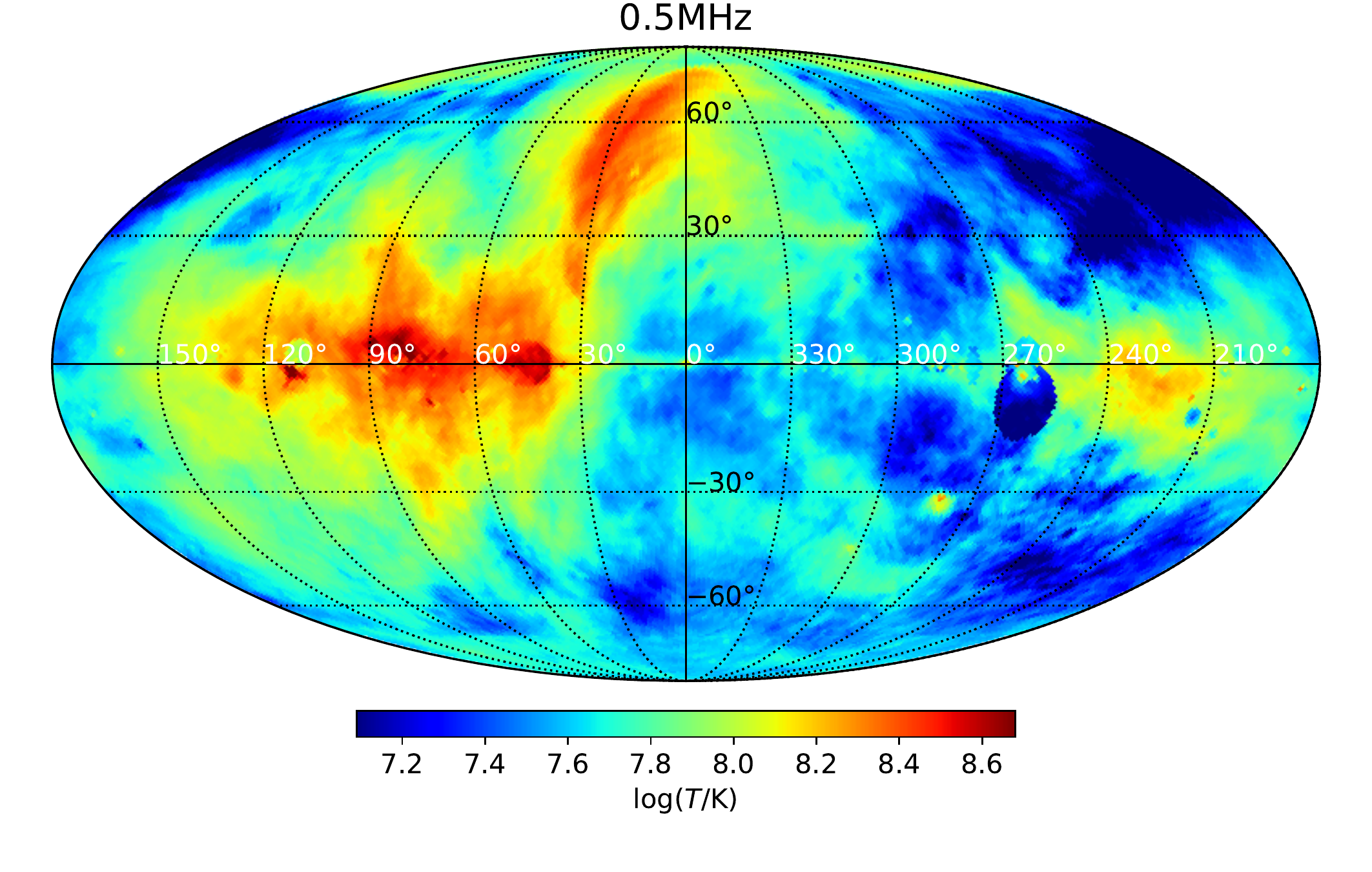},mouse,repeat,playerid=AAPL_QuickTime]{0.9\textwidth}{0.55\textwidth}{./map_evolution_lowres.mp4} 
		\caption{The evolution of the sky maps from 10 MHz to 0.1 MHz. 
			(An animation of this figure is available.)} 
	\end{center}
\end{figure*}

\subsection{Frequency dependence of the spectral index}\label{freq-depend}

So far we have only considered models with constant spectral indices for both extragalactic background and Galactic synchrotron radiation. However, the spectral index may vary with frequency. For example, the energy spectrum of interstellar cosmic ray electrons derived from the combination of PAMELA observations at near-Earth space and Voyager observations at distant space ($\gtrsim 100$ AU) 
can be described by a broken power-law form, with broken point at $\sim 0.1 - 1$ GeV \citep{Potgieter2013,Bisschoff2019}. Assuming a typical interstellar magnetic field $\sim 5~\mu$G, then the broken point corresponds to synchrotron critical frequency  $\sim1-100$ MHz. Inspired by this, in this section we investigate a model that the synchrotron spectral index depends on frequency.

 \begin{figure}[htbp]
	\centering
	\includegraphics[width=0.45\textwidth]{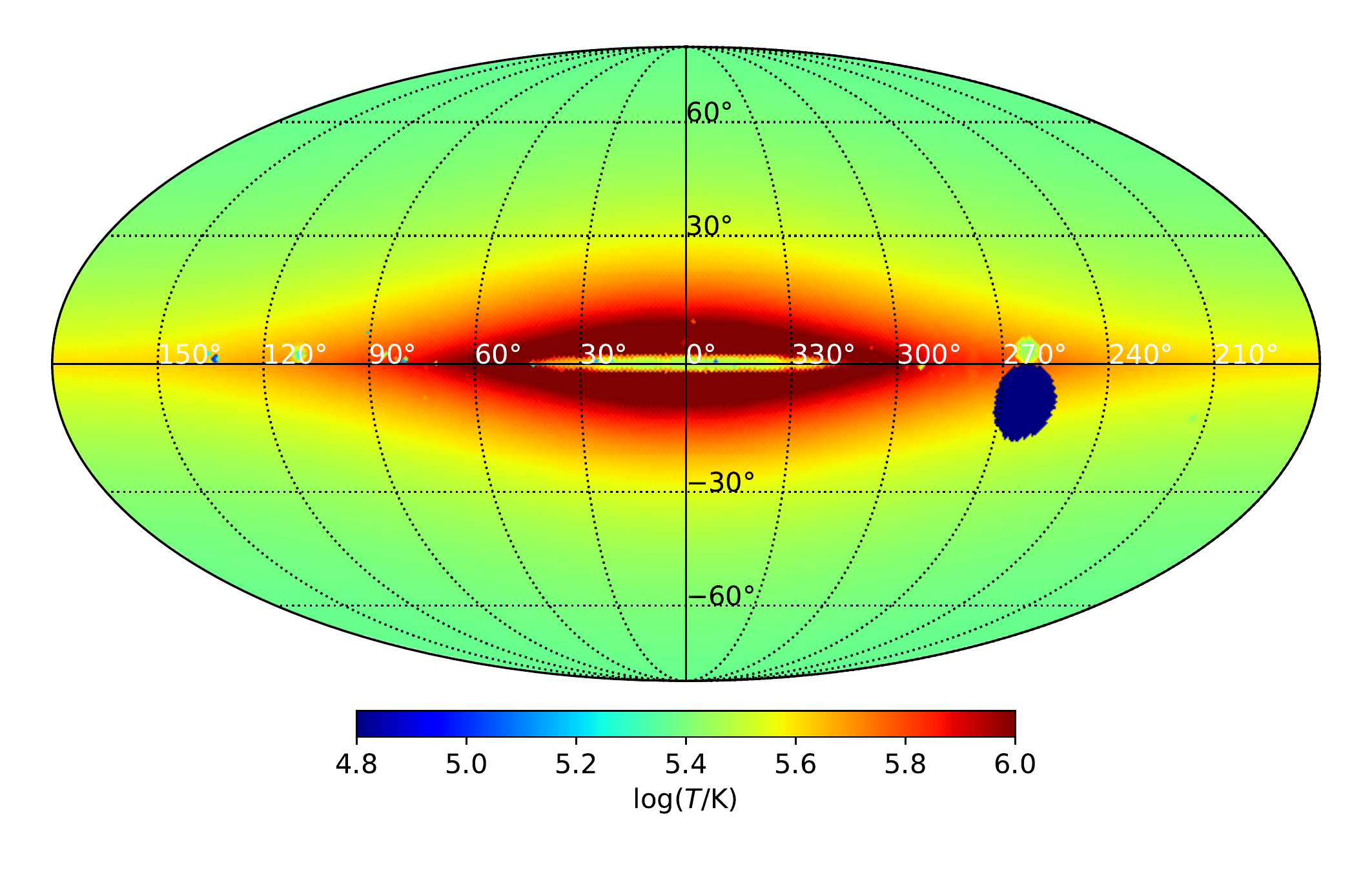}
	\includegraphics[width=0.45\textwidth]{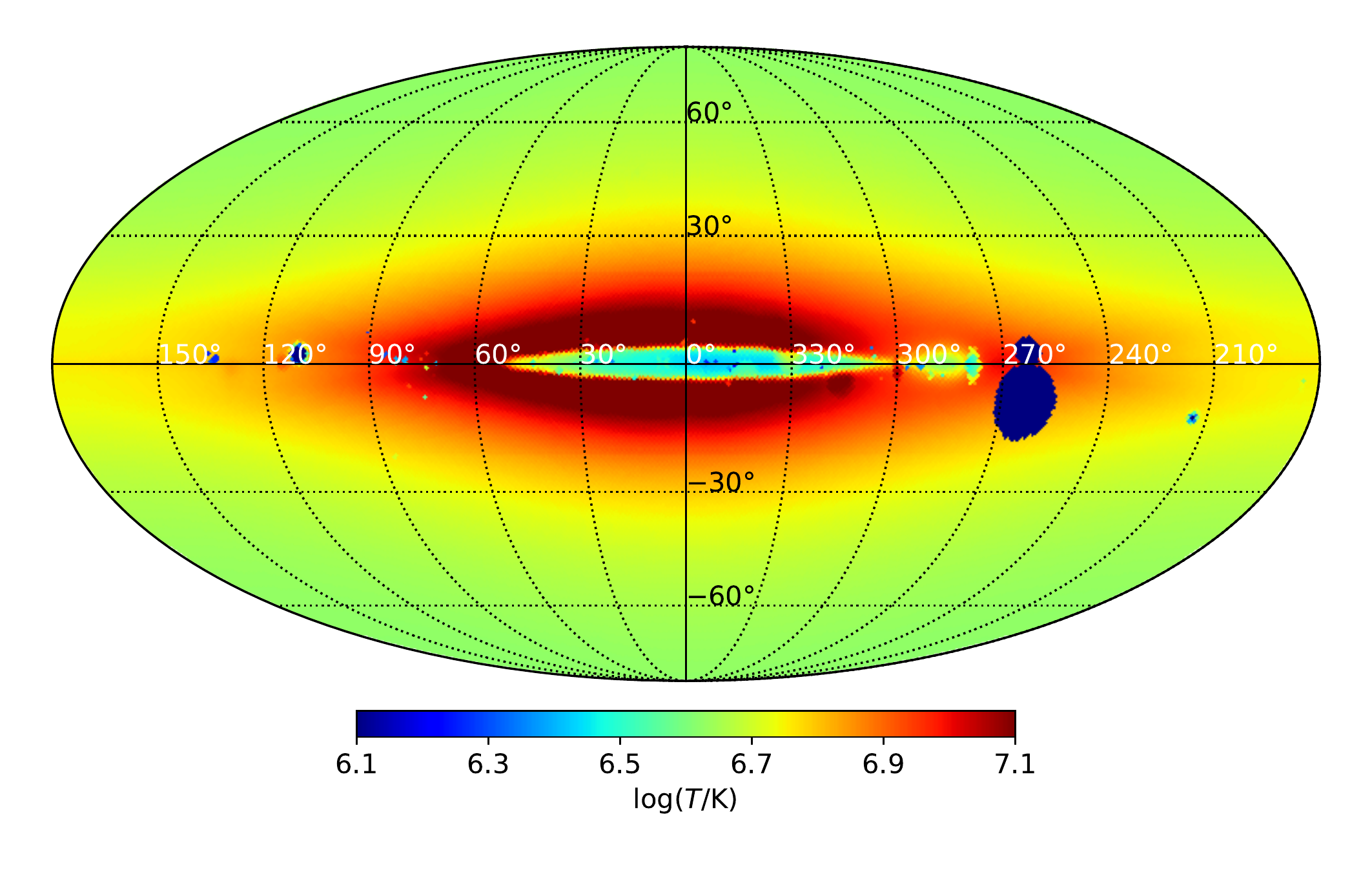}
	\includegraphics[width=0.45\textwidth]{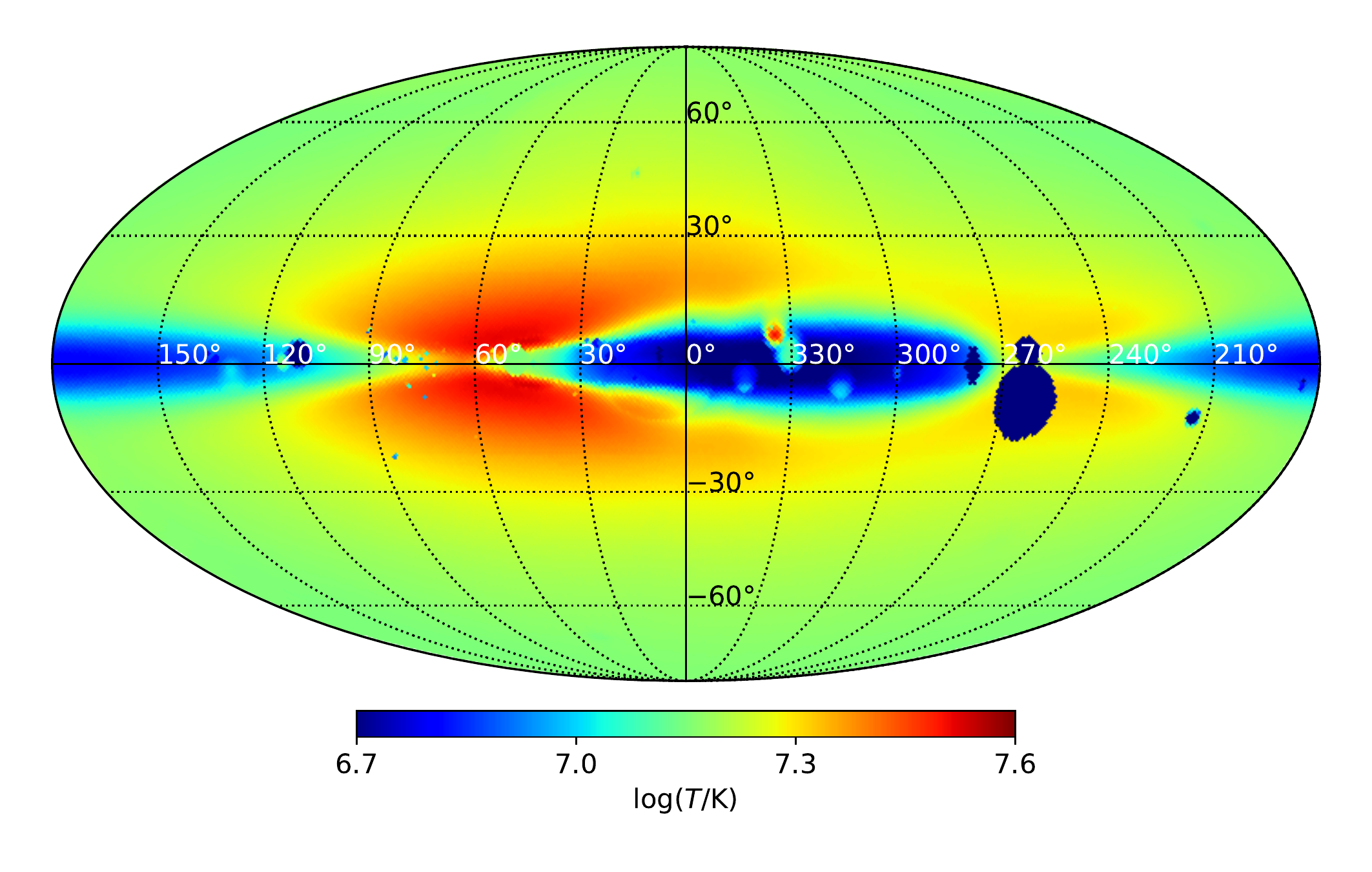}
	\caption{Similar to Fig. \ref{fig:standard_smooth_absorption_cf_1},  however here the synchrotron radiation has frequency-dependent spectrum index as Eq. (\ref{eq:beta}).
	\label{fig:freq_depend_smooth_absorption_cf_1}
		}
\end{figure}

The frequency dependence of the spectral index is not yet confirmed in current observations at $ \gtrsim10 \sim 22$ MHz. To proceed, 
we tentatively assume the following frequency-dependence for the spectral index, for both Galactic radiation and extragalactic background,
\begin{equation}
\beta=\beta_0 +\beta_1 {\rm exp}\left(-\frac{\nu}{\nu_1}\right),
\label{eq:beta}
\end{equation}
where $\beta_0 = -2.58\ \text{and} -2.51$ is the constant spectral index in our fiducial model for extragalactic background and Galactic radiation respectively.  $\nu_1$ is a critical frequency, for $\nu \gg \nu_1$, $\beta\approx \beta_0$; for $\nu \ll \nu_1$, $\beta\approx \beta_0+\beta_1$. For simplicity we adopt the same $\beta_1 = 0.7$ and $\nu_1 = 1.0~{\rm MHz}$
for both Galactic radiation and extragalactic background. 

The absorption-included sky maps with a frequency-dependent spectral index are shown  in 
Fig.~\ref{fig:freq_depend_smooth_absorption_cf_1}.
Here we adopt $F_1=0.18$, as the default of {\tt NE2001}.
Comparing them with the Fig. \ref{fig:smooth_absorption_cf_1}, we find that although the mean sky brightness is similar, the morphology is rather different, especially at the lower frequencies. Ultra-long wavelength observations would provide necessary information to distinguish these two models.

\begin{figure}
	\centering
	\includegraphics[width=0.45\textwidth]{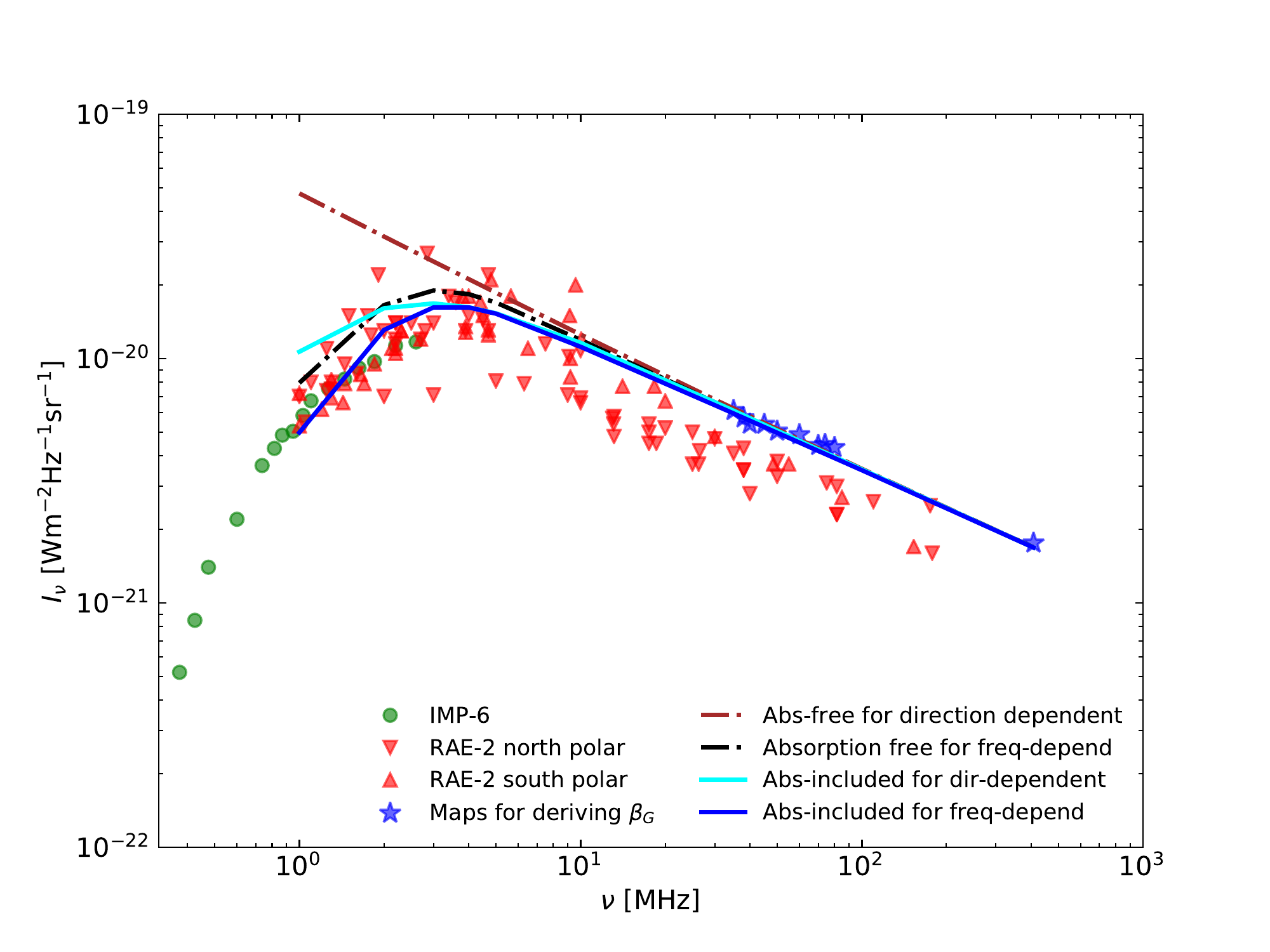}
	\caption{The  mean sky-brightness as a function of frequency in frequency-dependent and direction-dependent models. 
	The dot-dashed and solid curves show the absorption-free and absorption-included sky respectively, 
	for a frequency-dependent spectral index and a direction-dependent spectral index.
 \label{fig:meanskybrightnessabsorpfreefreqdepend}}
\end{figure}

With this frequency-dependent spectrum index, the tension on the fluctuation parameter discussed in Sec. \ref{sec:smooth_sky} can be mitigated, see Fig.~\ref{fig:meanskybrightnessabsorpfreefreqdepend} where we plot the mean sky brightness in this model.
Even for the fiducial fluctuation parameter $F_1=0.18$ in {\tt NE2001}, the spectrum is consistent with the 
ultra-long wavelength   observations.
 
 \subsection{Spatial variations of the spectral index}\label{sec:spatial-variation}

The spectral index could also have spatial variations.  However, at low frequency only the Haslam 408 MHz map covers the full sky.
 To derive the full-sky spectral index map, we first generate a spectral index map by combining the Haslam 408 MHz map with the  LWA maps,  and another one by combining the 408 MHz map with the Guzman 45 MHz map. These two maps are incomplete around  the south and north celestial pole respectively, but for sky region common to both maps, the two maps show similar structure, the 
 difference is about 13.6\% standard deviation.
We produce a combined map by connecting these two spectral index maps by a smooth function. 
The spectral index at declination $\delta$ is  
\begin{align}
\beta_{\rm G}(\delta) &= \frac{1}{2}\left[1 + \text{erf} \left(A\frac{\delta-\rm \delta_0}{\abs{\delta_0}}\right) \right]\beta_{\rm G1}(\delta) \nonumber \\ 
&+ \frac{1}{2}\left[1 - \text{erf} \left(A\frac{\delta-\delta_0}{\abs{\delta_0}}\right) \right] \beta_{\rm G2}(\delta),
\label{error_function}
\end{align}
where $\beta_{\rm G1}$ is the spectral index derived by Haslam 408 MHz map and LWA maps, while $\beta_{\rm G2}$ is the spectral index derived by Haslam 408 MHz map and Guzman 45 MHz map. 
When $\delta \gg \delta_0$, we have $\beta\approx \beta_{G1}$, and when $\delta \ll \delta_0$, $\beta\approx \beta_{G2}$.
We take $\delta_0=-15\degree$ and $A=1.7$.

\begin{figure}
	\centering
	\includegraphics[width=0.45\textwidth]{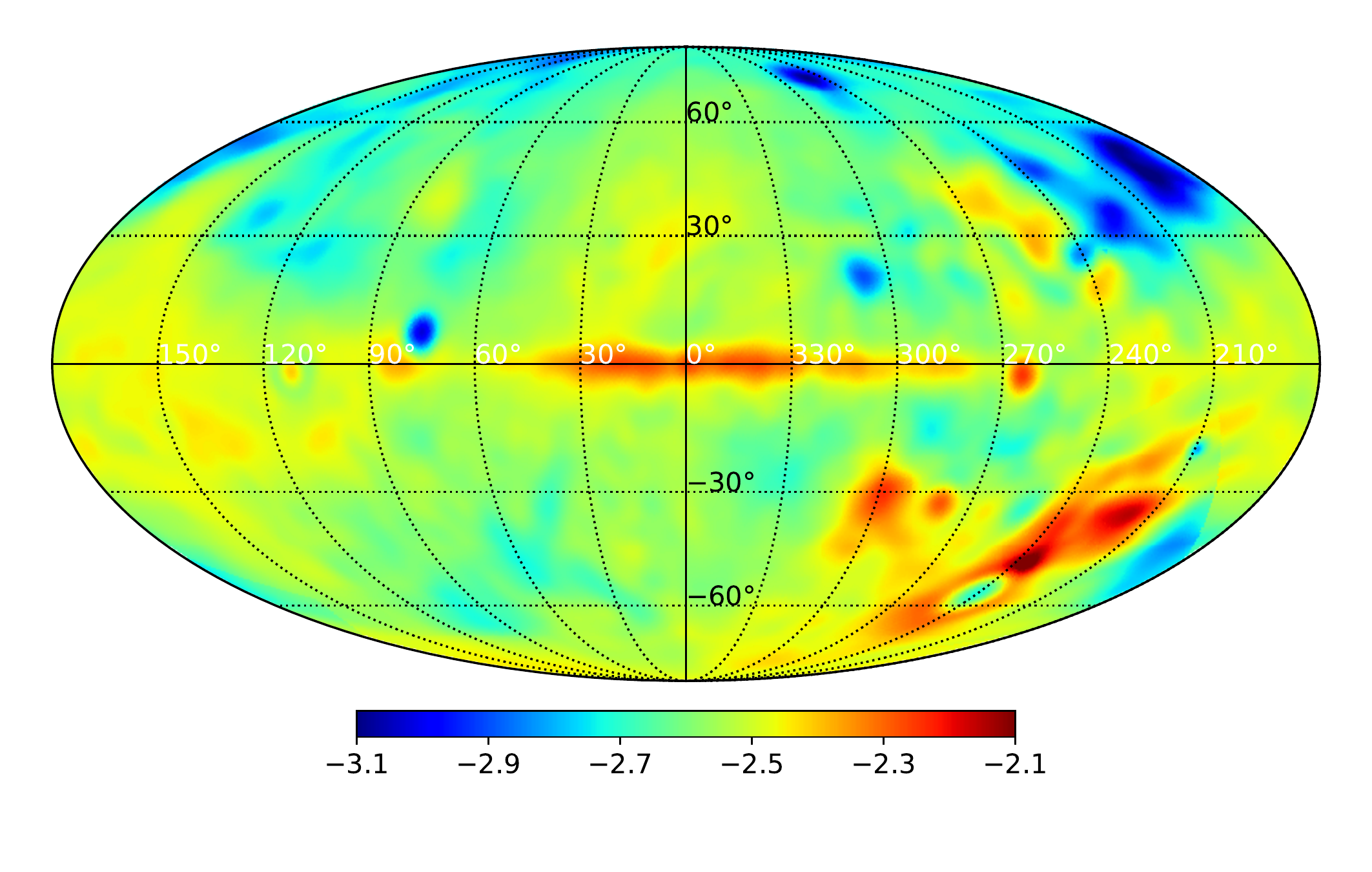}  
	\caption{The spectral index map obtained by combining the Haslam 408 MHz map with the LWA and Guzman maps.
	}
	\label{fig:spectral_index}
\end{figure}

All maps used here are smoothed by a Gaussian beam with FWHM = $5\degree$.  The resulting spectral index map is shown in Fig. \ref{fig:spectral_index}. At the Galactic plane the spectrum slope is shallower than the higher Galactic latitude, and the southern sky 
is also shallower than the northern sky. 
Note here the spectral index is for the frequency range of a few tens MHz to 408 MHz, where the free-free absorption 
is still small except for dense HII regions, so it reflects the intrinsic emission spectrum.

\begin{figure}[tbp]
\centering
\includegraphics[width=0.45\textwidth]{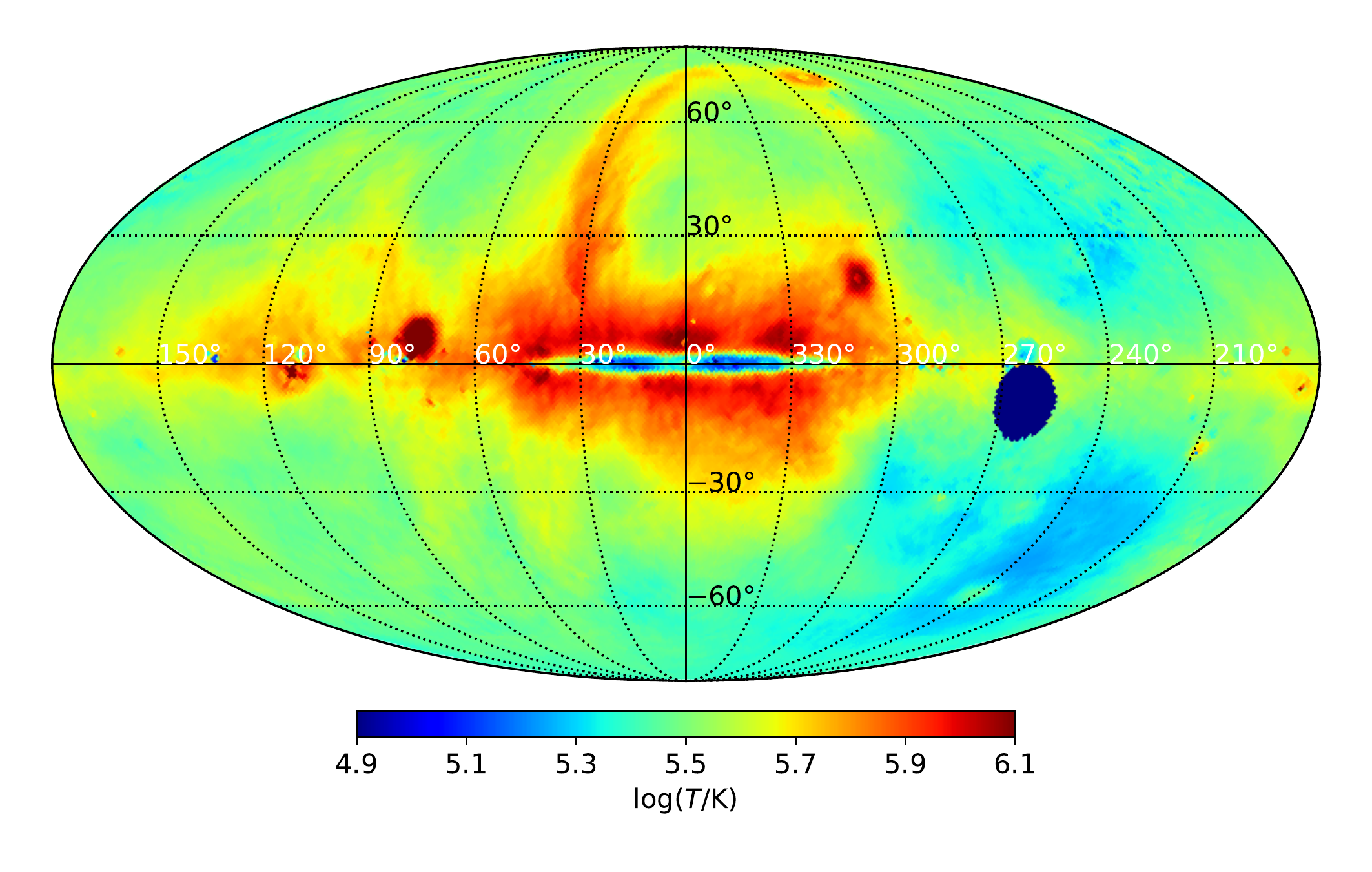}
\includegraphics[width=0.45\textwidth]{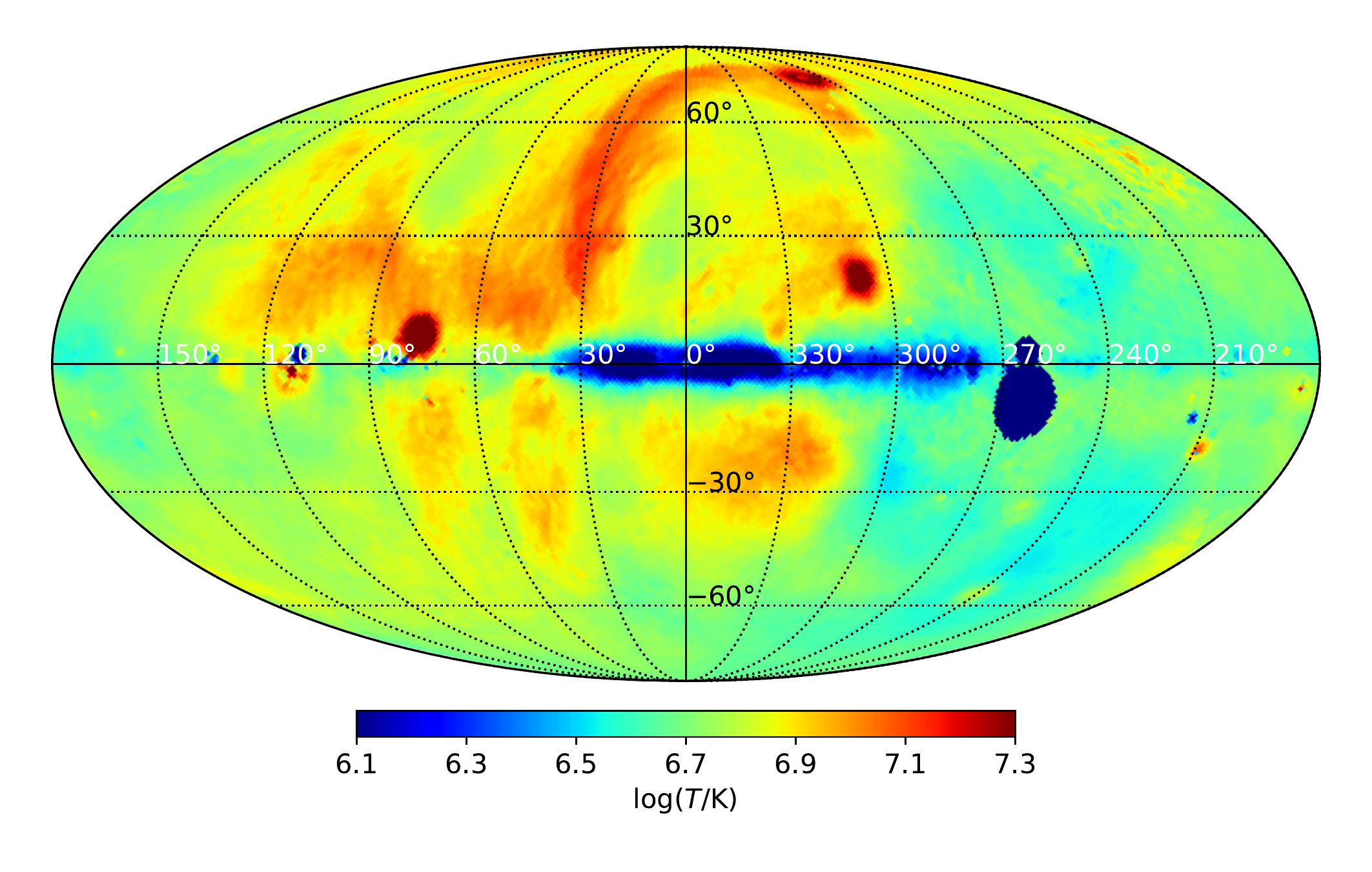}
\includegraphics[width=0.45\textwidth]{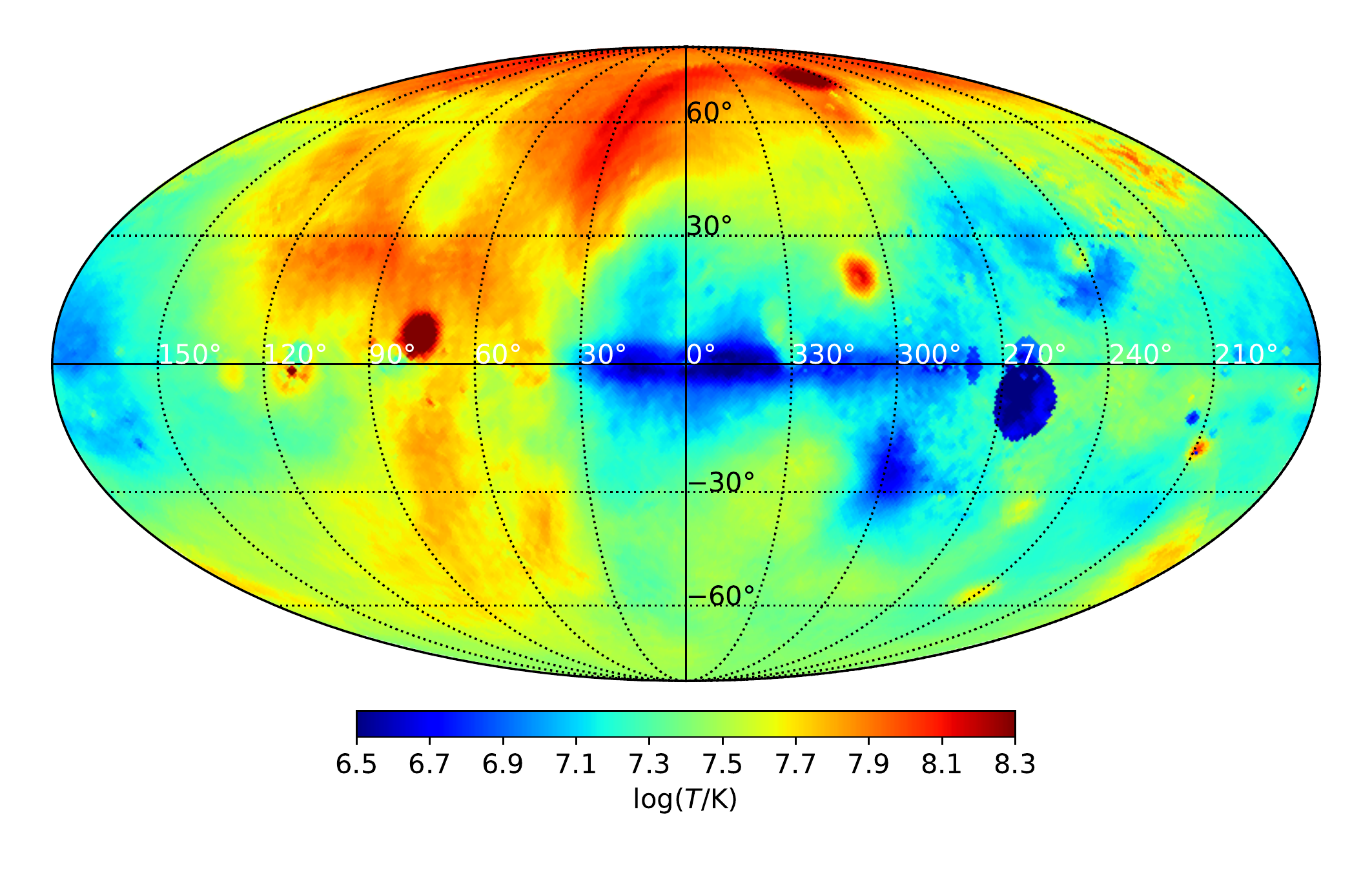}
\caption{Sky maps at (from top to bottom) 10, 3 and 1 MHz for the model with direction-dependent spectral index.
		\label{fig:fluctuation_absorption_pix_depend}}
\end{figure}

We then model the emissivity as
\begin{equation}
	\epsilon(\nu,R,Z) = A \left(\frac{R+r_1}{R_0}\right)^\alpha
	e^{-R/R_0} e^{-\abs{Z/Z_0}^{\gamma}}\left( \frac{\nu}{\nu_*} \right)^{\beta_{\rm G}(l,b)}.
	\label{eq:emissivity-spatial}
\end{equation}
It is formally same as Eq. (\ref{eq:emissivity}), except that now $\beta_{\rm G}$ is allowed to vary with direction (direction-dependent), and the value is taken from the spectrum index map obtained above.  Using the emissivity given above, and adding the free-free absorption and small-scale fluctuations
(here we also adopt $F_1=3.0$), we finally obtain the map shown in Fig. \ref{fig:fluctuation_absorption_pix_depend}. 
Compared with Fig.~\ref{fig:fluctuation_absorption}, the overall brightness is higher in these maps, because the pixels with steeper spectral index become brighter when extrapolated to the lower frequencies.

The mean sky brightness temperature is also plotted in Fig. \ref{fig:meanskybrightnessabsorpfreefreqdepend}. 
In this model, for same absorption strength ($F_1$ parameter) the sky brightness temperature is higher than 
the constant spectral index model, at frequencies below $\sim 3$ MHz. 
As we see in Fig. \ref{fig:spectral_index}, the spectrum slope at higher Galactic latitude is steeper than the lower Galactic 
latitude, and as a result, the absorption-free brightness at high Galactic latitude is larger than in the constant spectral index 
model. In addition, the free-free absorption at higher Galactic latitude is weaker. Consequently, the ultra-long wavelength sky 
 is brighter in the direction-dependent spectral index model.

\subsection{Absorption from Extended HII Region Envelopes (EHEs)}\label{sec:EHEs} 

As noted above, the {\tt NE2001} model with its fiducial  parameters may predict inadequate absorption to produce the observed downturn in the global radio spectrum. To overcome this difficulty we considered adopting a larger fluctuation parameter, i.e. 
by assuming there are more small scale fluctuations in the WIM. We now consider another more physical solution to this problem.

{\tt NE2001} includes 175 known HII objects (SNRs or HII regions) with enhanced electron densities ranging from 0.01 to  
$40 \cm^{-3}$. Most of them have density $\sim 0.5$ cm$^{-3}$ at the center, and are modeled with Gaussian profile 
of  width $\sim0.01 \kpc$. Of course this is not a complete list, there should be many more SNRs, probably between $1000 - 10000$ \citep{Berkhuijsen1984}, and HII regions (the WISE mission has already detected more than 8000, see \citealt{Anderson2014_WISE}). HII regions could be small and dense (compact or classical HII regions), or large and diffuse (diffuse HII regions, e.g. \citealt{Lockman1996,Anderson2018}). Around the classical HII regions there could be EHEs, with size extends to $\sim$ $0.05 - 0.2$ kpc, density $\sim$ $0.5 - 10$ cm$^{-3}$,  and temperature $3000 - 8000$ K, as found by the low frequency  radio recombination lines  \citep{Anantharamaiah1985a,Anantharamaiah1985b,Anantharamaiah1986,Roshi2001} or absorption of the radio continuum 
towards SNRs \citep{,Kassim1989b,Lacey2001}.

We now consider the contribution to free-free absorption if more absorbers are included, to see if they can account for the downturn of the observed global radio spectrum at $\lesssim3-5$ MHz.  Although the classical HII region is opaque for the low frequency radio flux, it is difficult for them to significantly reduce the global spectrum because their volume filling factor is too small, so here we focus only on the EHEs.

We assume the spatial distribution of EHEs in the Milky Way has a form similar to that of 
SNRs \citep{Strong1998}
\begin{equation}
q(R,Z)=q_0\left(\frac{R}{R_\odot}\right)^{\eta_{q}} {\rm exp}\left(-\xi\frac{R-R_\odot}{R_\odot}\right){\rm exp}\left(-\frac{|Z|}{Z_s}\right),
\label{eq:q}
\end{equation}   
where $Z_s=0.2$ kpc, $\eta_{q}=1.25$ and $\xi=3.56$ as constrained by the Fermi Galactic gamma-ray observations \citep{Trotta2011}. The normalization factor $q_0$ is set by 
\begin{equation}
N_{\rm EHE}=\int dZ \int dR~  2\pi R q(R,Z) ,
\end{equation}
where $N_{\rm EHE}$ is the total number of EHEs in our Milky Way, and is treated as a free parameter here.

 \begin{figure}[tbp]
	\centering
	\includegraphics[width=0.4\textwidth]{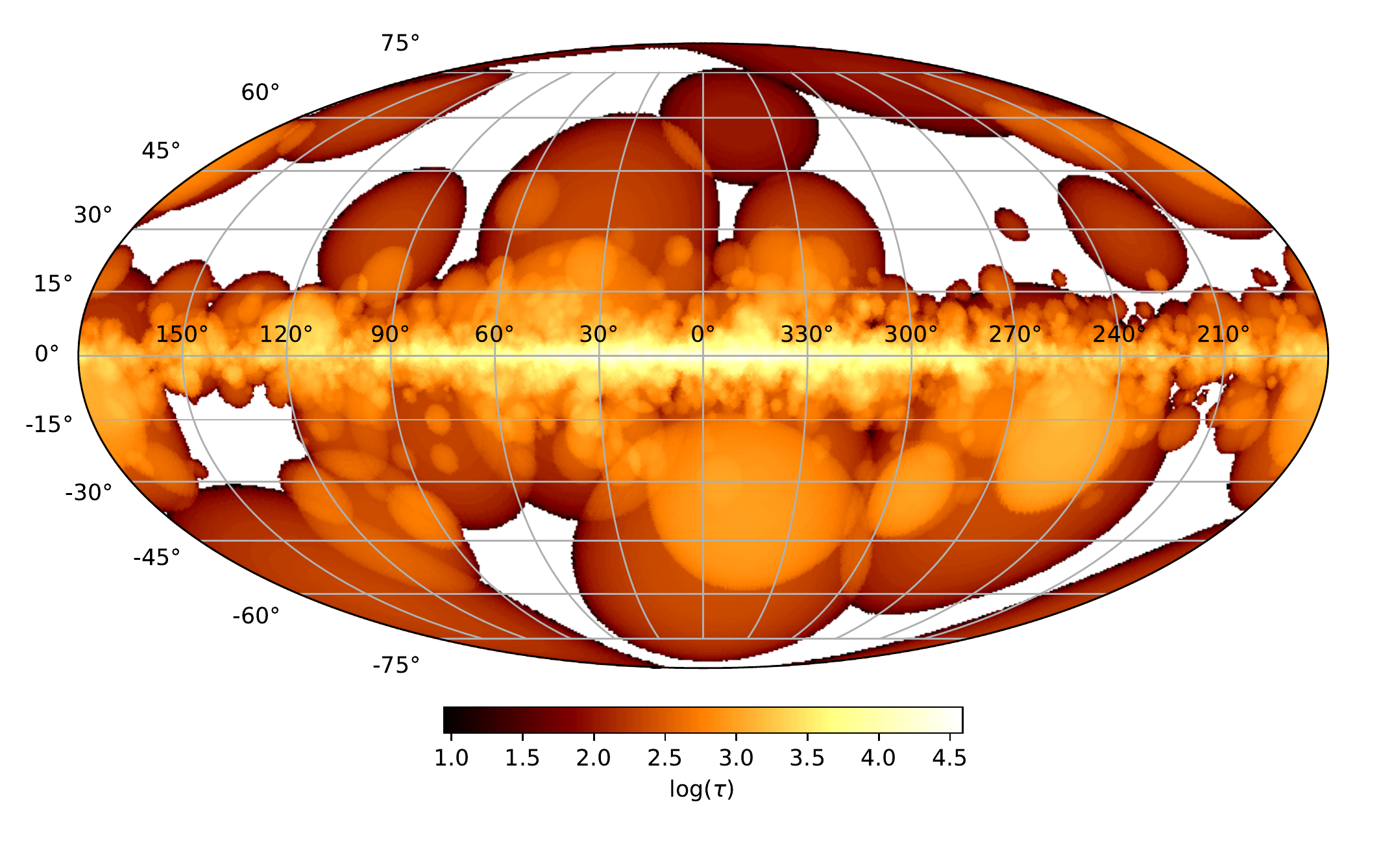}
	\includegraphics[width=0.4\textwidth]{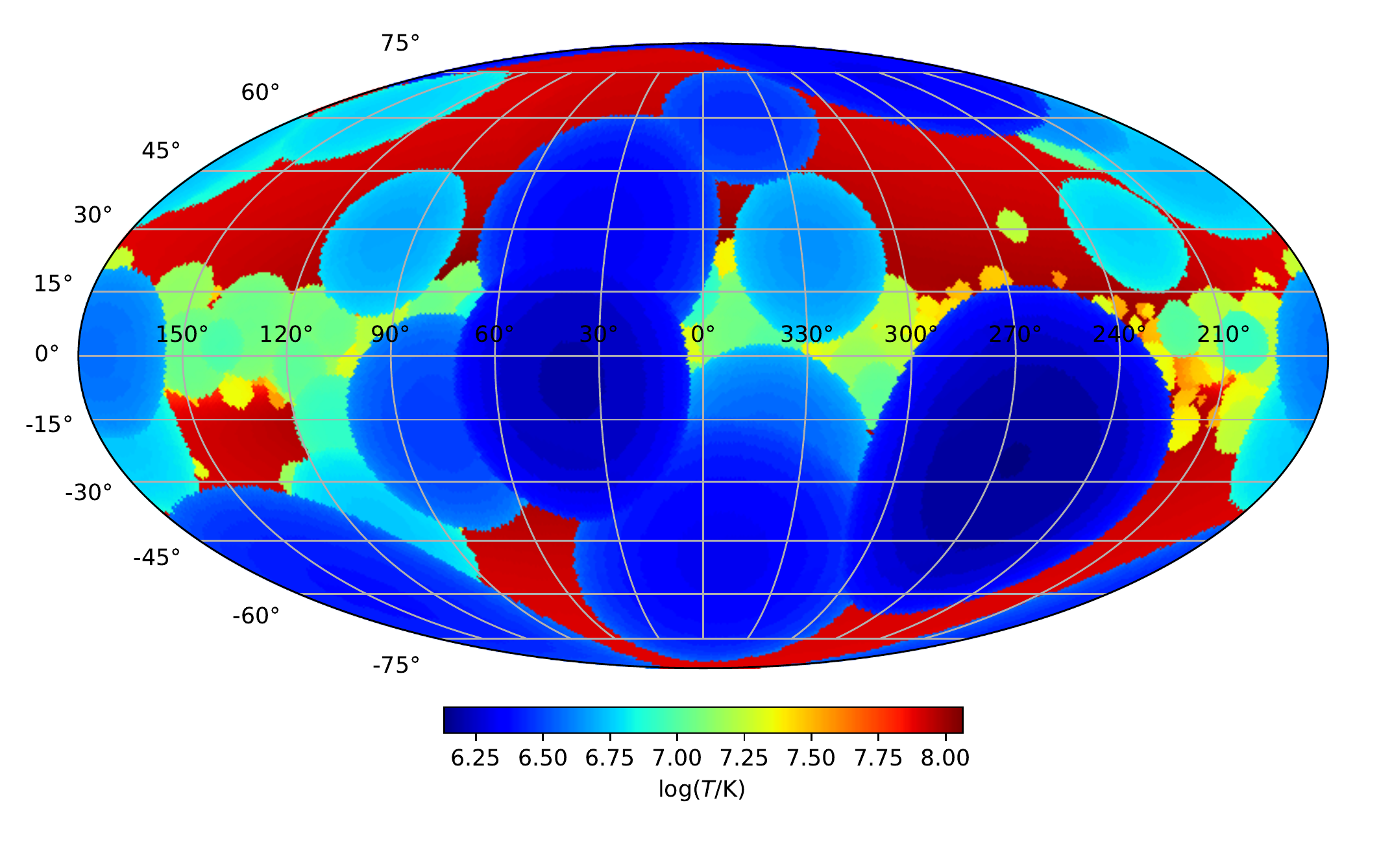}	
	\includegraphics[width=0.4\textwidth]{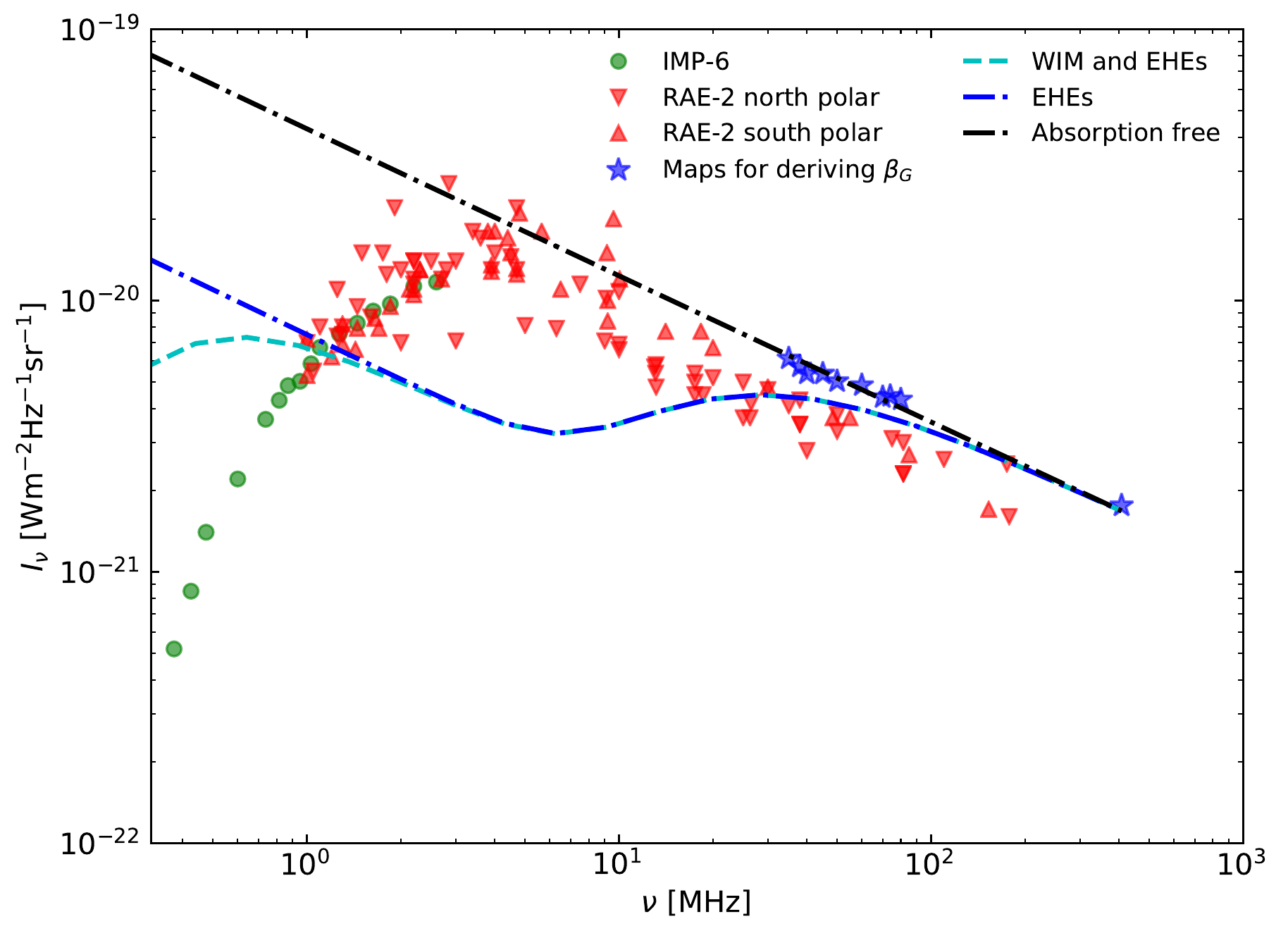}
	\caption{{\it Top}: the free-free optical depth of the mock EHEs; {\it Middle}:  the sky map; {\it Bottom}: global radio spectrum.
	\label{fig:mock_EHE}}
\end{figure}

We make random realizations of the EHEs that follow the spatial distribution of Eq. (\ref{eq:q}), and further assume that their size follows a log-normal distribution, with central value $\mean{ r_{\rm EHE}}=0.1 \kpc$ and variance $\sigma_{\log r_{\rm EHE}}=0.5$, 
 and limited in the range $0.05 - 0.2\kpc$. For simplicity, we model them with sharp edges, and within the edge the electron density is $1.0 \cm^{-3}$, to be consistent with the typical EHEs. We adopt the same density for all mock EHEs.
 Regarding the total number, we adopt $N_{\rm EHE}=20000$. We artificially remove one EHE that occasionally encompasses our Sun, it will not influence the statistics of the sample.

In the top panel of Fig. \ref{fig:mock_EHE} we plot the free-free absorption optical depth to extragalactic sources, {\it purely by our mock EHEs}, at 1 MHz. 
We then replace the HII clumps list  in  {\tt NE2001} with these mock EHEs, and then re-calculate the sky map. For all other parameters we still use the fiducial values in {\tt NE2001}, because we want to see if the mock EHEs could account for the downturn on the global radio spectrum without increasing the WIM absorption.

In the middle panel of  Fig. \ref{fig:mock_EHE} we plot the 1 MHz sky map,  and in the bottom panel we plot the global spectrum. 
For both of them the absorption from WIM and mock EHEs is involved. In the bottom panel we also plot the global spectrum if only the absorption from mock EHEs is involved.
Clearly, the mock EHEs cover $\sim80\%$ of the sky area and can absorb the ultra-long wavelength signal efficiently.
However, when we look through the shape of the global spectrum shown in the bottom panel of Fig. \ref{fig:mock_EHE}, we find that the predicted shape does not agree with the observations well.

In our random sample,  all EHEs are opaque for radio signal at 1 MHz and  they occupy $\sim80\%$ of the sky area. Therefore 
$\sim20\%$ of the flux leaks between gaps in the EHEs and can be detected unabsorbed. That is to say, the observed global spectrum is 
\begin{equation}
I(\nu)\approx(1-f_{\rm sky})I_0(\nu)+f_{\rm sky}I_0(\nu)\exp[-\tau(\nu)],
\end{equation}
where $I_0$ is the global spectrum without EHEs absorption, and $f_{\rm sky}$ is the sky coverage of the EHEs (not the volume filling factor, so it could be close to 1). Because here $f_{\rm sky}\sim 1$, so $1-f_{\rm sky}\ll 1$. When $\nu\gg 1$ MHz, the optical depth is still not that high. As a result the second term of the RHS dominates the observed global spectrum. When $\nu\sim1$ MHz, the EHEs  are opaque enough so that $I_0(\nu)\exp[-\tau(\nu)]\sim0$ and the first term of the RHS dominates. Therefore $I(\nu)$ shows complex trend with decreasing frequency, as shown by the bottom panel of Fig. \ref{fig:mock_EHE}.
The WIM absorption would change the $I_0(\nu)$, so it also changes the behavior of $I(\nu)$ a bit, but not significant, see the difference between the curve with and without WIM in the bottom panel of Fig. \ref{fig:mock_EHE}.

We further check that, to produce the downturn in such a model, we must have 
one low-density EHE encompass the Sun. If our Sun is  inside an EHE with a typical density of $\sim 0.1 \cm^{-3}$,  we can get the global spectrum to turn over at $\sim3 - 5 \MHz$ and be consistent with observations. However, we note that it is known that the Sun resides in a region known as the Local Bubble,  which has much lower density than above \citep{2011ARA&A..49..237F}.  
So if there are many EHEs they could indeed affect the global radio spectrum significantly, but to produce the downturn consistent with observations,  while also be consistent with the observed local density around the Sun, is still difficult.

\section{conclusions}\label{conclusion}

In this work,
we have developed an ultra-long wavelength radio sky model that is valid below $\sim10$ MHz. 
We first derive a cylindrical emissivity from the observed all-sky map at 408 MHz, extrapolate it to ultra-long wavelength 
using a power-law form, and then add the free-free absorption by the Galactic diffuse free electrons and some small-scale 
dense HII regions. The spectral index of the power-law is derived from observed multi-wavelength radio maps and the Galactic free electrons distribution is from the {\tt NE2001} model.

We found that if we use a constant spectral index $\beta_{\rm G}=-2.51$ for Galactic synchrotron radiation and adopt the fiducial fluctuation parameter for thick disk ($F_1=0.18$) in Galactic electron model {\tt NE2001}, we would over-predict the mean sky brightness  at $\lesssim3$ MHz compared with the observations by space satellites. The observed mean sky brightness spectrum (the specific intensity)  turns over at $\sim3$ MHz. To solve the problem, either the Galactic diffuse electrons have much larger small-scale fluctuations than modeled by {\tt NE2001}, or the intrinsic synchrotron radiation spectrum itself has shallower slope at $\lesssim3$ MHz. We investigated both solutions. The morphology of predicted sky maps in these two cases are quite different, therefore, in the future a high resolution ultra-long wavelength sky survey could potentially distinguish them.
Moreover, we also investigated a model in which the spectral index depends on direction.

In particular, we found that, at ultra-long wavelengths, the Galactic plane would be darker than higher Galactic latitude regions.
Moreover, one can see the shadows of spiral arms from the sky map, particularly at frequencies as low as $\sim1$ MHz. From the multi-frequency sky maps one can obtain the 3D information of the Galactic diffuse electrons and emissivity distribution.

Our model would be a useful tool for designing the upcoming ultra-long wavelength experiments. The first generation lunar-orbit and lunar-surface based interferometers would have limited number of antennas, hence a reasonable input sky model would be crucial. Our model can be used to test whether these interferometers can successfully and accurately recover the ultra-long wavelength sky, which has complex brightness distribution and changes dramatically with frequency. Moreover, when the new observational data is available, our model can generate low frequency sky maps, which can be used for imaging deconvolution.

We have made our model {\tt ULSA} publicly available, including the source code, the maps at different frequencies for different models, and some animated figures. They  could be downloaded at \href{https://doi.org/10.5281/zenodo.4454153}{DOI:10.5281/zenodo.4454153}.

\section*{ACKNOWLEDGMENTS}

We thank Dr. Jiaxin Wang for helpful discussions.
This work is supported by CAS Strategic Priority Research Program XDA15020200,
the National Natural Science Foundation of China (NSFC) grant 11973047, 11633004, 11653003, and the 
NSFC-ISF joint research program No. 11761141012.
BY also acknowledges the support by the NSFC-CAS  joint fund for space scientific satellites No. U1738125 and the CAS Bairen program.

\appendix
\numberwithin{equation}{section}
\numberwithin{figure}{section}

\section{The ultra-long wavelength sky model software}\label{sec:software}
\subsection{The structure of the code}
We developed a Python code package named ULSA (Ultra-Long wavelength Sky model with Absorption) to generate the 
sky map at very low frequencies, taking into account the free-free absorption effect by free electrons in both discrete HII clumps and the WIM. The code first sets up the model for the Galactic emissivity distribution 
by fitting available observational data at higher frequencies, and the free electron distribution, then generate the absorbed full-sky map.  The code files are organized as follows (all under the ULSA directory), with brief explanation in {\it italics}:
 
\begin{itemize}
		\item  {\tt spectral\_index\_fitting/} ~~~ {\it Spectral indices, three models are provided.} 
			\subitem {\tt spectral\_index\_constant.py}
			\subitem {\tt spectral\_index\_frequency\_dependent.py}
			\subitem {\tt spectral\_index\_direction\_dependent.py}
		\item  {\tt emissivity\_fitting/} ~~~ {\it Set up emissivity model}
			\subitem {\tt produce\_data\_for\_fitting.py}:  ~~~ {\it Pre-process the data for emissivity fitting,  removing nearby structures “NPS” and “Loop I” }
			\subitem {\tt fit\_emissivity\_params.py}:  ~~~   {\it Fit the emissivity parameters}.
		\item  {\tt NE2001/*}   ~~~ {\it The Galactic electron model}
		\item  {\tt sky\_map/} ~~~ {\it Make the sky map.}
			\subitem {\tt produce\_absorbed\_sky\_map.py}
\end{itemize}

The computations are done as the following three steps: 

\begin{enumerate}
\item Spectral Index Parameter Fitting.
As described in the text, we have developed three  models, with (i) constant spectral index; (ii) frequency-dependent spectral 
index (iii) direction-dependent spectral index. The user may choose one of these. In this step the code produces the spectral index parameters for use in the next step.  Once this is done, the sky map can be extrapolated from 408 MHz to lower frequencies. 
\item Galactic Emissivity Model Construction.
We then derive the emissivity model parameters using the code in the {\tt emissivity\_fitting/}  directory. 
First, we pre-process the data by removing local structures such as the NPS and Loop I from the 408 MHz map,  then 
the emissivity model parameters are obtained by fitting this map.
\item Low Frequency Sky Map Generation.
Finally,  the sky map is generated with the emissivity model 
constructed above and free-free absorption effect computed using the {\tt NE2001} model.  
\end{enumerate}

To run the code, one can simply call the {\tt absorption\_JRZ} function, which serves as the drive routine, with appropriate choice of various parameters, 
\begin{small} \\
{\tt >>> from ULSA.sky\_map.produce\_absorbed\_sky\_map import absorption\_JRZ \\
	>>> f = absorption\_JRZ(v, nside, index\_type, distance, using\_raw\_diffuse, using\_default\_params,\\
	critical\_dis,output\_absorp\_free\_skymap).mpi()\\
}
\end{small}\\
The parameters in the command are
\begin{itemize}
	\item {\bf v}(float): frequency in MHz for the output map;
	\item {\bf nside} (int): the Healpix NSIDE value for the sky map;
	\item {\bf index\_type} (str):  spectral index modeling option,  one of  `constant\_index', `freq\_dependent\_index', `direction\_dependent\_index';
	\item {\bf distance kpc}: maximum integration distance along line of sight, default is set to 50 kpc;
	\item {\bf using\_raw\_diffue} (bool): if False, the data will be smoothed by Gaussian kernel, otherwise use the raw data;
	\item {\bf v\_file\_dir} (dict):  a dictionary structure used to specify additional input map data. Specify the frequency of the map by
	the dictionary key, and the relative path of the map data as dictionary value. The  input sky map file should be in HDF5 format, such as \{XX:``/dir/xxx.hdf5"\}. if None, the spectral index is calculated with the existing data. 
	\item {\bf using\_default\_params} (bool): if True, use the default spectral index value in the code, or re-calculate the spectral index value if False.  
	\item{\bf input\_spectral\_index} (array): one can specify the spectral index value by putting in an array containing the spectral index map in the direction-dependent case, or containing one element for the constant or frequency dependent cases.
	\item {\bf params\_408} (list): the emissivity model parameters ($[A,R_0,\alpha,Z_0,\gamma]$) obtained by fitting the Haslam 408 MHz sky map.  If this parameter is omitted, the values given in Table \ref{table_params} will be used as defaults. One can also specify these parameters directly 
	by putting in the values, or force the code to re-fit by setting it to [0.,0.,0.,0.,0.].	
	\item {\bf critical\_dis} (bool): if True, calculating the half-brightness distance, otherwise this is not calculated. 
	\item {\bf output\_absorp\_free\_skymap} (bool): if True, produce absorption-free sky map at frequency {\bf v} as well.  
\end{itemize}

The function returns a two dimensional matrix, 
the first column is the {\tt Healpix} pixel number, and the second column gives the corresponding sky map. 
One can then plot this map using {\bf healpy.mollview}. The default coordinate system is set to the Galactic coordinate system, but 
one can transform it to the equatorial coordinate system by calling the Healpix function {\bf f.change\_coord(m,[``G",``C"])}. 
See the {\tt healpy} documentation for more information.  Additionally, some results are also automatically as an HDF5 file. 
The map is  saved in a file with name of the form {\tt freqMHz\_sky\_map\_with\_absorption.hdf5}, where {\tt freq} is the frequency in MHz. This file contains two keys which are respectively named ``spectral\_index" for spectral index map, and ``data" for the  sky map
at the given frequency.
	
An absorption-free sky map is computed as intermediate result though not saved automatically, but if the parameter 
{\tt output\_absorp\_free\_skymap} is set as True, it will be saved in a file named in the format {\tt freqMHz\_absor p\_free\_skymap.hdf5}.
Moreover, the half-brightness distance will be computed and saved as {\tt freqMHz\_critical\_dist.hdf5} if the parameter 
{\tt critical\_dist} is set as True.

\subsection{An example of using the code}
In the example shown below, sky maps are produced at 1, 2, ... 10 MHz with the constant spectral index model. 
 
\begin{lstlisting}
from ULSA.sky_map.produce_absorbed_sky_map import absorption_JRZ
import healpy as hp
import numpy as np
import matplotlib.pyplot as plt

def produce_sky_map():
	sky_map_list = []
	# initial params setting
	# NSIDE in healpix
	nside = 2**6
	# distance in kpc unit
	dist = 50.
	# calculate the skymap from 1 MHz to 10 Mhz
	for v in range(1,10,1):
		f = absorption_JRZ(v = v, nside = nside, index_type = 'constant_index', distance = dist,using_raw_diffuse = False,using_default_params=False,input_spectral_index=None,critical_dis = False,output_absorp_free_skymap = False)
		sky_map_list.append(f.mpi())
	# we got a list of sky_map with frequency from 1 MHz to 10 MHz with step 1 MHz.	
	return sky_map_list
	
# then plot the data using mollview
def plot():
	sky_map_list = produce_sky_map()
	plt.figure(1)
	for sky_map in sky_map_list:
		hp.mollview(np.log10(sky_map),cmap = plt.cm.jet)
		plt.show() 
		# or plt.savefig('xxx.eps',format='eps')
# calling the function to work		
plot()
\end{lstlisting}

\subsection{Adding new input map data}

All the observation maps used in this work are under the directory  {\tt obs\_sky\_data/}, one can also  add  new map data (e.g. at additional frequency) there. The map file should be in the HDF5 format, with first column HEALPIX pixel number, second column sky temperature values in equatorial  
coordinate. Most observed data in their original form is in this coordinate; if not, one must convert it into equatorial coordinate. 
To use such new data, first, put the new data under {\tt obs\_sky\_data/}; then, when calling the our function, register the new data by a dictionary whose key is the frequency of this new input map, and the value is the path relative to {\tt obs\_sky\_data/}. For example, if there is a new map at 22 MHz, one first creates a file  {\tt 22MHz\_sky\_map.hdf5}, putting it under {\tt obs\_sky\_data/22MHz/22MHz\_sky\_map.hdf5}; then, when calling the {\tt absorption\_JRZ} function, provide  {\tt v\_file\_dir = \{22:"/22MHz/22MHz\_sky\_map.hdf5"\}}. The code will then add the new data when calculating the spectral index. 

\bibliographystyle{aasjournal}

\bibliography{refe}

\begin{thebibliography}{}
\expandafter\ifx\csname natexlab\endcsname\relax\def\natexlab#1{#1}\fi

\bibitem[{{Ackermann} {et~al.}(2012){Ackermann}, {Ajello}, {Atwood}, {Baldini},
  {Ballet}, {Barbiellini}, {Bastieri}, {Bechtol}, {Bellazzini}, {Berenji},
  {Blandford}, {Bloom}, {Bonamente}, {Borgland}, {Brandt}, {Bregeon},
  {Brigida}, {Bruel}, {Buehler}, {Buson}, {Caliandro}, {Cameron}, {Caraveo},
  {Cavazzuti}, {Cecchi}, {Charles}, {Chekhtman}, {Chiang}, {Ciprini}, {Claus},
  {Cohen-Tanugi}, {Conrad}, {Cutini}, {de Angelis}, {de Palma}, {Dermer},
  {Digel}, {Silva}, {Drell}, {Drlica-Wagner}, {Falletti}, {Favuzzi}, {Fegan},
  {Ferrara}, {Focke}, {Fortin}, {Fukazawa}, {Funk}, {Fusco}, {Gaggero},
  {Gargano}, {Germani}, {Giglietto}, {Giordano}, {Giroletti}, {Glanzman},
  {Godfrey}, {Grove}, {Guiriec}, {Gustafsson}, {Hadasch}, {Hanabata},
  {Harding}, {Hayashida}, {Hays}, {Horan}, {Hou}, {Hughes}, {J{\'o}hannesson},
  {Johnson}, {Johnson}, {Kamae}, {Katagiri}, {Kataoka}, {Kn{\"o}dlseder},
  {Kuss}, {Lande}, {Latronico}, {Lee}, {Lemoine-Goumard}, {Longo}, {Loparco},
  {Lott}, {Lovellette}, {Lubrano}, {Mazziotta}, {McEnery}, {Michelson},
  {Mitthumsiri}, {Mizuno}, {Monte}, {Monzani}, {Morselli}, {Moskalenko},
  {Murgia}, {Naumann-Godo}, {Norris}, {Nuss}, {Ohsugi}, {Okumura}, {Omodei},
  {Orlando}, {Ormes}, {Paneque}, {Panetta}, {Parent}, {Pesce-Rollins},
  {Pierbattista}, {Piron}, {Pivato}, {Porter}, {Rain{\`o}}, {Rando}, {Razzano},
  {Razzaque}, {Reimer}, {Reimer}, {Sadrozinski}, {Sgr{\`o}}, {Siskind},
  {Spandre}, {Spinelli}, {Strong}, {Suson}, {Takahashi}, {Tanaka}, {Thayer},
  {Thayer}, {Thompson}, {Tibaldo}, {Tinivella}, {Torres}, {Tosti}, {Troja},
  {Usher}, {Vandenbroucke}, {Vasileiou}, {Vianello}, {Vitale}, {Waite}, {Wang},
  {Winer}, {Wood}, {Wood}, {Yang}, {Ziegler}, \& {Zimmer}}]{Ackermann2012}
{Ackermann}, M., {Ajello}, M., {Atwood}, W.~B., {et~al.} 2012, \apj, 750, 3

\bibitem[{{Alexander} {et~al.}(1969){Alexander}, {Brown}, {Clark}, {Stone}, \&
  {Weber}}]{Alexander1969}
{Alexander}, J.~K., {Brown}, L.~W., {Clark}, T.~A., {Stone}, R.~G., \& {Weber},
  R.~R. 1969, \apjl, 157, L163

\bibitem[{{Alexander} {et~al.}(1975){Alexander}, {Kaiser}, {Novaco}, {Grena},
  \& {Weber}}]{Alexander1975}
{Alexander}, J.~K., {Kaiser}, M.~L., {Novaco}, J.~C., {Grena}, F.~R., \&
  {Weber}, R.~R. 1975, \aap, 40, 365

\bibitem[{{Alexander} \& {Novaco}(1974)}]{Alexander1974}
{Alexander}, J.~K., \& {Novaco}, J.~C. 1974, \aj, 79, 777

\bibitem[{{Alexander} \& {Stone}(1965)}]{Alexander1965}
{Alexander}, J.~K., \& {Stone}, R.~G. 1965, \apj, 142, 1327

\bibitem[{{Anantharamaiah}(1985{\natexlab{a}})}]{Anantharamaiah1985a}
{Anantharamaiah}, K.~R. 1985{\natexlab{a}}, \japa, 6, 177

\bibitem[{{Anantharamaiah}(1985{\natexlab{b}})}]{Anantharamaiah1985b}
---. 1985{\natexlab{b}}, \japa, 6, 203

\bibitem[{{Anantharamaiah}(1986)}]{Anantharamaiah1986}
---. 1986, \japa, 7, 131

\bibitem[{{Anderson} {et~al.}(2018){Anderson}, {Armentrout}, {Luisi}, {Bania},
  {Balser}, \& {Wenger}}]{Anderson2018}
{Anderson}, L.~D., {Armentrout}, W.~P., {Luisi}, M., {et~al.} 2018, \apjs, 234,
  33

\bibitem[{{Anderson} {et~al.}(2014){Anderson}, {Bania}, {Balser}, {Cunningham},
  {Wenger}, {Johnstone}, \& {Armentrout}}]{Anderson2014_WISE}
{Anderson}, L.~D., {Bania}, T.~M., {Balser}, D.~S., {et~al.} 2014, \apjs, 212,
  1

\bibitem[{{Berkhuijsen}(1984)}]{Berkhuijsen1984}
{Berkhuijsen}, E.~M. 1984, \aap, 140, 431

\bibitem[{{Bisschoff} {et~al.}(2019){Bisschoff}, {Potgieter}, \&
  {Aslam}}]{Bisschoff2019}
{Bisschoff}, D., {Potgieter}, M.~S., \& {Aslam}, O.~P.~M. 2019, \apj, 878, 59

\bibitem[{{Bridle}(1969)}]{Bridle1969}
{Bridle}, A.~H. 1969, \nat, 221, 648

\bibitem[{{Brown}(1973)}]{Brown1973}
{Brown}, L.~W. 1973, \apj, 180, 359

\bibitem[{{Burns} {et~al.}(2019){Burns}, {Hallinan}, {Lux}, {Teitelbaum},
  {Kocz}, {MacDowall}, {Bradley}, {Rapetti}, {Wu}, {Furlanetto}, {Austin},
  {Romero-Wolf}, {Chang}, {Bowman}, {Kasper}, {Anderson}, {Zhen}, {Pober}, \&
  {Mirocha}}]{FARSIDE}
{Burns}, J.~O., {Hallinan}, G., {Lux}, J., {et~al.} 2019, arXiv e-prints,
  arXiv:1911.08649

\bibitem[{{Cane}(1979)}]{Cane1979}
{Cane}, H.~V. 1979, \mnras, 189, 465

\bibitem[{{Cane} \& {Whitham}(1977)}]{Cane1977}
{Cane}, H.~V., \& {Whitham}, P.~S. 1977, \mnras, 179, 21

\bibitem[{{Caswell}(1976)}]{Caswell1976_10MHz}
{Caswell}, J.~L. 1976, \mnras, 177, 601

\bibitem[{{Chen} {et~al.}(2019){Chen}, {Burns}, {Koopmans}, {Rothkaehi},
  {Silk}, {Wu}, {Boonstra}, {Cecconi}, {Chiang}, {Chen}, {Deng}, {Falanga},
  {Falcke}, {Fan}, {Fang}, {Fialkov}, {Gurvits}, {Ji}, {Kasper}, {Li}, {Mao},
  {Mckinley}, {Monsalve}, {Peterson}, {Ping}, {Subrahmanyan}, {Vedantham},
  {Klein Wolt}, {Wu}, {Xu}, {Yan}, \& {Yue}}]{ChenXL2019DSL}
{Chen}, X., {Burns}, J., {Koopmans}, L., {et~al.} 2019, arXiv e-prints,
  arXiv:1907.10853

\bibitem[{Condon \& Ransom(2016)}]{condon2016essential}
Condon, J.~J., \& Ransom, S.~M. 2016, Essential radio astronomy, Vol.~2
  (Princeton University Press)

\bibitem[{{Cordes} \& {Lazio}(1991)}]{cordes1991interstellar}
{Cordes}, J.~M., \& {Lazio}, T.~J. 1991, \apj, 376, 123

\bibitem[{Cordes \& Lazio(2002)}]{cordes2002ne2001}
Cordes, J.~M., \& Lazio, T. J.~W. 2002, arXiv preprint astro-ph/0207156

\bibitem[{Cordes \& Lazio(2003)}]{cordes2003ne2001}
---. 2003, arXiv preprint astro-ph/0301598

\bibitem[{Cordes {et~al.}(1991)Cordes, Weisberg, Frail, Spangler, \&
  Ryan}]{cordes1991galactic}
Cordes, J.~M., Weisberg, J., Frail, D., Spangler, S., \& Ryan, M. 1991, \nat,
  354, 121

\bibitem[{{Cordes} {et~al.}(1985){Cordes}, {Weisberg}, \&
  {Boriakoff}}]{cordes1985small}
{Cordes}, J.~M., {Weisberg}, J.~M., \& {Boriakoff}, V. 1985, \apj, 288, 221

\bibitem[{{Danny C. Price}(2016)}]{PyGSM2016}
{Danny C. Price}. 2016, {PyGSM: Python interface to the Global Sky Model}, , ,
  ascl:1603.013

\bibitem[{{de Avillez} {et~al.}(2012){de Avillez}, {Asgekar}, {Breitschwerdt},
  \& {Spitoni}}]{deAvillez2012}
{de Avillez}, M.~A., {Asgekar}, A., {Breitschwerdt}, D., \& {Spitoni}, E. 2012,
  \mnras, 423, L107

\bibitem[{{de Oliveira-Costa} {et~al.}(2008){de Oliveira-Costa}, {Tegmark},
  {Gaensler}, {Jonas}, {Landecker}, \& {Reich}}]{deOliveira-Costa2008GSM}
{de Oliveira-Costa}, A., {Tegmark}, M., {Gaensler}, B.~M., {et~al.} 2008,
  \mnras, 388, 247

\bibitem[{{Dickinson}(2018)}]{Dickinson2018}
{Dickinson}, C. 2018, \galaxies, 6, 56

\bibitem[{{Dickinson} {et~al.}(2003){Dickinson}, {Davies}, \&
  {Davis}}]{Dickinson2003}
{Dickinson}, C., {Davies}, R.~D., \& {Davis}, R.~J. 2003, \mnras, 341, 369

\bibitem[{{Dowell} \& {Taylor}(2018)}]{Dowell2018}
{Dowell}, J., \& {Taylor}, G.~B. 2018, \apjl, 858, L9

\bibitem[{{Dowell} {et~al.}(2017){Dowell}, {Taylor}, {Schinzel}, {Kassim}, \&
  {Stovall}}]{dowell2017lwa1}
{Dowell}, J., {Taylor}, G.~B., {Schinzel}, F.~K., {Kassim}, N.~E., \&
  {Stovall}, K. 2017, \mnras, 469, 4537

\bibitem[{{Draine}(2011)}]{Draine2011}
{Draine}, B.~T. 2011, {Physics of the Interstellar and Intergalactic Medium}

\bibitem[{{Ellis} \& {Hamilton}(1966)}]{Ellis1966}
{Ellis}, G.~R.~A., \& {Hamilton}, P.~A. 1966, \apj, 143, 227

\bibitem[{{Ferri{\`e}re}(2001)}]{Ferriere2001}
{Ferri{\`e}re}, K.~M. 2001, \rvmp, 73, 1031

\bibitem[{{Frisch} {et~al.}(2011){Frisch}, {Redfield}, \&
  {Slavin}}]{2011ARA&A..49..237F}
{Frisch}, P.~C., {Redfield}, S., \& {Slavin}, J.~D. 2011, \araa, 49, 237

\bibitem[{{Gaensler} {et~al.}(2008){Gaensler}, {Madsen}, {Chatterjee}, \&
  {Mao}}]{gaensler2008vertical}
{Gaensler}, B.~M., {Madsen}, G.~J., {Chatterjee}, S., \& {Mao}, S.~A. 2008,
  \pasa, 25, 184

\bibitem[{{George} {et~al.}(2015){George}, {Orchiston}, {Slee}, \&
  {Wielebinski}}]{George2015}
{George}, M., {Orchiston}, W., {Slee}, B., \& {Wielebinski}, R. 2015, \jahh,
  18, 14

\bibitem[{{Gervasi} {et~al.}(2008){Gervasi}, {Tartari}, {Zannoni}, {Boella}, \&
  {Sironi}}]{Gervasi2008}
{Gervasi}, M., {Tartari}, A., {Zannoni}, M., {Boella}, G., \& {Sironi}, G.
  2008, \apj, 682, 223

\bibitem[{{Ghisellini}(2013)}]{Ghisellini2013}
{Ghisellini}, G. 2013, {Radiative Processes in High Energy Astrophysics}, Vol.
  873, doi:10.1007/978-3-319-00612-3

\bibitem[{{G{\'o}mez} {et~al.}(2001){G{\'o}mez}, {Benjamin}, \&
  {Cox}}]{gomez2001reexamination}
{G{\'o}mez}, G.~C., {Benjamin}, R.~A., \& {Cox}, D.~P. 2001, \aj, 122, 908

\bibitem[{{Guzm{\'a}n} {et~al.}(2011){Guzm{\'a}n}, {May}, {Alvarez}, \&
  {Maeda}}]{guzman2011all}
{Guzm{\'a}n}, A.~E., {May}, J., {Alvarez}, H., \& {Maeda}, K. 2011, \aap, 525,
  A138

\bibitem[{{Haslam} {et~al.}(1981){Haslam}, {Klein}, {Salter}, {Stoffel},
  {Wilson}, {Cleary}, {Cooke}, \& {Thomasson}}]{haslam1981408}
{Haslam}, C.~G.~T., {Klein}, U., {Salter}, C.~J., {et~al.} 1981, \aap, 100, 209

\bibitem[{{Haslam} {et~al.}(1982){Haslam}, {Salter}, {Stoffel}, \&
  {Wilson}}]{haslam1982408}
{Haslam}, C.~G.~T., {Salter}, C.~J., {Stoffel}, H., \& {Wilson}, W.~E. 1982,
  \aaps, 47, 1

\bibitem[{{Haslam} {et~al.}(1974){Haslam}, {Wilson}, {Graham}, \&
  {Hunt}}]{haslam1974further}
{Haslam}, C.~G.~T., {Wilson}, W.~E., {Graham}, D.~A., \& {Hunt}, G.~C. 1974,
  \aaps, 13, 359

\bibitem[{{Hindson} {et~al.}(2016){Hindson}, {Johnston-Hollitt},
  {Hurley-Walker}, {Callingham}, {Su}, {Morgan}, {Bell}, {Bernardi}, {Bowman},
  {Briggs}, {Cappallo}, {Deshpande}, {Dwarakanath}, {For}, {Gaensler},
  {Greenhill}, {Hancock}, {Hazelton}, {Kapi{\'n}ska}, {Kaplan}, {Lenc},
  {Lonsdale}, {Mckinley}, {McWhirter}, {Mitchell}, {Morales}, {Morgan},
  {Oberoi}, {Offringa}, {Ord}, {Procopio}, {Prabu}, {Shankar}, {Srivani},
  {Staveley-Smith}, {Subrahmanyan}, {Tingay}, {Wayth}, {Webster}, {Williams},
  {Williams}, {Wu}, \& {Zheng}}]{Hindson2016}
{Hindson}, L., {Johnston-Hollitt}, M., {Hurley-Walker}, N., {et~al.} 2016,
  \pasa, 33, e020

\bibitem[{{Hinshaw} {et~al.}(2009){Hinshaw}, {Weiland}, {Hill}, {Odegard},
  {Larson}, {Bennett}, {Dunkley}, {Gold}, {Greason}, {Jarosik}, {Komatsu},
  {Nolta}, {Page}, {Spergel}, {Wollack}, {Halpern}, {Kogut}, {Limon}, {Meyer},
  {Tucker}, \& {Wright}}]{WMAP5yr}
{Hinshaw}, G., {Weiland}, J.~L., {Hill}, R.~S., {et~al.} 2009, \apjs, 180, 225

\bibitem[{{Huang} {et~al.}(2019){Huang}, {Wu}, \& {Chen}}]{Huang2019SSM}
{Huang}, Q., {Wu}, F., \& {Chen}, X. 2019, \scpma, 62, 989511

\bibitem[{{Hurley-Walker} {et~al.}(2017){Hurley-Walker}, {Callingham},
  {Hancock}, {Franzen}, {Hindson}, {Kapi{\'n}ska}, {Morgan}, {Offringa},
  {Wayth}, {Wu}, {Zheng}, {Murphy}, {Bell}, {Dwarakanath}, {For}, {Gaensler},
  {Johnston-Hollitt}, {Lenc}, {Procopio}, {Staveley-Smith}, {Ekers}, {Bowman},
  {Briggs}, {Cappallo}, {Deshpande}, {Greenhill}, {Hazelton}, {Kaplan},
  {Lonsdale}, {McWhirter}, {Mitchell}, {Morales}, {Morgan}, {Oberoi}, {Ord},
  {Prabu}, {Shankar}, {Srivani}, {Subrahmanyan}, {Tingay}, {Webster},
  {Williams}, \& {Williams}}]{hurley2017galactic}
{Hurley-Walker}, N., {Callingham}, J.~R., {Hancock}, P.~J., {et~al.} 2017,
  \mnras, 464, 1146

\bibitem[{{Iliev} {et~al.}(2007){Iliev}, {Mellema}, {Shapiro}, \&
  {Pen}}]{Iliev2007}
{Iliev}, I.~T., {Mellema}, G., {Shapiro}, P.~R., \& {Pen}, U.-L. 2007, \mnras,
  376, 534

\bibitem[{{Jester} \& {Falcke}(2009)}]{Jester2009}
{Jester}, S., \& {Falcke}, H. 2009, \nar, 53, 1

\bibitem[{{Kassim}(1988)}]{Kassim1988}
{Kassim}, N.~E. 1988, \apjs, 68, 715

\bibitem[{{Kassim}(1989)}]{Kassim1989b}
---. 1989, \apj, 347, 915

\bibitem[{{Keshet} {et~al.}(2004){Keshet}, {Waxman}, \&
  {Loeb}}]{2004ApJ...617..281K}
{Keshet}, U., {Waxman}, E., \& {Loeb}, A. 2004, \apj, 617, 281

\bibitem[{{Kim} {et~al.}(2018){Kim}, {Liu}, \& {Switzer}}]{eGSM2018}
{Kim}, D., {Liu}, A., \& {Switzer}, E. 2018, in 2018AAS~~~231, 153.09

\bibitem[{{Kogut}(2012)}]{Kogut2012}
{Kogut}, A. 2012, \apj, 753, 110

\bibitem[{{Kogut} {et~al.}(2011){Kogut}, {Fixsen}, {Levin}, {Limon}, {Lubin},
  {Mirel}, {Seiffert}, {Singal}, {Villela}, {Wollack}, \&
  {Wuensche}}]{Kogut2011}
{Kogut}, A., {Fixsen}, D.~J., {Levin}, S.~M., {et~al.} 2011, \apj, 734, 4

\bibitem[{{Lacey} {et~al.}(2001){Lacey}, {Lazio}, {Kassim}, {Duric}, {Briggs},
  \& {Dyer}}]{Lacey2001}
{Lacey}, C.~K., {Lazio}, T. J.~W., {Kassim}, N.~E., {et~al.} 2001, \apj, 559,
  954

\bibitem[{{Lian} {et~al.}(2020){Lian}, {Xu}, {Zhu}, \& {Hu}}]{Lian2020}
{Lian}, X., {Xu}, H., {Zhu}, Z., \& {Hu}, D. 2020, \mnras, 496, 1232

\bibitem[{{Lockman} {et~al.}(1996){Lockman}, {Pisano}, \&
  {Howard}}]{Lockman1996}
{Lockman}, F.~J., {Pisano}, D.~J., \& {Howard}, G.~J. 1996, \apj, 472, 173

\bibitem[{{McKinley} {et~al.}(2018){McKinley}, {Bernardi}, {Trott}, {Line},
  {Wayth}, {Offringa}, {Pindor}, {Jordan}, {Sokolowski}, {Tingay}, {Lenc},
  {Hurley-Walker}, {Bowman}, {Briggs}, \& {Webster}}]{mckinley2018measuring}
{McKinley}, B., {Bernardi}, G., {Trott}, C.~M., {et~al.} 2018, \mnras, 481,
  5034

\bibitem[{{Nava} {et~al.}(2017){Nava}, {Benyamin}, {Piran}, \&
  {Shaviv}}]{Nava2017}
{Nava}, L., {Benyamin}, D., {Piran}, T., \& {Shaviv}, N.~J. 2017, \mnras, 466,
  3674

\bibitem[{{Ni{\c{t}}u} {et~al.}(2021){Ni{\c{t}}u}, {Bevins}, {Bray}, \&
  {Scaife}}]{Nitu2021}
{Ni{\c{t}}u}, I.~C., {Bevins}, H.~T.~J., {Bray}, J.~D., \& {Scaife}, A.~M.~M.
  2021, \aph, 126, 102532

\bibitem[{{Nord} {et~al.}(2006){Nord}, {Henning}, {Rand}, {Lazio}, \&
  {Kassim}}]{Nord2006}
{Nord}, M.~E., {Henning}, P.~A., {Rand}, R.~J., {Lazio}, T. J.~W., \& {Kassim},
  N.~E. 2006, \aj, 132, 242

\bibitem[{{Novaco} \& {Brown}(1978)}]{Novaco1978}
{Novaco}, J.~C., \& {Brown}, L.~W. 1978, \apj, 221, 114

\bibitem[{{Odegard}(1986)}]{Odegard1986}
{Odegard}, N. 1986, \apj, 301, 813

\bibitem[{{Orlando} \& {Strong}(2013)}]{Orlando2013}
{Orlando}, E., \& {Strong}, A. 2013, \mnras, 436, 2127

\bibitem[{{Paladini} {et~al.}(2004){Paladini}, {Davies}, \& {De
  Zotti}}]{Paladini2004}
{Paladini}, R., {Davies}, R.~D., \& {De Zotti}, G. 2004, \mnras, 347, 237

\bibitem[{{Peterson} \& {Webber}(2002)}]{peterson2002interstellar}
{Peterson}, J.~D., \& {Webber}, W.~R. 2002, \apj, 575, 217

\bibitem[{{Planck Collaboration} {et~al.}(2016{\natexlab{a}}){Planck
  Collaboration}, {Adam}, {Ade}, {Aghanim}, {Akrami}, {Alves}, {Arg{\"u}eso},
  {Arnaud}, {Arroja}, {Ashdown}, {Aumont}, {Baccigalupi}, {Ballardini}, {Band
  ay}, {Barreiro}, {Bartlett}, {Bartolo}, {Basak}, {Battaglia}, {Battaner},
  {Battye}, {Benabed}, {Beno{\^\i}t}, {Benoit-L{\'e}vy}, {Bernard},
  {Bersanelli}, {Bertincourt}, {Bielewicz}, {Bikmaev}, {Bock}, {B{\"o}hringer},
  {Bonaldi}, {Bonavera}, {Bond}, {Borrill}, {Bouchet}, {Boulanger}, {Bucher},
  {Burenin}, {Burigana}, {Butler}, {Calabrese}, {Cardoso}, {Carvalho},
  {Casaponsa}, {Castex}, {Catalano}, {Challinor}, {Chamballu}, {Chary},
  {Chiang}, {Chluba}, {Chon}, {Christensen}, {Church}, {Clemens}, {Clements},
  {Colombi}, {Colombo}, {Combet}, {Comis}, {Contreras}, {Couchot}, {Coulais},
  {Crill}, {Cruz}, {Curto}, {Cuttaia}, {Danese}, {Davies}, {Davis}, {de
  Bernardis}, {de Rosa}, {de Zotti}, {Delabrouille}, {Delouis}, {D{\'e}sert},
  {Di Valentino}, {Dickinson}, {Diego}, {Dolag}, {Dole}, {Donzelli},
  {Dor{\'e}}, {Douspis}, {Ducout}, {Dunkley}, {Dupac}, {Efstathiou},
  {Eisenhardt}, {Elsner}, {En{\ss}lin}, {Eriksen}, {Falgarone}, {Fantaye},
  {Farhang}, {Feeney}, {Fergusson}, {Fernandez-Cobos}, {Feroz}, {Finelli},
  {Florido}, {Forni}, {Frailis}, {Fraisse}, {Franceschet}, {Franceschi},
  {Frejsel}, {Frolov}, {Galeotta}, {Galli}, {Ganga}, {Gauthier},
  {G{\'e}nova-Santos}, {Gerbino}, {Ghosh}, {Giard}, {Giraud-H{\'e}raud},
  {Giusarma}, {Gjerl{\o}w}, {Gonz{\'a}lez-Nuevo}, {G{\'o}rski}, {Grainge},
  {Gratton}, {Gregorio}, {Gruppuso}, {Gudmundsson}, {Hamann}, {Handley},
  {Hansen}, {Hanson}, {Harrison}, {Heavens}, {Helou}, {Henrot-Versill{\'e}},
  {Hern{\'a}ndez-Monteagudo}, {Herranz}, {Hildebrandt}, {Hivon}, {Hobson},
  {Holmes}, {Hornstrup}, {Hovest}, {Huang}, {Huffenberger}, {Hurier},
  {Ili{\'c}}, {Jaffe}, {Jaffe}, {Jin}, {Jones}, {Juvela}, {Karakci},
  {Keih{\"a}nen}, {Keskitalo}, {Khamitov}, {Kiiveri}, {Kim}, {Kisner},
  {Kneissl}, {Knoche}, {Knox}, {Krachmalnicoff}, {Kunz}, {Kurki-Suonio},
  {Lacasa}, {Lagache}, {L{\"a}hteenm{\"a}ki}, {Lamarre}, {Langer}, {Lasenby},
  {Lattanzi}, {Lawrence}, {Le Jeune}, {Leahy}, {Lellouch}, {Leonardi},
  {Le{\'o}n-Tavares}, {Lesgourgues}, {Levrier}, {Lewis}, {Liguori}, {Lilje},
  {Lilley}, {Linden-V{\o}rnle}, {Lindholm}, {Liu}, {L{\'o}pez-Caniego},
  {Lubin}, {Ma}, {Mac{\'\i}as-P{\'e}rez}, {Maggio}, {Maino}, {Mak},
  {Mandolesi}, {Mangilli}, {Marchini}, {Marcos-Caballero}, {Marinucci},
  {Maris}, {Marshall}, {Martin}, {Martinelli}, {Mart{\'\i}nez-Gonz{\'a}lez},
  {Masi}, {Matarrese}, {Mazzotta}, {McEwen}, {McGehee}, {Mei}, {Meinhold},
  {Melchiorri}, {Melin}, {Mendes}, {Mennella}, {Migliaccio}, {Mikkelsen},
  {Millea}, {Mitra}, {Miville-Desch{\^e}nes}, {Molinari}, {Moneti}, {Montier},
  {Moreno}, {Morgante}, {Mortlock}, {Moss}, {Mottet}, {M{\"u}nchmeyer},
  {Munshi}, {Murphy}, {Narimani}, {Naselsky}, {Nastasi}, {Nati}, {Natoli},
  {Negrello}, {Netterfield}, {N{\o}rgaard-Nielsen}, {Noviello}, {Novikov},
  {Novikov}, {Olamaie}, {Oppermann}, {Orlando}, {Oxborrow}, {Paci}, {Pagano},
  {Pajot}, {Paladini}, {Pandolfi}, {Paoletti}, {Partridge}, {Pasian},
  {Patanchon}, {Pearson}, {Peel}, {Peiris}, {Pelkonen}, {Perdereau}, {Perotto},
  {Perrott}, {Perrotta}, {Pettorino}, {Piacentini}, {Piat}, {Pierpaoli},
  {Pietrobon}, {Plaszczynski}, {Pogosyan}, {Pointecouteau}, {Polenta}, {Popa},
  {Pratt}, {Pr{\'e}zeau}, {Prunet}, {Puget}, {Rachen}, {Racine}, {Reach},
  {Rebolo}, {Reinecke}, {Remazeilles}, {Renault}, {Renzi}, {Ristorcelli},
  {Rocha}, {Roman}, {Romelli}, {Rosset}, {Rossetti}, {Rotti}, {Roudier},
  {Rouill{\'e} d'Orfeuil}, {Rowan-Robinson}, {Rubi{\~n}o-Mart{\'\i}n},
  {Ruiz-Granados}, {Rumsey}, {Rusholme}, {Said}, {Salvatelli}, {Salvati},
  {Sandri}, {Sanghera}, {Santos}, {Saunders}, {Sauv{\'e}}, {Savelainen},
  {Savini}, {Schaefer}, {Schammel}, {Scott}, {Seiffert}, {Serra}, {Shellard},
  {Shimwell}, {Shiraishi}, {Smith}, {Souradeep}, {Spencer}, {Spinelli},
  {Stanford}, {Stern}, {Stolyarov}, {Stompor}, {Strong}, {Sudiwala}, {Sunyaev},
  {Sutter}, {Sutton}, {Suur-Uski}, {Sygnet}, {Tauber}, {Tavagnacco}, {Terenzi},
  {Texier}, {Toffolatti}, {Tomasi}, {Tornikoski}, {Tramonte}, {Tristram},
  {Troja}, {Trombetti}, {Tucci}, {Tuovinen}, {T{\"u}rler}, {Umana},
  {Valenziano}, {Valiviita}, {Van Tent}, {Vassallo}, {Vibert}, {Vidal}, {Viel},
  {Vielva}, {Villa}, {Wade}, {Walter}, {Wand elt}, {Watson}, {Wehus},
  {Welikala}, {Weller}, {White}, {White}, {Wilkinson}, {Yvon}, {Zacchei},
  {Zibin}, \& {Zonca}}]{Planck2015_I}
{Planck Collaboration}, {Adam}, R., {Ade}, P.~A.~R., {et~al.}
  2016{\natexlab{a}}, \aap, 594, A1

\bibitem[{{Planck Collaboration} {et~al.}(2016{\natexlab{b}}){Planck
  Collaboration}, {Adam}, {Ade}, {Aghanim}, {Alves}, {Arnaud}, {Ashdown},
  {Aumont}, {Baccigalupi}, {Banday}, {Barreiro}, {Bartlett}, {Bartolo},
  {Battaner}, {Benabed}, {Beno{\^\i}t}, {Benoit-L{\'e}vy}, {Bernard},
  {Bersanelli}, {Bielewicz}, {Bock}, {Bonaldi}, {Bonavera}, {Bond}, {Borrill},
  {Bouchet}, {Boulanger}, {Bucher}, {Burigana}, {Butler}, {Calabrese},
  {Cardoso}, {Catalano}, {Challinor}, {Chamballu}, {Chary}, {Chiang},
  {Christensen}, {Clements}, {Colombi}, {Colombo}, {Combet}, {Couchot},
  {Coulais}, {Crill}, {Curto}, {Cuttaia}, {Danese}, {Davies}, {Davis}, {de
  Bernardis}, {de Rosa}, {de Zotti}, {Delabrouille}, {D{\'e}sert}, {Dickinson},
  {Diego}, {Dole}, {Donzelli}, {Dor{\'e}}, {Douspis}, {Ducout}, {Dupac},
  {Efstathiou}, {Elsner}, {En{\ss}lin}, {Eriksen}, {Falgarone}, {Fergusson},
  {Finelli}, {Forni}, {Frailis}, {Fraisse}, {Franceschi}, {Frejsel},
  {Galeotta}, {Galli}, {Ganga}, {Ghosh}, {Giard}, {Giraud-H{\'e}raud},
  {Gjerl{\o}w}, {Gonz{\'a}lez-Nuevo}, {G{\'o}rski}, {Gratton}, {Gregorio},
  {Gruppuso}, {Gudmundsson}, {Hansen}, {Hanson}, {Harrison}, {Helou},
  {Henrot-Versill{\'e}}, {Hern{\'a}ndez-Monteagudo}, {Herranz}, {Hildebrandt},
  {Hivon}, {Hobson}, {Holmes}, {Hornstrup}, {Hovest}, {Huffenberger}, {Hurier},
  {Jaffe}, {Jaffe}, {Jones}, {Juvela}, {Keih{\"a}nen}, {Keskitalo}, {Kisner},
  {Kneissl}, {Knoche}, {Kunz}, {Kurki-Suonio}, {Lagache},
  {L{\"a}hteenm{\"a}ki}, {Lamarre}, {Lasenby}, {Lattanzi}, {Lawrence}, {Le
  Jeune}, {Leahy}, {Leonardi}, {Lesgourgues}, {Levrier}, {Liguori}, {Lilje},
  {Linden-V{\o}rnle}, {L{\'o}pez-Caniego}, {Lubin}, {Mac{\'\i}as-P{\'e}rez},
  {Maggio}, {Maino}, {Mandolesi}, {Mangilli}, {Maris}, {Marshall}, {Martin},
  {Mart{\'\i}nez-Gonz{\'a}lez}, {Masi}, {Matarrese}, {McGehee}, {Meinhold},
  {Melchiorri}, {Mendes}, {Mennella}, {Migliaccio}, {Mitra},
  {Miville-Desch{\^e}nes}, {Moneti}, {Montier}, {Morgante}, {Mortlock}, {Moss},
  {Munshi}, {Murphy}, {Naselsky}, {Nati}, {Natoli}, {Netterfield},
  {N{\o}rgaard-Nielsen}, {Noviello}, {Novikov}, {Novikov}, {Orlando},
  {Oxborrow}, {Paci}, {Pagano}, {Pajot}, {Paladini}, {Paoletti}, {Partridge},
  {Pasian}, {Patanchon}, {Pearson}, {Perdereau}, {Perotto}, {Perrotta},
  {Pettorino}, {Piacentini}, {Piat}, {Pierpaoli}, {Pietrobon}, {Plaszczynski},
  {Pointecouteau}, {Polenta}, {Pratt}, {Pr{\'e}zeau}, {Prunet}, {Puget},
  {Rachen}, {Reach}, {Rebolo}, {Reinecke}, {Remazeilles}, {Renault}, {Renzi},
  {Ristorcelli}, {Rocha}, {Rosset}, {Rossetti}, {Roudier},
  {Rubi{\~n}o-Mart{\'\i}n}, {Rusholme}, {Sandri}, {Santos}, {Savelainen},
  {Savini}, {Scott}, {Seiffert}, {Shellard}, {Spencer}, {Stolyarov}, {Stompor},
  {Strong}, {Sudiwala}, {Sunyaev}, {Sutton}, {Suur-Uski}, {Sygnet}, {Tauber},
  {Terenzi}, {Toffolatti}, {Tomasi}, {Tristram}, {Tucci}, {Tuovinen}, {Umana},
  {Valenziano}, {Valiviita}, {Van Tent}, {Vielva}, {Villa}, {Wade}, {Wandelt},
  {Wehus}, {Wilkinson}, {Yvon}, {Zacchei}, \& {Zonca}}]{Planck2015_X}
---. 2016{\natexlab{b}}, \aap, 594, A10

\bibitem[{{Planck Collaboration} {et~al.}(2020){Planck Collaboration},
  {Aghanim}, {Akrami}, {Ashdown}, {Aumont}, {Baccigalupi}, {Ballardini},
  {Banday}, {Barreiro}, {Bartolo}, {Basak}, {Battye}, {Benabed}, {Bernard},
  {Bersanelli}, {Bielewicz}, {Bock}, {Bond}, {Borrill}, {Bouchet}, {Boulanger},
  {Bucher}, {Burigana}, {Butler}, {Calabrese}, {Cardoso}, {Carron},
  {Challinor}, {Chiang}, {Chluba}, {Colombo}, {Combet}, {Contreras}, {Crill},
  {Cuttaia}, {de Bernardis}, {de Zotti}, {Delabrouille}, {Delouis}, {Di
  Valentino}, {Diego}, {Dor{\'e}}, {Douspis}, {Ducout}, {Dupac}, {Dusini},
  {Efstathiou}, {Elsner}, {En{\ss}lin}, {Eriksen}, {Fantaye}, {Farhang},
  {Fergusson}, {Fernandez-Cobos}, {Finelli}, {Forastieri}, {Frailis},
  {Fraisse}, {Franceschi}, {Frolov}, {Galeotta}, {Galli}, {Ganga},
  {G{\'e}nova-Santos}, {Gerbino}, {Ghosh}, {Gonz{\'a}lez-Nuevo}, {G{\'o}rski},
  {Gratton}, {Gruppuso}, {Gudmundsson}, {Hamann}, {Handley}, {Hansen},
  {Herranz}, {Hildebrandt}, {Hivon}, {Huang}, {Jaffe}, {Jones}, {Karakci},
  {Keih{\"a}nen}, {Keskitalo}, {Kiiveri}, {Kim}, {Kisner}, {Knox},
  {Krachmalnicoff}, {Kunz}, {Kurki-Suonio}, {Lagache}, {Lamarre}, {Lasenby},
  {Lattanzi}, {Lawrence}, {Le Jeune}, {Lemos}, {Lesgourgues}, {Levrier},
  {Lewis}, {Liguori}, {Lilje}, {Lilley}, {Lindholm}, {L{\'o}pez-Caniego},
  {Lubin}, {Ma}, {Mac{\'\i}as-P{\'e}rez}, {Maggio}, {Maino}, {Mandolesi},
  {Mangilli}, {Marcos-Caballero}, {Maris}, {Martin}, {Martinelli},
  {Mart{\'\i}nez-Gonz{\'a}lez}, {Matarrese}, {Mauri}, {McEwen}, {Meinhold},
  {Melchiorri}, {Mennella}, {Migliaccio}, {Millea}, {Mitra},
  {Miville-Desch{\^e}nes}, {Molinari}, {Montier}, {Morgante}, {Moss}, {Natoli},
  {N{\o}rgaard-Nielsen}, {Pagano}, {Paoletti}, {Partridge}, {Patanchon},
  {Peiris}, {Perrotta}, {Pettorino}, {Piacentini}, {Polastri}, {Polenta},
  {Puget}, {Rachen}, {Reinecke}, {Remazeilles}, {Renzi}, {Rocha}, {Rosset},
  {Roudier}, {Rubi{\~n}o-Mart{\'\i}n}, {Ruiz-Granados}, {Salvati}, {Sandri},
  {Savelainen}, {Scott}, {Shellard}, {Sirignano}, {Sirri}, {Spencer},
  {Sunyaev}, {Suur-Uski}, {Tauber}, {Tavagnacco}, {Tenti}, {Toffolatti},
  {Tomasi}, {Trombetti}, {Valenziano}, {Valiviita}, {Van Tent}, {Vibert},
  {Vielva}, {Villa}, {Vittorio}, {Wandelt}, {Wehus}, {White}, {White},
  {Zacchei}, \& {Zonca}}]{Planck2018_VI}
{Planck Collaboration}, {Aghanim}, N., {Akrami}, Y., {et~al.} 2020, \aap, 641,
  A6

\bibitem[{{Platania} {et~al.}(1998){Platania}, {Bensadoun}, {Bersanelli}, {De
  Amici}, {Kogut}, {Levin}, {Maino}, \& {Smoot}}]{platania1998determination}
{Platania}, P., {Bensadoun}, M., {Bersanelli}, M., {et~al.} 1998, \apj, 505,
  473

\bibitem[{{Polderman} {et~al.}(2020){Polderman}, {Haverkorn}, \&
  {Jaffe}}]{Polderman2020}
{Polderman}, I.~M., {Haverkorn}, M., \& {Jaffe}, T.~R. 2020, \aap, 636, A2

\bibitem[{{Polderman} {et~al.}(2019){Polderman}, {Haverkorn}, {Jaffe}, \&
  {Alves}}]{Polderman2019}
{Polderman}, I.~M., {Haverkorn}, M., {Jaffe}, T.~R., \& {Alves}, M.~I.~R. 2019,
  \aap, 621, A127

\bibitem[{{Potgieter} \& {Nndanganeni}(2013)}]{Potgieter2013}
{Potgieter}, M.~S., \& {Nndanganeni}, R.~R. 2013, \aph, 48, 25

\bibitem[{{Protheroe} \& {Biermann}(1996)}]{Protheroe1996}
{Protheroe}, R.~J., \& {Biermann}, P.~L. 1996, \aph, 6, 45

\bibitem[{{Reich} \& {Reich}(1986)}]{Reich1986}
{Reich}, P., \& {Reich}, W. 1986, \aaps, 63, 205

\bibitem[{{Reich} {et~al.}(2001){Reich}, {Testori}, \& {Reich}}]{Reich2001}
{Reich}, P., {Testori}, J.~C., \& {Reich}, W. 2001, \aap, 376, 861

\bibitem[{{Reich}(1982)}]{Reich1982}
{Reich}, W. 1982, \aaps, 48, 219

\bibitem[{{Remazeilles} {et~al.}(2015){Remazeilles}, {Dickinson}, {Banday},
  {Bigot-Sazy}, \& {Ghosh}}]{remazeilles2015improved}
{Remazeilles}, M., {Dickinson}, C., {Banday}, A.~J., {Bigot-Sazy}, M.~A., \&
  {Ghosh}, T. 2015, \mnras, 451, 4311

\bibitem[{{Reynolds}(1990)}]{Reynolds1990}
{Reynolds}, R.~J. 1990, {The Low Density Ionized Component of the Interstellar
  Medium and Free-Free Absorption at High Galactic Latitudes}, ed. N.~E.
  {Kassim} \& K.~W. {Weiler}, Vol. 362, 121

\bibitem[{{Roger} {et~al.}(1999){Roger}, {Costain}, {Landecker}, \&
  {Swerdlyk}}]{roger1999radio}
{Roger}, R.~S., {Costain}, C.~H., {Landecker}, T.~L., \& {Swerdlyk}, C.~M.
  1999, \aaps, 137, 7

\bibitem[{{Roshi} \& {Anantharamaiah}(2001)}]{Roshi2001}
{Roshi}, D.~A., \& {Anantharamaiah}, K.~R. 2001, \apj, 557, 226

\bibitem[{{Sathyanarayana Rao} {et~al.}(2017){Sathyanarayana Rao},
  {Subrahmanyan}, {Udaya Shankar}, \& {Chluba}}]{GMOSS2017}
{Sathyanarayana Rao}, M., {Subrahmanyan}, R., {Udaya Shankar}, N., \& {Chluba},
  J. 2017, \aj, 153, 26

\bibitem[{{Scheuer} \& {Ryle}(1953)}]{scheuer1953}
{Scheuer}, P.~A.~G., \& {Ryle}, M. 1953, \mnras, 113, 3

\bibitem[{{Schnitzeler}(2012)}]{schnitzeler2012modelling}
{Schnitzeler}, D.~H.~F.~M. 2012, \mnras, 427, 664

\bibitem[{{Seiffert} {et~al.}(2011){Seiffert}, {Fixsen}, {Kogut}, {Levin},
  {Limon}, {Lubin}, {Mirel}, {Singal}, {Villela}, {Wollack}, \&
  {Wuensche}}]{seiffert2011interpretation}
{Seiffert}, M., {Fixsen}, D.~J., {Kogut}, A., {et~al.} 2011, \apj, 734, 6

\bibitem[{{Shain}(1959)}]{shain1959iau}
{Shain}, C.~A. 1959, in 1959IAUS~~~~9, ed. R.~N. {Bracewell}, 451

\bibitem[{{Shaw} {et~al.}(2014){Shaw}, {Sigurdson}, {Pen}, {Stebbins}, \&
  {Sitwell}}]{Shaw2014}
{Shaw}, J.~R., {Sigurdson}, K., {Pen}, U.-L., {Stebbins}, A., \& {Sitwell}, M.
  2014, \apj, 781, 57

\bibitem[{{Smith}(1965)}]{Smith1965}
{Smith}, F.~G. 1965, \mnras, 131, 145

\bibitem[{{Strong} \& {Moskalenko}(1998{\natexlab{a}})}]{strong1998propagation}
{Strong}, A.~W., \& {Moskalenko}, I.~V. 1998{\natexlab{a}}, \apj, 509, 212

\bibitem[{{Strong} \& {Moskalenko}(1998{\natexlab{b}})}]{Strong1998}
---. 1998{\natexlab{b}}, \apj, 509, 212

\bibitem[{{Su} {et~al.}(2017){Su}, {Hurley-Walker}, {Jackson},
  {McClure-Griffiths}, {Tingay}, {Hindson}, {Hancock}, {Wayth}, {Gaensler},
  {Staveley-Smith}, {Morgan}, {Johnston-Hollitt}, {Lenc}, {Bell}, {Callingham},
  {Dwarkanath}, {For}, {Kapi{\'n}ska}, {McKinley}, {Offringa}, {Procopio},
  {Wu}, \& {Zheng}}]{Su2017}
{Su}, H., {Hurley-Walker}, N., {Jackson}, C.~A., {et~al.} 2017, \mnras, 465,
  3163

\bibitem[{{Su} {et~al.}(2018){Su}, {Macquart}, {Hurley-Walker},
  {McClure-Griffiths}, {Jackson}, {Tingay}, {Tian}, {Gaensler}, {McKinley},
  {Kapi{\'n}ska}, {Hindson}, {Hancock}, {Wayth}, {Staveley-Smith}, {Morgan},
  {Johnston-Hollitt}, {Lenc}, {Bell}, {Callingham}, {Dwarkanath}, {For},
  {Offringa}, {Procopio}, {Wu}, \& {Zheng}}]{Su2018}
{Su}, H., {Macquart}, J.~P., {Hurley-Walker}, N., {et~al.} 2018, \mnras, 479,
  4041

\bibitem[{{Subrahmanyan} \& {Cowsik}(2013)}]{Subrahmanyan2013}
{Subrahmanyan}, R., \& {Cowsik}, R. 2013, \apj, 776, 42

\bibitem[{{Sun} {et~al.}(2008){Sun}, {Reich}, {Waelkens}, \&
  {En{\ss}lin}}]{sun2008radio}
{Sun}, X.~H., {Reich}, W., {Waelkens}, A., \& {En{\ss}lin}, T.~A. 2008, \aap,
  477, 573

\bibitem[{{Taylor} \& {Cordes}(1993)}]{taylor1993pulsar}
{Taylor}, J.~H., \& {Cordes}, J.~M. 1993, \apj, 411, 674

\bibitem[{{Trotta} {et~al.}(2011){Trotta}, {J{\'o}hannesson}, {Moskalenko},
  {Porter}, {Ruiz de Austri}, \& {Strong}}]{Trotta2011}
{Trotta}, R., {J{\'o}hannesson}, G., {Moskalenko}, I.~V., {et~al.} 2011, \apj,
  729, 106

\bibitem[{{van de Voort} {et~al.}(2019){van de Voort}, {Springel}, {Mandelker},
  {van den Bosch}, \& {Pakmor}}]{Voort2018}
{van de Voort}, F., {Springel}, V., {Mandelker}, N., {van den Bosch}, F.~C., \&
  {Pakmor}, R. 2019, \mnras, 482, L85

\bibitem[{{Webber} {et~al.}(2008){Webber}, {Cummings}, {McDonald}, {Stone},
  {Heikkila}, \& {Lal}}]{webber2008limits}
{Webber}, W.~R., {Cummings}, A.~C., {McDonald}, F.~B., {et~al.} 2008, \jgra,
  113, A10108

\bibitem[{{Woermann} {et~al.}(2001){Woermann}, {Gaylard}, \&
  {Otrupcek}}]{woermann2001kinematics}
{Woermann}, B., {Gaylard}, M.~J., \& {Otrupcek}, R. 2001, \mnras, 325, 1213

\bibitem[{{Wolleben}(2007)}]{wolleben2007new}
{Wolleben}, M. 2007, \apj, 664, 349

\bibitem[{{Yao} {et~al.}(2017){Yao}, {Manchester}, \& {Wang}}]{YMW16}
{Yao}, J.~M., {Manchester}, R.~N., \& {Wang}, N. 2017, \apj, 835, 29

\bibitem[{{Zheng} {et~al.}(2017){Zheng}, {Tegmark}, {Dillon}, {Kim}, {Liu},
  {Neben}, {Jonas}, {Reich}, \& {Reich}}]{Zheng2017}
{Zheng}, H., {Tegmark}, M., {Dillon}, J.~S., {et~al.} 2017, \mnras, 464, 3486

\bibitem[{{Zuo} {et~al.}(2019){Zuo}, {Chen}, {Ansari}, \& {Lu}}]{Zuo2019}
{Zuo}, S., {Chen}, X., {Ansari}, R., \& {Lu}, Y. 2019, \aj, 157, 4

\end{thebibliography}

\end{document}